\documentclass[12pt]{elsarticle} 

\usepackage{amssymb}

\usepackage{amsmath}
\usepackage{amsfonts}

\usepackage{enumerate}
\usepackage{here}
\usepackage{indentfirst}
\usepackage{bm}
\usepackage{url}
\usepackage{color}
\usepackage{ulem}
\usepackage{here}

\usepackage{caption}
\usepackage[subrefformat=parens]{subcaption}

\newcommand{\Add}[1]{{\color{black}#1}}
\newcommand{\Erase}[1]{}	

\newcommand{\EraseRe}[1]{}

\def\XXint#1#2#3{{\setbox0=\hbox{$#1{#2#3}{\int}$ }
		\vcenter{\hbox{$#2#3$ }}\kern-.6\wd0}}

\def\x{\bm{x}}
\def\d{\text{d}}

\def\eq#1{Eq. \eqref{eq:#1}}
\def\TD{\mathcal{D}}
\def\Lag{\mathcal{L}}
\def\and{\text{~and~}}
\def\strain{\mathcal{E}}

\newcommand{\argmin}{\mathop{\rm arg~min}\limits}

\journal{Computer Methods in Applied Mechanics and Engineering,~}

\begin{document}

\begin{frontmatter}



\title{Orientation Optimization Based on \\Topological Derivatives in Cooperation with \\Multi-Material Topology Optimization \\Based on Extended Level Set Method}

\author[label1]{Masaki~Noda}
\author[label1]{Kei~Matsushima}
\author[label1,label2]{Takayuki~Yamada \corref{cor1}}
\ead{t.yamada@mech.t.u-tokyo.ac.jp}
\cortext[cor1]{Corresponding author.
Tel.: +81-3-5841-0294;
Fax: +81-3-5841-0294.}

\address[label1]{Department of Mechanical Engineering, Graduate School of Engineering, The University of Tokyo, Yayoi 2-11-16, Bunkyo-ku, Tokyo 113-8656, Japan.}

\address[label2]{Department of Strategic Studies, Institute of Engineering Innovation, Graduate School of Engineering, The University of Tokyo, Yayoi 2-11-16, Bunkyo-ku, Tokyo 113-8656, Japan.}

\begin{abstract}
	This paper provides an orientation angle optimization method for the design of fiber-reinforced composite materials using topology optimization.
    The orientation angle optimization is based on a topological derivative, which measures the sensitivity of an objective function with respect to a topological change of anisotropic materials.
	The sensitivity is incorporated into a new gradient-based optimization algorithm.
	This method allows us to avoid local optima and seek a global optimal solution.
	We provide some numerical examples and verify the effectiveness of the proposed method.
\end{abstract}


\end{frontmatter}


\section{Introduction}\label{sec:4 Intro}
In recent years, multi-material design has been attracting attention in automobile and aircraft design, in which multiple materials are placed in the appropriate places to achieve higher mechanical performance. High performance can be achieved especially by combining conventional metal materials with fiber-reinforced materials, which have higher stiffness and strength per weight \cite{soutis2005carbon}.
Extensive research has been conducted on fiber-reinforced materials, including investigations into their material properties \cite{rajak2019fiber, kasahara2021evaluation, hashin1990thermoelastic}, manufacturing technologies \cite{riveiro2012experimental, tekinalp2014highly, geier2019advanced}, bonding techniques with metallic materials \cite{balle2009ultrasonic, goushegir2014friction, kah2014techniques}, and recycling methods \cite{pimenta2011recycling}.

While the use of multi-material design and fiber-reinforced materials can greatly improve mechanical performance, it also presents some challenges. These multi-material configurations are more complex to design than conventional structures made with a single material, which can increase design costs significantly. Additionally, fiber-reinforced materials are anisotropic, meaning that their mechanical properties are directionally dependent. This means that the orientation angle of the fibers must be carefully determined in order to maximize the benefits of utilizing fiber-reinforced materials.

To solve these problems, structural optimization methods have been used to design structures based on mathematical modeling and numerical techniques. Topology optimization, one of the structural optimization methods, offers the greatest design flexibility and allows radical optimization, including the presence or absence of holes \cite{bendsoe1988generating,sigmund1997design,allaire2004structural,yamada2010topology}.
Topology optimization can be broadly classified into three types based on how the structure is represented: homogenization, density, and level set methods.
Homogenization methods regard infinitely fine structures as the optimal solution, in theory, \cite{bendsoe1988generating}, while density methods allow for an ideal material with an intermediate density as the optimal solution \cite{bendsoe1989optimal}. On the other hand, the level-set method has the advantage of obtaining smooth structures with distinct boundaries \cite{allaire2004structural, yamada2010topology}. It has been attracting attention in recent years and applied to several problems, e.g., elasticity \cite{emmendoerfer2016topology, emmendoerfer2020stress}, thermal \cite{yamada2011level, jahangiry2019combination}, acoustic \cite{isakari2014topology, isakari2017level, lanznaster2021level}, electromagnetic \cite{otomori2012topology, jung2021multi,matsushima2022unidirectional}, fluid \cite{yaji2014topology}, current control \cite{fujii2019dc}, thermoelectric \cite{fujii2019optimizing}, manufacturability \cite{sato2017manufacturability, hur2020level, yamada2022topology,feng2023multi}, and other optimal design problems.

In addition, recent studies have proposed multi-material design methods based on level set-based topology optimization.
The piecewise-constant level set method is a method to represent multiple regions using a single level set function, providing a simple multiphase representation with minimal data requirements \cite{wei2009piecewise,luo2009design}. Color level set and multi-material level set methods use multiple level set functions to represent multiple phases \cite{Wang2004color, Wang2015multi, cui2016level, kishimoto2017optimal, onco2020robust}. The vector-valued level set method is a method to represent multiple materials using different value ranges of a vector-valued function and has the advantage of being less likely to fall into local optima \cite{gangl2020multi, MasakiNODA202120-00412}.
The topology optimization method based on X-LS (EXtended Level Set) uses multiple level set functions that correspond to the combination of two materials to represent the boundaries between multiple materials. By considering topological changes between all materials, this method can reduce the possibility of local optima. Furthermore, as each boundary is represented by an individual level set function, it enables convenient individual control of boundary shape complexity \cite{noda2022extended, feng2023multi}.



%
Several studies have been undertaken to examine the orientation of anisotropic materials and their concurrent optimization in conjunction with topology.
\cite{pedersen1989optimal, bendsoe1994analytical, nomura2015general, nomura2019inverse}. 
In combination with multi-material topology optimization, topology and orientation optimization of multiple materials, including anisotropic materials, has also been performed \cite{stegmann2005discrete, desai2021topological}.
A comprehensive review of the literature on the optimization of fiber-reinforced materials can be found in \cite{nikbakt2018review}.
Among these, 
methods that attempt to derive the optimal solution algebraically can be very difficult to achieve, depending on the complexity of the optimization problem \cite{pedersen1989optimal, bendsoe1994analytical, desai2021topological}.
Genetic algorithms possess the potential to yield an optimal solution if only provided that the evaluation function is ascertained; however, they necessitate a substantial quantity of numerical analyses to be performed prior to obtaining a satisfactory optimal outcome.

The gradient-based methods obtain an optimal solution by iteratively updating the design variables based on the gradient of the objective function with respect to the design variables. Derivation of the gradient is easier than deriving the optimal solution, and sensitivity analysis using the adjoint variable method has the advantage of converging to the optimal solution in a relatively small number of numerical analyses.
Some studies have proposed optimization methods for the orientation direction such as continuous fiber angle optimization (CFAO) \cite{bruyneel2002composite, setoodeh2005combined}, discrete material optimization (DMO) \cite{stegmann2005discrete, yu2020topology, noda2022domain}, vector field- or tensor field-based optimization methods \cite{nomura2015general, nomura2019inverse}.
CFAO is a simple method that updates the orientation based on the gradient of the objective function with respect to the orientation; however, Steagmann and Mund pointed out the drawback that CFAO is prone to local optima \cite{stegmann2005discrete}. 
DMO is a very practical method that reduces the possibility of local optimums, but the distribution of orientations obtained is piecewise constant, and discrete.
Vector field- or tensor field-based optimization methods require
introducing intermediate material constants, which may lead to a local optimum when these constants cannot be set appropriately.
The above methods treat the orientation as
scalar, vector, or tensor fields distributed in space.
Some research defines the orientation in terms of the streamlines and optimizes the line shapes \cite{shimoda2023shape}.
Although such a method has the advantage of obtaining an optimal solution with continuous fiber lines and high engineering practicality, it is very difficult to avoid local optimums.


This study proposes a new optimization method to design multi-material topology and fiber orientation.
In the proposed method, the orientation is optimized by a topological derivative with respect to a topological perturbation from an anisotropic material to another anisotropic material. 
By taking into account the topological derivative for all orientations, this method effectively circumvents local optima.
Also, the topology of multiple materials is optimized based on the X-LS method.

In the following sections, the design problem is first formulated in terms of a domain representation using the X-LS method and an orientation representation using vector-valued variables. Next, design sensitivity is constructed based on topological derivatives. The numerical implementation of the optimization is then described. Finally, the validity and usefulness of the proposed method are verified via the stiffness maximization problem.

\section{Formulation}\label{sec:4 Formula}
This section describes the formulation of the orientation optimization problem. In this optimization, the orientation distribution, which defines the attributes of the structure composed of anisotropic materials, is the design variable to be optimized. The domain of each material is also optimized simultaneously.
This formulation is widely applicable to optimizing elastic materials, solving heat conduction problems, and mitigating electromagnetic problems. The numerical examples presented later in this paper focus on the problem of linear elasticity to demonstrate the specific utility of this method. For simplicity and clarity, the discussion that follows will be limited to the two-dimensional problem.

\subsection{Optimization problem}
\begin{figure}
	\centering
	\includegraphics[width=8cm]{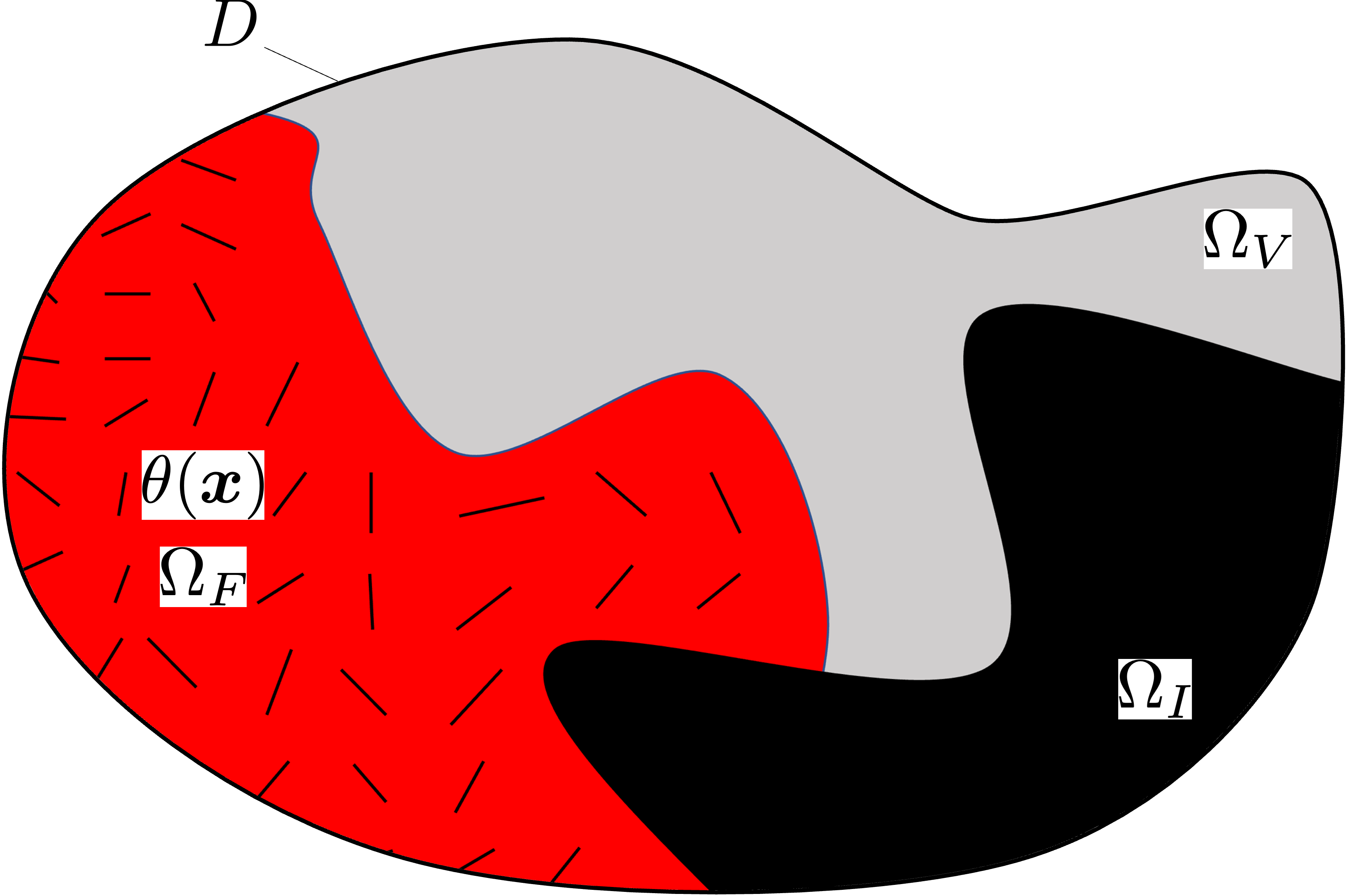}
	\caption{
		Optimizing domains and orientation.
	}
	\label{fig:4domains angle}       
\end{figure}

The general optimization problem of domains and orientation is formulated in a two-dimensional case as follows:
\begin{eqnarray}
	\inf_{\bm\Omega,\theta} &J(\bm\Omega,\theta, U),&\\
	\text{subject~to} &G(\bm\Omega,\theta, U)&=0,\\
	&g_k(\bm\Omega,\theta, U)&\le 0 ~~\forall k,\\
	&h_l(\bm\Omega,\theta, U)&= 0 ~~\forall l,\\
	&\Omega_V\cup\Omega_I\cup\Omega_F&=D, \label{eq:4chi constraint or}\\
	&\Omega_a\cap\Omega_b&=\varnothing\text{~for~} b\ne a ,\label{eq:4chi constraint and}
\end{eqnarray}
where $\bm\Omega$ denotes the domains $\Omega_V,\Omega_I,\Omega_F\subset D$ for a void, isotropic material, and fiber-reinforced material, respectively,
and $\theta(\x)$ is the direction of fiber orientation at each position $\x$. For $k=1,2,\ldots,n^\text{ineq}$ and $l= 1,2,\ldots,n^\text{eq}$, the functionals $g_k$ and $h_l$ denote inequality and equality constraint functions, respectively, where $n^\text{ineq}$ and $n^\text{eq}$ are the number of constraints.
Constraint equations (\ref{eq:4chi constraint or}) and (\ref{eq:4chi constraint and}) are necessary for every position in the design domain $D$ to be assigned to one and only one of the domains $\Omega_V,\Omega_I,\Omega_F\subset D$.
Fig.~\ref{fig:4domains angle} shows an example of a design domain $D$, domains $\Omega_V,\Omega_I,\Omega_F\subset D$, and the distribution of orientation $\theta(\x)$.

\subsection{Representation of multi-material topology and orientation}
In this study, we use the X-LS (EXtended Level Set) method\cite{noda2022extended} to represent the domains $\Omega_V,\Omega_I,\and \Omega_F$. 
In the X-LS method, we first define the extended level set functions $\phi_{ab}$ for each combination of $a\in\{V,I,F\}$ and $b\in\{V,I,F\}$ with $a\neq b$, which implicitly represent the domains as follows: 
\begin{align}
	\chi_V
	=H(\phi_{VI})H(\phi_{VF}),\\
	\chi_I
	=H(\phi_{IV})H(\phi_{IF}),\\
	\chi_F
	=H(\phi_{FI})H(\phi_{FV}),
\end{align}
where $\chi_m~(m\in\{V,I,F\})$ are the characteristic functions defined as follows:
\begin{align}
	\chi_m
	&=\begin{cases}
		1~~\x \in \Omega_m\\
		0~~\x \notin \Omega_m\\
	\end{cases},
\end{align}
and $H(s)$ is the Heaviside function defined as follows:
\begin{align}
	H(s)
	&=\begin{cases}
		1~~s\ge0\\
		0~~s<0\\
	\end{cases}.
\end{align}

To continuously represent the orientation, which is a periodic parameter, we use the relaxed Cartesian representation, which is a method to represent an angle \cite{nomura2015general}.
In the relaxed Cartesian representation, the orientation of anisotropic materials $\theta\in[0,\pi/n_\text{sym})$ ($n_\text{sym}$ is the number of symmetricity) are represented by auxiliary variables $\xi$ and $\eta$ as follows:
\begin{align}
	\theta=\frac{1}{2 n_\text{sym}}
	\arctan\left({\eta}/{\xi}\right ),\label{eq:4theta}\\
	\xi^2+\eta^2\le 1.\label{eq:4theta const}
\end{align}
%
If the constraint is replaced with $\xi^2+\eta^2 = 1$, that representation is the ``Cartesian representation" of the orientation. Relaxed formulation enables a smoother transition of the auxiliary variables, thus design variables can be changed gradually. Therefore, an oscillating optimization process can be avoided.
%
%
In the following, we consider the number to be set to $n_\text{sym}=1$ for simplicity.

\subsection*{Remark}
To consider the cases $n_\text{sym}\ge 2$, we only need to insert $n_\text{sym}$ appropriately in the following formulations.

\section{Constructions of Sensitivity}
The topology optimization problem is difficult to solve directly and thus is solved iteratively in this paper. That is, initial values of design variables are given and updated using the sensitivity to improve the objective function. In this section, the sensitivity is constructed.

\subsection{Multi-material topological derivative considering anisotropic materials}

\begin{figure}
	\centering
	\includegraphics[width=0.7\linewidth]{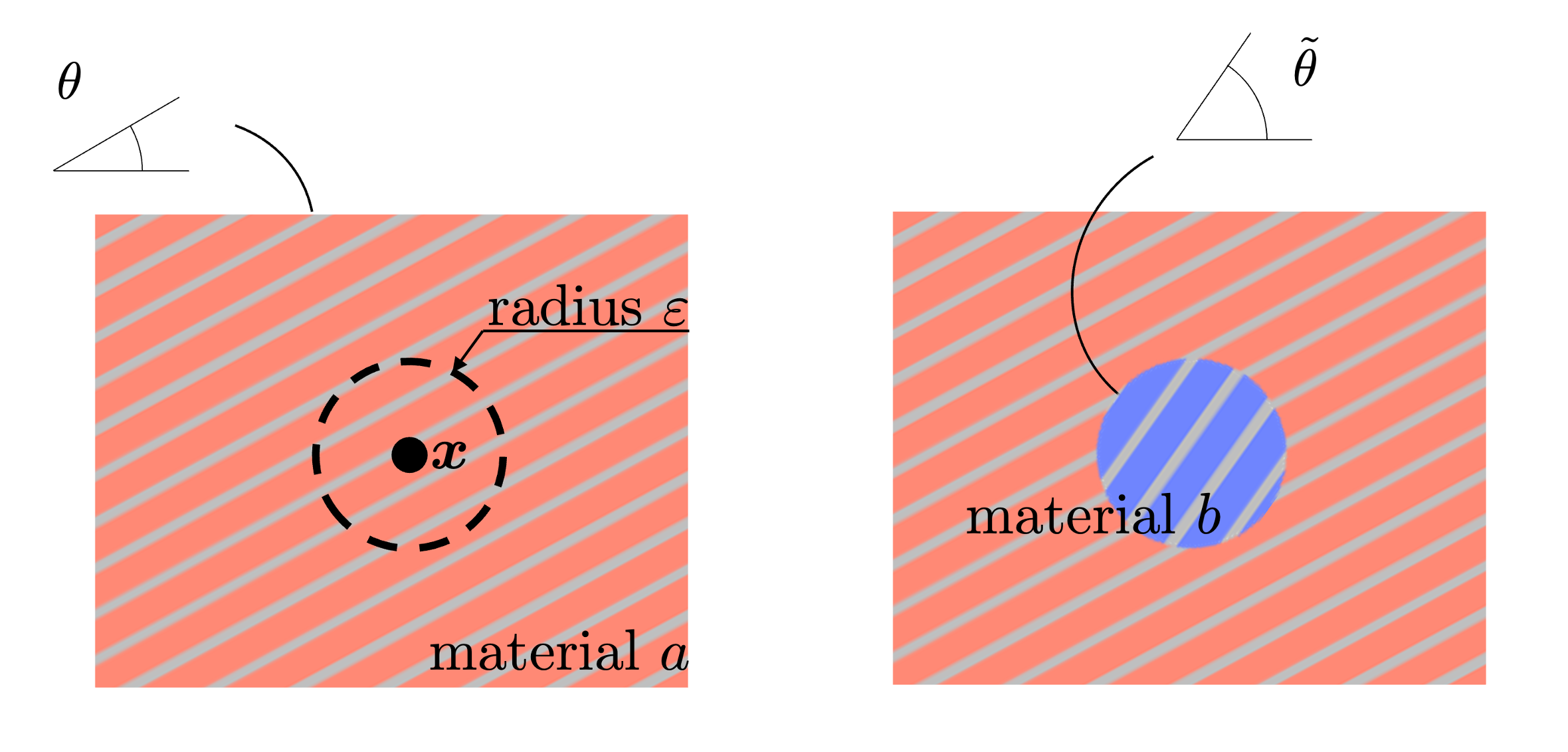}
	\caption{Illustration of multi-material topological derivative considering anisotropic materials. Left: material $a$ oriented in angular $\theta$ without inclusion. Right: with the inclusion of material $b$ oriented in angular $\tilde\theta$ with a circular area of center $\x$, radius $\varepsilon$.}
	\label{fig:4anisotd}
\end{figure}
Here, we define the multi-material topological derivative considering anisotropic material.
The topological derivative is a sensitivity to insert an infinitesimal small circular domain with one material into another material domain.
The topological derivative 
$D_{a\to b,\tilde\theta}J(\x)$
for material $a$ oriented in $\theta(\x)$ to material $b$ oriented in $\tilde\theta$ of objective function $J$ at position $\x$ is defined as follows:
\begin{align}
	D_{a\to b,\tilde\theta}J(\x)
	=\begin{cases}
		\lim_{\varepsilon\to 0}\dfrac{J_{b,\tilde\theta}(\Omega_{\varepsilon,\x})-J}{\text{meas}(\Omega_{\varepsilon,\x})}&\text{~for~}\x\in \Omega_a\\
		0&\text{~for~}\x\notin \Omega_a\\
	\end{cases},
	\label{eq:4def aniso td}
\end{align}
where $J_{b,\tilde\theta}(\Omega_{\varepsilon,\x})$ is the objective function when a small domain $\Omega_\varepsilon$, whose center is $\x$ and radius is $\varepsilon$, is replaced from material $a$ to material $b$ oriented in $\tilde\theta$, as shown in Fig.~\ref{fig:4anisotd}. The function $\text{meas}(\omega)$ is the measure of the domain $\omega$, that is, the area in two-dimensional cases.
If the inserted material $b$ is isotropic, the topological derivative is independent of orientation $\tilde\theta$, so we abbreviate the notation as follows:
\begin{align}
	D_{a\to b}J(\x)
	&\equiv D_{a\to b,\tilde\theta}J(\x)~~~\text{if}~ b\in \{I,V\}.
\end{align}

\subsection{Optimally oriented topological derivative}
An optimally oriented topological derivative is defined as follows:
\begin{align}
	D_{a\to b}^*J(\x)
	=
	\min_{\tilde\theta}D_{a\to b,\tilde\theta}J(\x).\label{eq:4td to theta}
\end{align}
Eq.~\eqref{eq:4td to theta} is meant to evaluate the phase derivative when an anisotropic material inclusion with optimal orientation is inserted.

Note that the sensitivity to multiple objective and constraint functions cannot be linearly combined because the topological derivatives in this definition involve a 
minimizing calculation.
For example, 
\begin{align}
	D_{a\to b}^*(J_1+J_2)
	\ge D_{a\to b}^*J_1+D_{a\to b}^*J_2.
\end{align}

In addition, if the inserted material is isotropic, the optimally oriented topological derivative coincides with the multi-material topological derivative defined in Eq.~\eqref{eq:4def aniso td}, as follows:
\begin{align}
	D_{a\to b}^*J(\x)
	\equiv
	D_{a\to b}J(\x)~~~\text{if}~ b\in \{I,V\}.
\end{align}


\subsection{Update of level set function}\label{sec:4Update of level set function}
In this study, the topology optimization is performed using the concept of topological derivative, as follows:
\begin{align}
	\phi_{ab}^\text{new} + \tau_{ab}\nabla^2 \phi_{ab}^\text{new}
	=\phi_{ab}-\alpha_{ab}\TD_{ab}\Lag,
\end{align}
where $\phi_{ab}^\text{new}$ are the updated level set functions, the term
$\tau_{ab}\nabla^2\phi_{ab}^\text{new}$ is the regularization term to avoid oscillatory distribution of $\phi_{ab}^\text{new}$ (discussed in \cite{yamada2010topology,sigmund2013topology}), $\Lag$ is the the Lagrangian. 
The coefficient $\alpha_{ab}$ is the step size, and $\TD_{ab}\Lag$ is the extended topological derivative, which is the sensitivity for the level set function $\phi_{ab}$, defined as follows:
\begin{align}
	\TD_{ab}\Lag=-\TD_{ba}\Lag= D_{a\to b}^*\Lag - D_{b\to a}^*\Lag + D_{c\to b}^*\Lag - D_{c\to a}^*\Lag,
\end{align}
where $c\in\{I,V,F\}\backslash \{a,b\}$.




\subsection{Update of orientations}
\begin{figure}
	\centering
	
	\begin{minipage}[t]{0.4\linewidth}
		\centering
		\includegraphics[width=\linewidth]{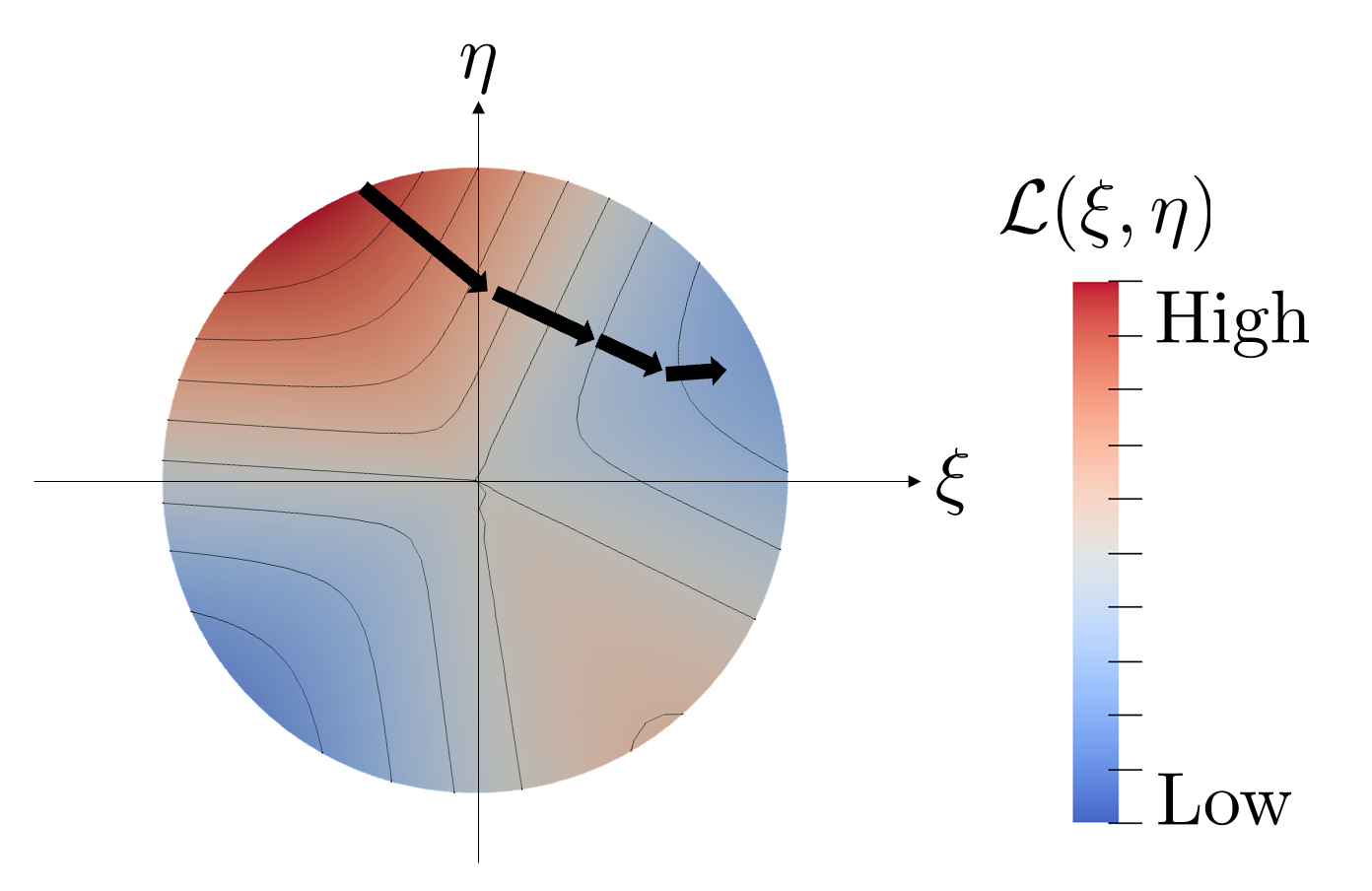}
		\caption{Update using a gradient, which can fail into local optima.}
		\label{fig:4update-nabla}
	\end{minipage}
	\qquad
	\begin{minipage}[t]{0.4\linewidth}
		\centering
		\includegraphics[width=\linewidth]{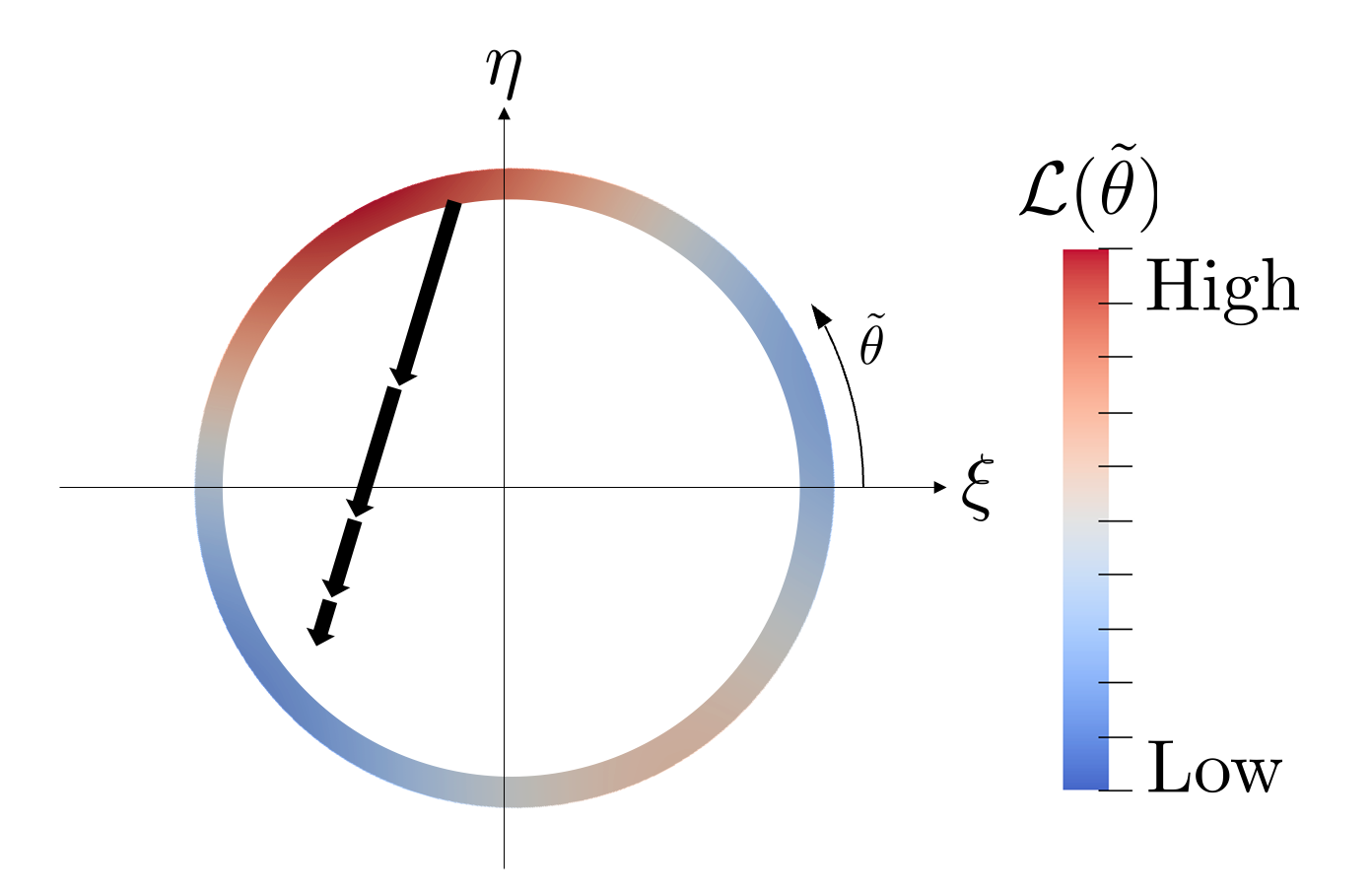}
		\caption{Update using estimated optimal orientation.}
		\label{fig:4update-thetaopt}
	\end{minipage}
	
	\end{figure}
	In this study, we represent orientations of anisotropic materials using design variables $\xi$ and $\eta$ defined in \eq{4theta}.
	A possible method to optimize these values is to update them using a gradient-based method as follows:
	\begin{align}
		\xi^\text{new}=\xi+\alpha_\theta\frac{\partial \Lag}{\partial \xi},\label{eq:4xi-ordinal-update}\\
		\eta^\text{new}=\eta+\alpha_\theta\frac{\partial \Lag}{\partial \eta},\label{eq:4eta-ordinal-update}
	\end{align}
	where $\Lag$ is the Lagrangian and $\alpha_\theta$ is the step size.
	However, this update method needs a fictitious intermediate material property 
	when the variables $\xi$ and $\eta$ lie within the unit circle, i.e., $\xi^2+\eta^2<1$. In some cases, the update can fail into local optima, as shown in Fig.~\ref{fig:4update-nabla}, or most domains have the fictitious intermediate property, depending on the given fictitious material properties.
	
	To exploit this, we consider the following update scheme:
	\begin{align}
		\xi^\text{new}+\tau_\theta\nabla^2\xi^\text{new}&=(1-\alpha_\theta)\xi+
		\alpha_\theta\chi_F\cos(2\theta^*)
		,\label{eq:4xi-proposing-update}\\
		\eta^\text{new}+\tau_\theta\nabla^2\eta^\text{new}&=(1-\alpha_\theta)\eta+
		\alpha_\theta\chi_F\sin(2\theta^*)
		.\label{eq:4eta-proposing-update}
	\end{align}
	where $\tau_\theta$ is the regularization factor, $\alpha_\theta$ is a step size and $\theta^*$ is the estimated optimal orientation estimated by the method shown in the following subsection.
	If the regularization term is not considered, the auxiliary variables $(\xi,\eta)$ gradually move closer to $(\cos(2\theta^*),\sin(2\theta^*))$, as shown in Fig.~\ref{fig:4update-thetaopt}, meaning the orientation $\theta$ becomes $\theta^*$.
	By updating in this way, the design variables move toward the globally optimal solution regardless of the worsening of the objective function at angles along the way, making it less likely to fall into a local optimum.

	\subsection{Topological derivative for orientations}
	\begin{figure}
		\centering
		\includegraphics[width=0.6\linewidth]{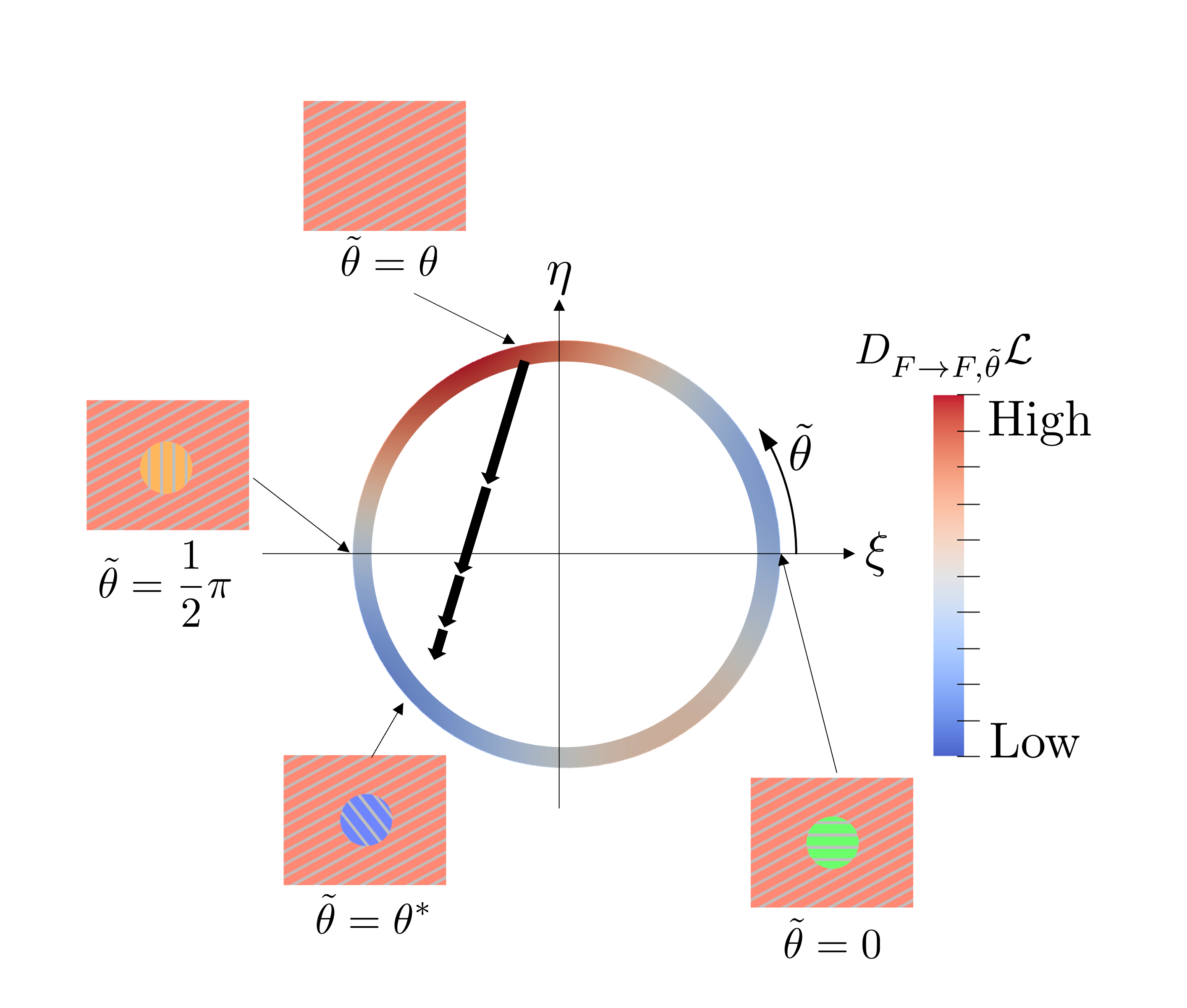}
		\caption{Concept of estimation of optimal orientation. The estimated optimal orientation $\theta^*$ minimizes the topological derivative $D_{F\to F,\tilde{\theta}} \mathcal{L}$ defined in Eq. \eqref{eq:4def aniso td}.}
		\label{fig:4thetatd}
	\end{figure}
	
	In order to perform the updates described in the previous subsection, it is necessary to know or estimate the optimal orientation at each position. For this purpose, 
	we consider a domain of an anisotropic material, and at the position $\x$, where the orientation of the anisotropic material is $\theta(\x)$, the same anisotropic material with another orientation $\tilde\theta$ is inserted, as shown in Fig.~\ref{fig:4anisotd}. In the figure, inserted material is different from the background material, but here we consider that the same material but in a different orientation is inserted.
	
	If the optimal orientation at that position is $\theta$, the objective function will not be improved no matter what the value of $\tilde\theta$ is. On the other hand, if the optimal orientation at the position is not $\theta$, the objective function will be improved by inserting anisotropic material with a certain orientation $\tilde\theta\in[0,\pi)$. Therefore, the orientation that improves the objective function the most is estimated by the following calculation:
	\begin{align}
		\theta^{*}(\x) = \argmin_{\tilde{\theta}\in[0,\pi)} D_{F\to F,\tilde{\theta}} \mathcal{L}(\x)
	\end{align}
	where $D_{F\to F,\tilde{\theta}}(\x)$ is the topological derivative between anisotropic materials when the same anisotropic material oriented in another direction $\tilde\theta$ is inserted in a domain of anisotropic material oriented in the direction $\theta(\x)$ at position $\x$, that is, the amount of improvement in the Lagrangian normalized as a ratio to the measure of the inserted domain. 
	This orientation $\theta^{*}$ will be referred to as the estimated optimal orientation.
	Fig.~\ref{fig:4thetatd} shows the concept of estimated optimal orientation.
	
	\section{Numerical implementation}\label{sec:4Numerical implementation}
	
	\subsection{Approximation of topological derivative for continuous orientations using discrete orientations}
	In some cases, the topological derivative for anisotropic material defined in Eq. \eqref{eq:4def aniso td} can be expressed as follows:
	\begin{align}
		D_{a\to b,\tilde\theta}\Lag=f(\strain(\bm u),A(a,\theta,b,\tilde\theta),\strain(\bm v)),
	\end{align}
	where $f$ is a real-valued function, $\bm u$ and $\bm v$ are state and adjoint variables, respectively, that depend on the objective function and governing equation. The tensor $\strain(\bm u)$ has components that depend on $\bm u$, but are independent of $a,\theta,b,\and\tilde\theta$. The tensor $A(a,\theta,b,\tilde\theta)$ has components that depend on $a,\theta,b,\and\tilde\theta$, but are independent of $\bm u$ and $\bm v$.
	If the expression of $A(a,\theta,b,\tilde\theta)$ is very complicated, c.f., nonlinear in terms of $\theta \and \tilde\theta$, the calculation of the multi-material topological derivative considering anisotropic material $D_{a\to b,\tilde\theta}\Lag$ costs much calculation time, and of course the estimated optimal orientation $\theta^*$ even further so.
	
	Therefore, in this study, we calculate the tensor $A(a,\theta,b,\tilde\theta)$ for a number of $n$ discretized orientations, $\theta_i,\tilde\theta_i=\frac{i}{n}\pi,i\in\{0,1,\ldots,n-1\}$, prior to the optimization.
	
	In the optimization steps, 
	we approximately calculate
	the multi-material topological derivative considering anisotropic material $D_{a\to b,\tilde\theta}\Lag$ and the estimated optimal orientation $\theta^*$ as shown bellow.
	
	First, the multi-material topological derivative $D_{F\to b}\Lag$ from anisotropic material oriented in the direction $\theta$ to isotropic material $b\in\{I,V\}$ is interpolated using the piecewise quadratic functions, from the function $f$ for discrete angles nearby $\theta$.
	
	Similarly, we calculate the multi-material topological derivative to anisotropic materials $D_{F\to F,\tilde\theta_i}\Lag$ for $n$ discrete angles $\tilde\theta_i=\frac{i}{n}\pi,i\in\{0,1,\ldots,n-1\}$.
	
	Also, we calculate the multi-material topological derivative from isotropic materials $D_{a\to F,\tilde\theta_i}\Lag,a\in\{I,V\}$.
	
	Then, the optimally oriented topological derivatives $D^*_{a\to F}\Lag, a\in\{I,V,F\}$ is interpolated using the piecewise quadratic functions, from the calculated derivatives $D_{a\to F,\tilde\theta_i}\Lag$.

	Finally, the estimated optimal orientation $\theta^*$ is calculated as the minimizing point for $D^*_{a\to F}\Lag, a\in\{I,V,F\}$.

\begin{figure}
		\centering
		
		\begin{minipage}[t]{0.34\linewidth}
			\centering
			\includegraphics[height=5cm]{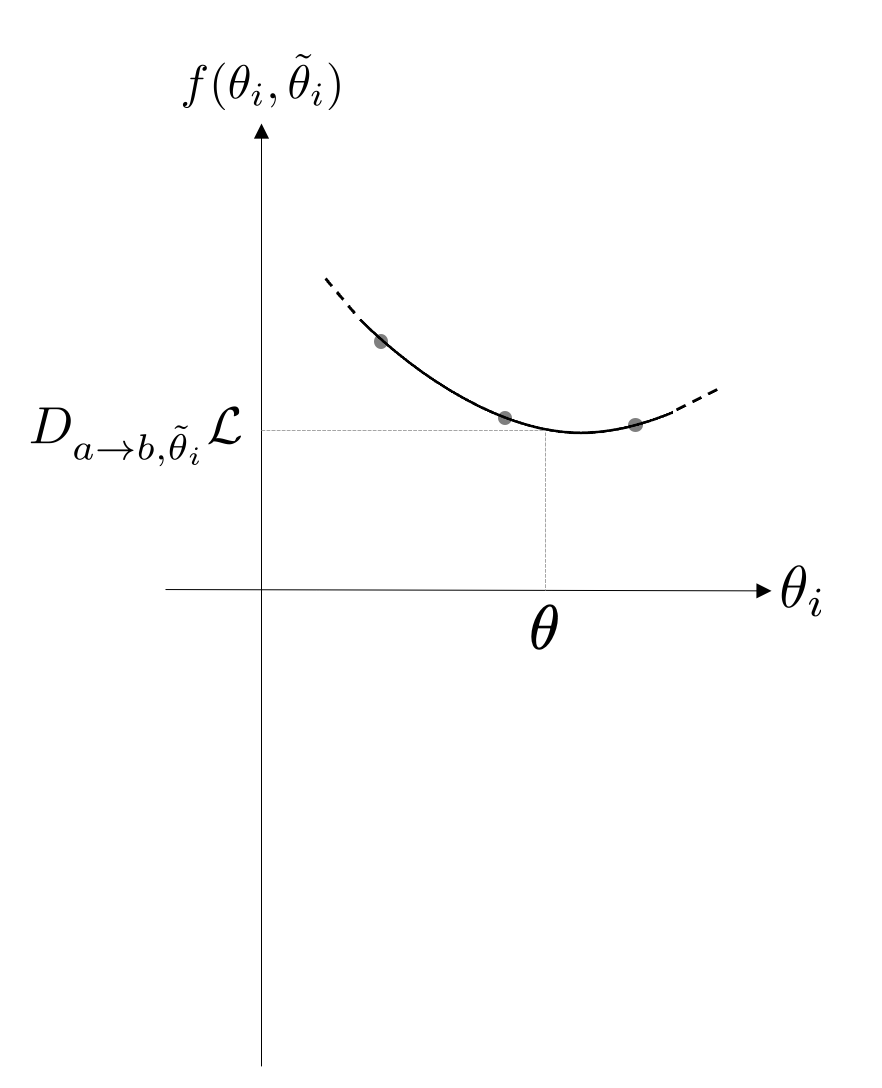}
		\end{minipage}
		\begin{minipage}[t]{0.65\linewidth}
			\centering
			\includegraphics[height=5cm]{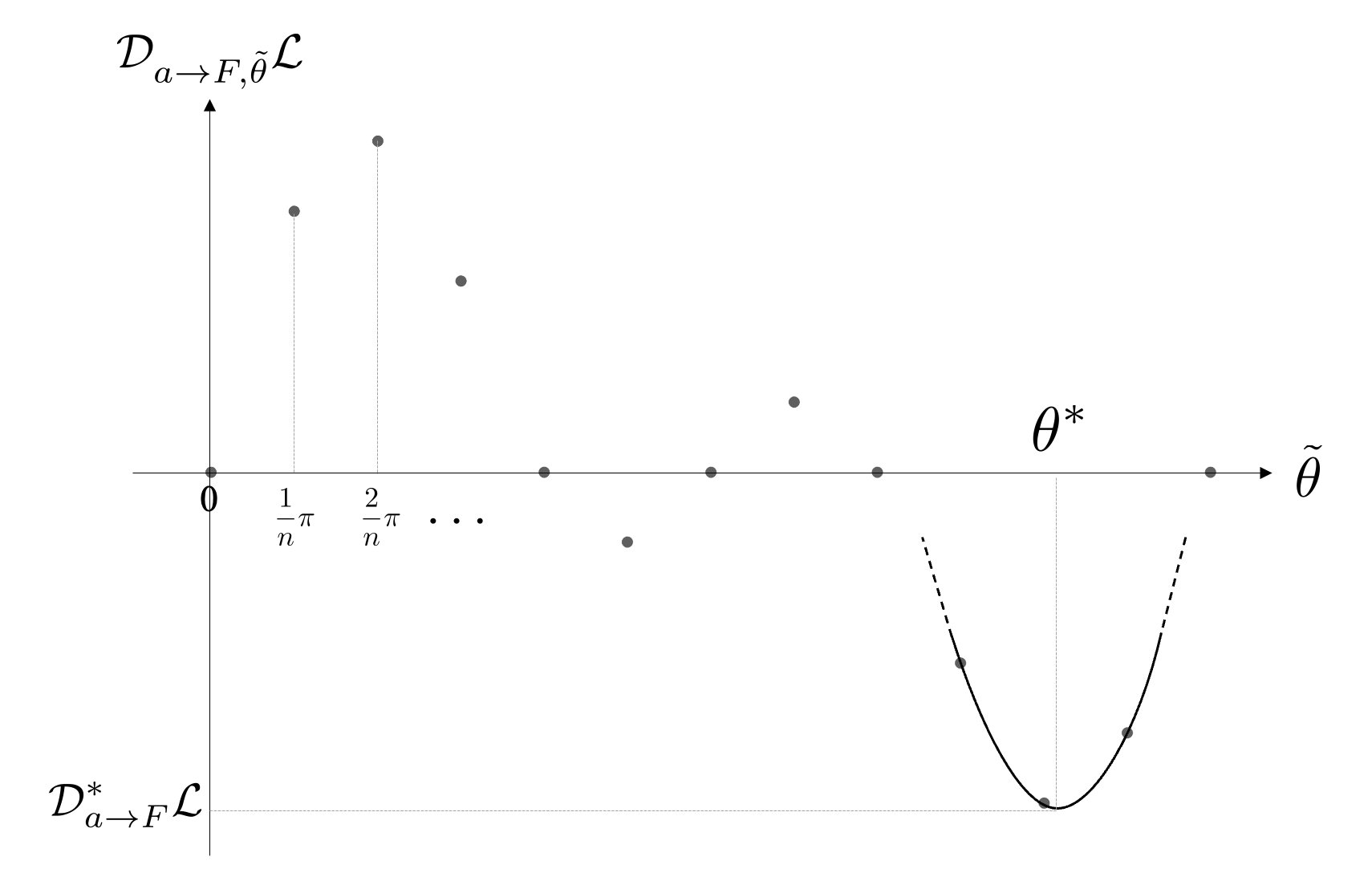}
		\end{minipage}
		\caption{Approximately calculated topological derivative from anisotropic material (left) and to anisotropic material (right). } 
		
		\label{fig:4approximated calculation of thetaopt}       
	\end{figure}

	Fig.~\ref{fig:4approximated calculation of thetaopt} shows an example of the approximated calculation.

	\subsubsection*{Remark}
	Actually, topological derivative $D_{F \to a}\Lag$ can be calculated in the straightway described below, not using the interpolation scheme, but that calculation needs time-consuming calculation, so is approximated as described above.
	
	\subsection{Approximation of level set functions}\label{sec:4Approximation}
	As discussed in \cite{noda2022extended}, the X-LS function needs to satisfy the following constraint equation:
	\begin{align}
		\sum_{m=0}^{M-1}\chi_m=1.\label{eq:4sum const}
	\end{align}
	In the update method as described in subsection \ref{sec:4Update of level set function}, this constraint is not considered, so the level set functions do not always satisfy \eq{4sum const}.
	Instead of updating the X-LS functions with satisfying the constraint, we construct the constrained X-LS functions $\tilde{\phi}_{ba}$ from the level set functions updated in the above method $\phi_{ba}$ as follows:
	\begin{align}
		\psi_a &= \prod_{b\in \{0,1,\ldots,M-1\}\backslash a} (\phi_{ba}+1)/2\label{eq:4 tilde chi}\\
		\tilde{\phi}_{ba}&=\psi_a - \psi_b,\label{eq:4 tilde phi}
	\end{align}
	\subsection{Smoothing of Characteristic Functions}
	In the Ersatz material approach \cite{allaire2004structural}, which implicitly represents each material boundary as a smooth transition of material properties at the boundary, the material properties and phase derivatives are represented by smoothed characteristic functions.
	
	This approach eliminates the computational time and effort of regenerating finite elements and ensures the numerical stability of the finite element analysis.
	
	In this study, smoothed characteristic function $\tilde\chi$ is calculated as follows:
	\begin{align}
		\chi'_a&=\prod_{b\ne a} \tilde{H}(\phi_{ab}/w_m),\\
		\chi''_a&=\chi'_a+\varepsilon_\chi\prod_{b\ne a} (1-\chi'_b),\\
		\tilde\chi_a&=\frac{\chi''_a}{\sum_b \chi''_b}.
	\end{align}
	where $0<w_m<1$ is a constant that determines the width of the material transition and $0<\varepsilon_\chi\ll 1$ is a sufficiently small positive number. This parameter $\varepsilon_\chi$ is introduced for computational stability. 
	Function $\tilde H$ is the approximated Heaviside function defined as follows:
	\begin{align}
		\tilde{H}(s) &=
		\begin{cases}
			0 &(s<-1)\\
			\frac{1}{2}+s
			\left[\frac{15}{16}-s^2 \left(\frac{5}{8}-\frac{3}{16}s^2 \right) \right] \qquad&(-1 \le s \le 1)\\
			1 &(s>1)
		\end{cases}.\label{eq:4 tilde heaviside}
	\end{align}

	\subsection{Updating scheme for Lagrange multipliers}\label{sec:4control}
	
	If we consider the case where the constraint function is a single inequality constraint, then the Lagrangian $\Lag$ is defined as follows:
	\begin{align}
		\Lag=J+\lambda g.\label{eq:4Lagrangian}
	\end{align}
	Note that the extended topological derivatives are calculated with $\lambda$ being fixed.
	The multiplier $\lambda$ is the control multiplier, which is to ensure that constraints are satisfied at the end of optimization and coincides with the Lagrange multiplier at the end of optimization.
	
	Following \cite{tovar2006topology} and other studies, the control multiplier
	for $i$th optimization step $\lambda^i$ was determined under the proportional-integral-differential control concept as follows:
	\begin{align}
		\lambda^i&= \max\left(K_{\text{P}} g^i,0\right)+{K_\text{D}} \dot{g}^i
		+\Lambda^i,\label{eq:4pid}
	\end{align}
	where $\dot{g}^i$ and $\Lambda^i$ are 
	defined as follows:
	\begin{align}
		\dot{g}^i &=g^i-g^{(i-1)},\\
		\Lambda^i &= \max\left(\Lambda^{(i-1)} + \left({K_\text{IP}} g^i+{K_\text{ID}}\dot{g}^i\right),0\right), \label{eq:4pid2}\\
		\Lambda^0&=0,\\
		\dot{g}^0&=0
	\end{align}
	In Eqs.~(\ref{eq:4pid}) and (\ref{eq:4pid2}), ${K_\text{P}}\in\mathbb R$, ${K_\text{D}}\in\mathbb R$, and ${K_\text{IP}}\in\mathbb{R}$ are the proportional, differential, and integral gains, respectively, and ${K_\text{ID}}\in\mathbb R$ is a gain which improves the numerical stability.
	
	\subsection{Limitation of X-LS functions}
	In order to concentrate the effects of diffusion near the boundary, the X-LS functions were limited as follows:
	\begin{equation}
		\phi_{ab}(\x,t+\Delta t) = \max(-1,\min(1,\hat\phi_{ab}(\x,t+\Delta t))).\label{eq:4limnum}
	\end{equation}

	\subsection{Optimization algorithm}
	The optimization problems are solved using the following optimization processes: 
	\begin{description}
		\item[step 1] Initialize the design variables $\phi_{ab}, \xi, \eta$
		\item[step 2] Solve the governing and adjoint equations using a finite element method
		\item[step 3] Evaluate the objective and constraint functions
		\item[step 4] Calculate the sensitivities using the solutions of governing and adjoint equations
		\item[step 5] Update the design variables
		\item[step 6] If  the solution has converged, terminate the optimization calculation. Otherwise, return to \textbf{step 2}.
	\end{description}
	We performed the finite element method using the finite-element analysis software FreeFEM++ \cite{hecht2012new}.
	
	\subsection{Order estimation of the amount of calculation}
	Let $m$ and $n$ be the numbers of finite elements and discretized orientation, respectively. The computational cost required for each step is estimated as follows:
	
	step 1: $\mathcal{O}(m)$,
	
	step 2: $\mathcal{O}(m^2)$,
	
	step 3: $\mathcal{O}(m)$,
	
	step 4: $\mathcal{O}(Kmn)$,
	
	step 5: $\mathcal{O}(m^2)$,
	
	step 6: $\mathcal{O}(1)$,
	\\
	where $\mathcal{O}$ is the Landau symbol. Note that step 2 can be larger or smaller depending on the problem. Here, we assume linear elasticity, which is the case in this study.
	The computational cost of the partial-differential equation is evaluated as the square of the degrees of freedom.
	Note that the constant number $K$ in step 4 is very large, including many conditional branches; therefore step 4 sometimes costs more time than the finite element analysis in step 2 or step 5.

	\section{Application to linear elasticity problem}\label{sec:4 linear elasticity}
	To verify proposing method, it is applied to mean compliance minimization considering linear elasticity with the maximum weight constraint. 
	\subsection{Problem formulation}
\begin{figure}
		\centering
		\includegraphics[width=6cm]{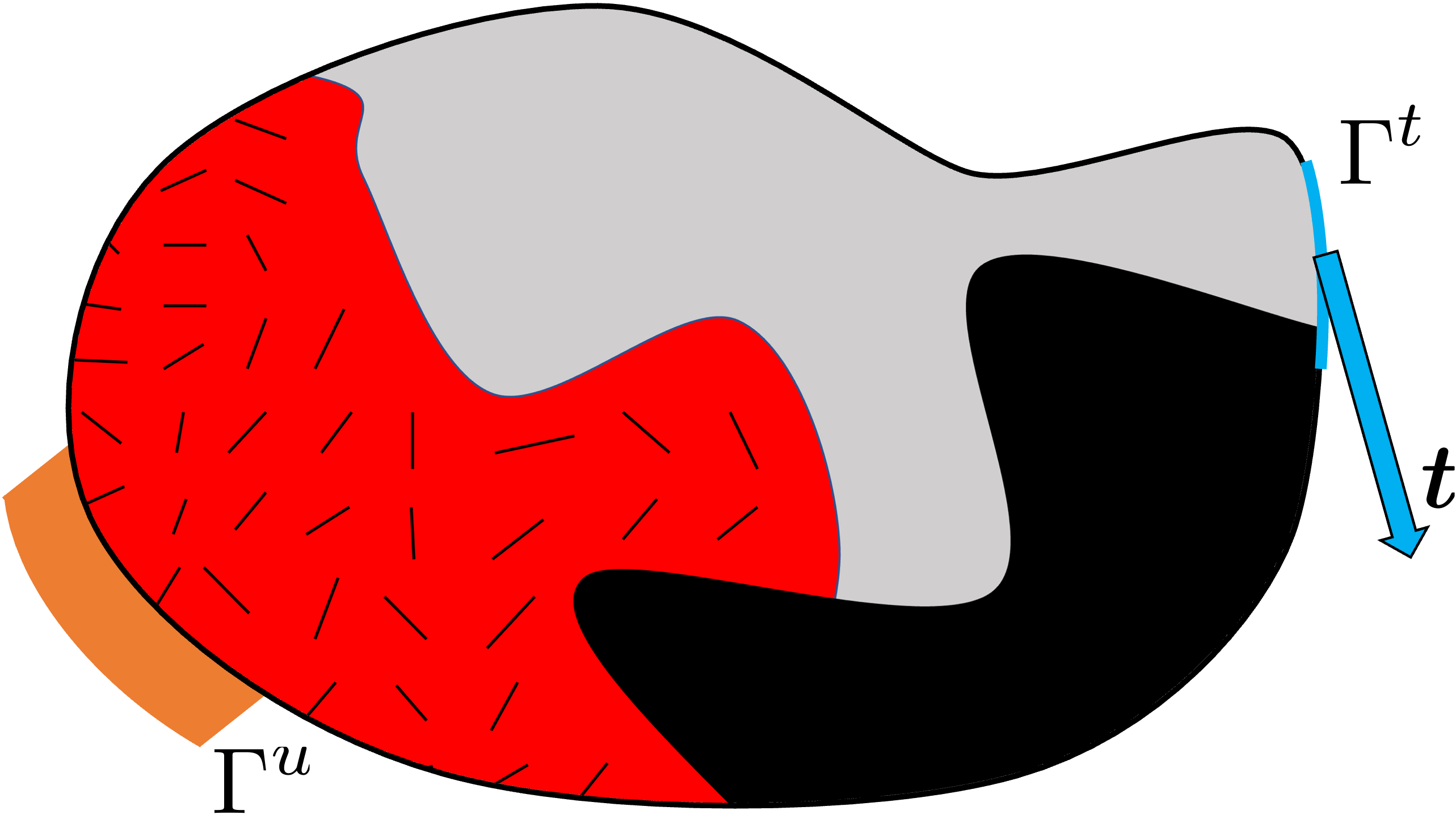}
		\caption{Boundary conditions}
		\label{fig:4 stiff}
	\end{figure}

	Fig.~\ref{fig:4 stiff} shows the boundary conditions in this problem.
	$\Gamma^u$ is a boundary given a fixed displacement. $\Gamma^t$ is a boundary imposed a traction vector $\bm t \in \mathbb{R}^{d}$, whose $i$-th component is denoted by $t_i$. 
	The displacement field was denoted by $u_i$ in the static equilibrium state.
	
	The minimum mean compliance problem was then formulated as follows:
	\begin{equation}
		\begin{array}{lll}
			\displaystyle\inf_{\bm\Omega,\theta(\x)} \qquad& \displaystyle J_C=\int_{\Gamma^t} t_i u_i \d\Gamma,&\\
			\mathrm{subject~to}\qquad
			& C_{ijkl}u_{k,lj}=0\qquad&\mathrm{in}~D, \\
			&u_{i}=0  &\mathrm{on}~\Gamma^{u}, \\
			&\sigma_{i j}\Add{(u)} n_{j}=t_{i} &\mathrm{on}~\Gamma^{t},\\
			&g_W = \displaystyle\sum_{m\in\{V,I,F\}}\int_{D} \rho_m\chi_m \text{d}\Omega - {W^\text{max}} \le0.
		\end{array}\label{eq:4 problem comp minim}
	\end{equation}
	Here, the indices $i,j,k$, and $l\in\{x,y\}$ follow the summation convention and the indices after the comma denote the partial derivative of the coordinate components. $\sigma_{ij}(u)=C_{ijkl}u_{k,l}$ is the stress tensor, $J$ is the objective function, and $g_W$ is the volume constraint function of material $m$. $W^\mathrm{max}$ is the maximum weight.
	$C_{ijkl}(\bm x)$ is the elasticity tensor for multiple materials, defined as follows in terms of the characteristic function $\chi_a$ for each material $a$ and the single-material elasticity tensor $C^{a}_{ijkl}$.
	\begin{align}
		C_{ijkl}&=\sum_{a\in\{V,I,F\}}\chi_a C^{a}_{ijkl}\\
		C^F_{ijkl}(\theta)&=C^{Fx}_{pqmn}R_{pi}(\theta)R_{qj}(\theta)R_{mk}(\theta)R_{nl}(\theta)
	\end{align}
	
	\subsection{Multi-material topological derivative including anisotropic material}
	Elastic anisotropic topological derivatives for the 2D problem have been analyzed theoretically in multiple studies. In this study, we use the formulation by 
	Bonnet and Delgado \cite{bonnet2013topological},
	which is based on Eshelby's theory. 
	The topological derivative $D_{a\to b} J_C$ is given by
	\begin{equation}
		D_{a\to b} J_C=-\strain_{ij}A^{ab}_{ijkl}\strain_{kl},
	\end{equation}
	where $\strain_{ij}$ is strain tensor, defined as $\strain_{ij}=\frac{1}{2}(u_{i,j}+u_{j,i})$, and $A^{ab}_{ijkl}$ is the elastic moment tensor, expressed as follows:
	\begin{eqnarray}
		A^{ab}_{ijkl}=C^b_{ijkl}(\tilde{C}^{ab}_{klmn})^{-1}(C^a_{mnpq}-C^b_{mnpq}),\\
		\tilde{C}^{ab}_{ijkl}=C^b_{ijkl}+(C^a_{ijmn}-C^b_{ijmn})S^{\text{int}~ab}_{mnkl}.
	\end{eqnarray}
	The fourth order tensor $S^{\text{int}~ab}_{mnkl}$ is the interior Eshelby tensor, given by
	\begin{equation}
		S^{\text {int }ab}_{ijmn}= \frac{1}{4\pi}C^b_{klmn} \int_{-\pi / 2}^{\pi / 2}  N_{ik}(\boldsymbol{\alpha}(\theta)) \alpha_l(\theta)\alpha_j(\theta) \mathrm{d} \theta,
	\end{equation}
	where $\boldsymbol{\alpha}(\theta)$ and $N_{ik}(\boldsymbol{\alpha}(\theta))$ are defined as follows:
	\begin{eqnarray}
		\boldsymbol{\alpha}(\theta)=(\cos(\theta),\sin(\theta)),\\
		N_{ik}(\boldsymbol{\alpha}(\theta))=(C^b_{ijkl} \alpha_{j} \alpha_{l})^{-1}.
	\end{eqnarray}
	Note that the inverse $(X_{ijmn})^{-1}$ is defined such that\\$(X_{ijmn})^{-1}X_{mnkl}=\delta_{ik}\delta_{jl}$.
	
	The topological derivative for weight constraint $g_W$ is as follows:
	\begin{align}
		D_{a\to b} g_W(\x) = \rho_b-\rho_a.
	\end{align}

	\section{Numerical examples}\label{sec:4 numerical examples}
\begin{figure}
		\centering
		\includegraphics[width=0.4\linewidth]{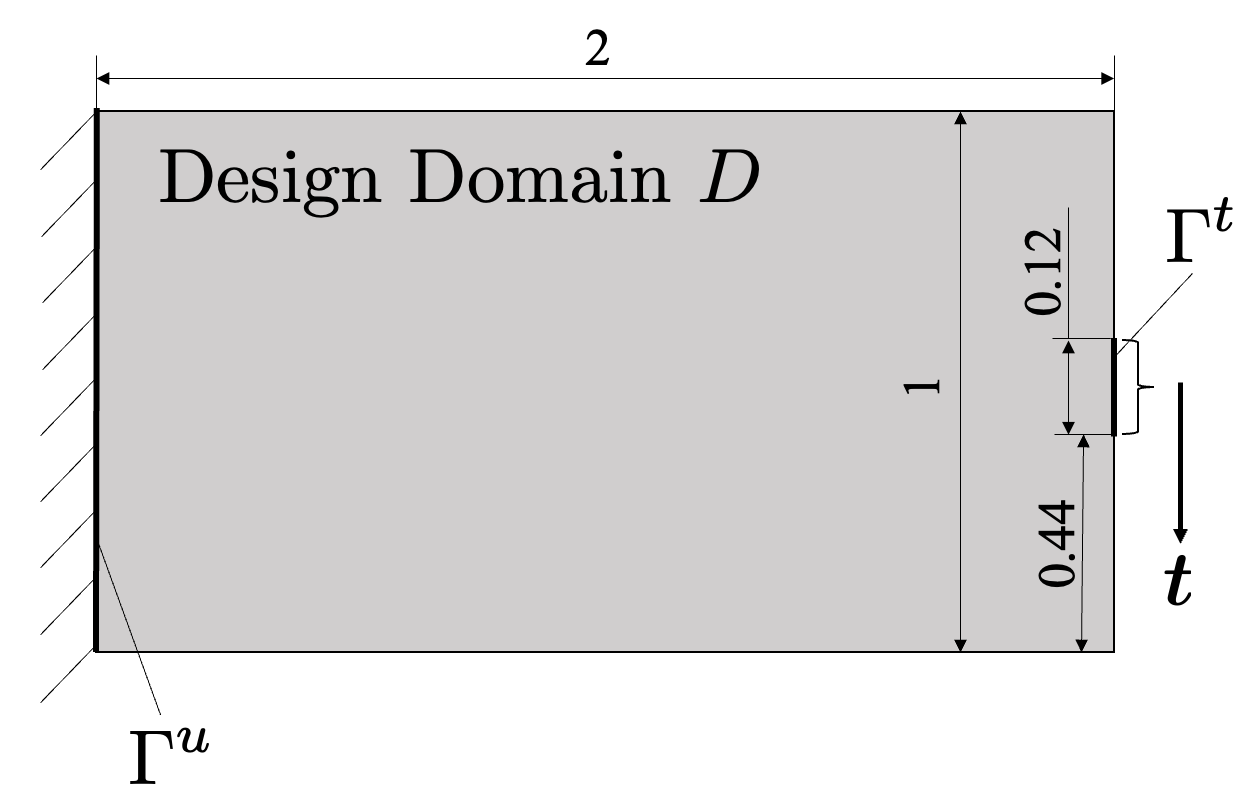}
		\caption{Design domain and boundary conditions.}\label{fig:4domainboundary}
	\end{figure}

	In this section, several numerical examples are provided to verify the utility of the proposed method.
	Design domain and boundary conditions are shown in Fig.~\ref{fig:4domainboundary}.
	
\begin{figure}
		\centering
		\includegraphics[width=0.4\linewidth]{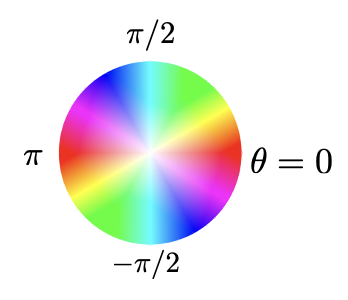}
		\caption{Coloring of auxiliary variables  $\xi$ and $\eta$.}\label{fig:4scale}
	\end{figure}

	To illustrate the orientation, the figures are colored by the auxiliary variables $\xi\and\eta$ as shown in Fig.~\ref{fig:4scale}.
	
	First, for each material $I,V,\and F$, elasticity tensors are set as follows:
	\begin{align}
		\nu^I=\nu^V=\nu^F&=0.3,\nonumber\\
		E^I&=80~[\text{GPa}],\nonumber\\
		C^{I}_{ijkl}&=\begin{cases}
			E^{I}/(1-(\nu^I)^2)\quad&\text{if}\quad i=j=k=l\\
			E^{I}\nu^I/(1-(\nu^I)^2)\quad&\text{if}\quad i=j,j\ne k,k=l\\
			E^I/(2(1+\nu^I))\quad&\text{if}\quad i\ne j,k\ne l\\
		\end{cases},\nonumber\\
		E^V&=0.01~[\text{GPa}],\nonumber\\
		C^{V}&=\begin{cases}
			E^{V}/(1-(\nu^V)^2)\quad&\text{if}\quad i=j=k=l\\
			E^{V}\nu^V/(1-(\nu^V)^2)\quad&\text{if}\quad i=j,j\ne k,k=l\\
			E^V/(2(1+\nu^V))\quad&\text{if}\quad i\ne j,k\ne l\\
		\end{cases},\nonumber\\
		E^\text{fib}&=100~[\text{GPa}],\nonumber\\
		E^\text{back}/E^\text{fib}&=0.2,\nonumber\\
		C^{Fx}_{ijkl}&=\begin{cases}
			E^\text{fib}/(1-(\nu^F)^2)\quad&\text{if}\quad i=j=k=l=x\\
			E^\text{back}/(1-(\nu^F)^2)\quad&\text{if}\quad i=j=k=l=y\\
			E^\text{back}\nu^F/(1-(\nu^F)^2)\quad&\text{if}\quad i=j,j\ne k,k=l\\
			E^\text{back}/(2(1+\nu^F))\quad&\text{if}\quad i\ne j,k\ne l\\
		\end{cases}.\label{eq:4props}
	\end{align}
	
	For various initial designs, optimization is processed. 
	Let initial design A be when initial values of design variables are set to as follows:
	\begin{align}
		\phi_{IV}=\phi_{VI}=0,\nonumber\\
		\phi_{VF}=\phi_{FV}=0,\nonumber\\
		\phi_{FI}=\phi_{IF}=0,\nonumber\\
		\xi=0,\nonumber\\
		\eta=0.\label{eq:4init0}
	\end{align}
	
\begin{figure}
		\centering
		\begin{minipage}[b]{0.3\linewidth}
			\centering
			\includegraphics[width=\linewidth]{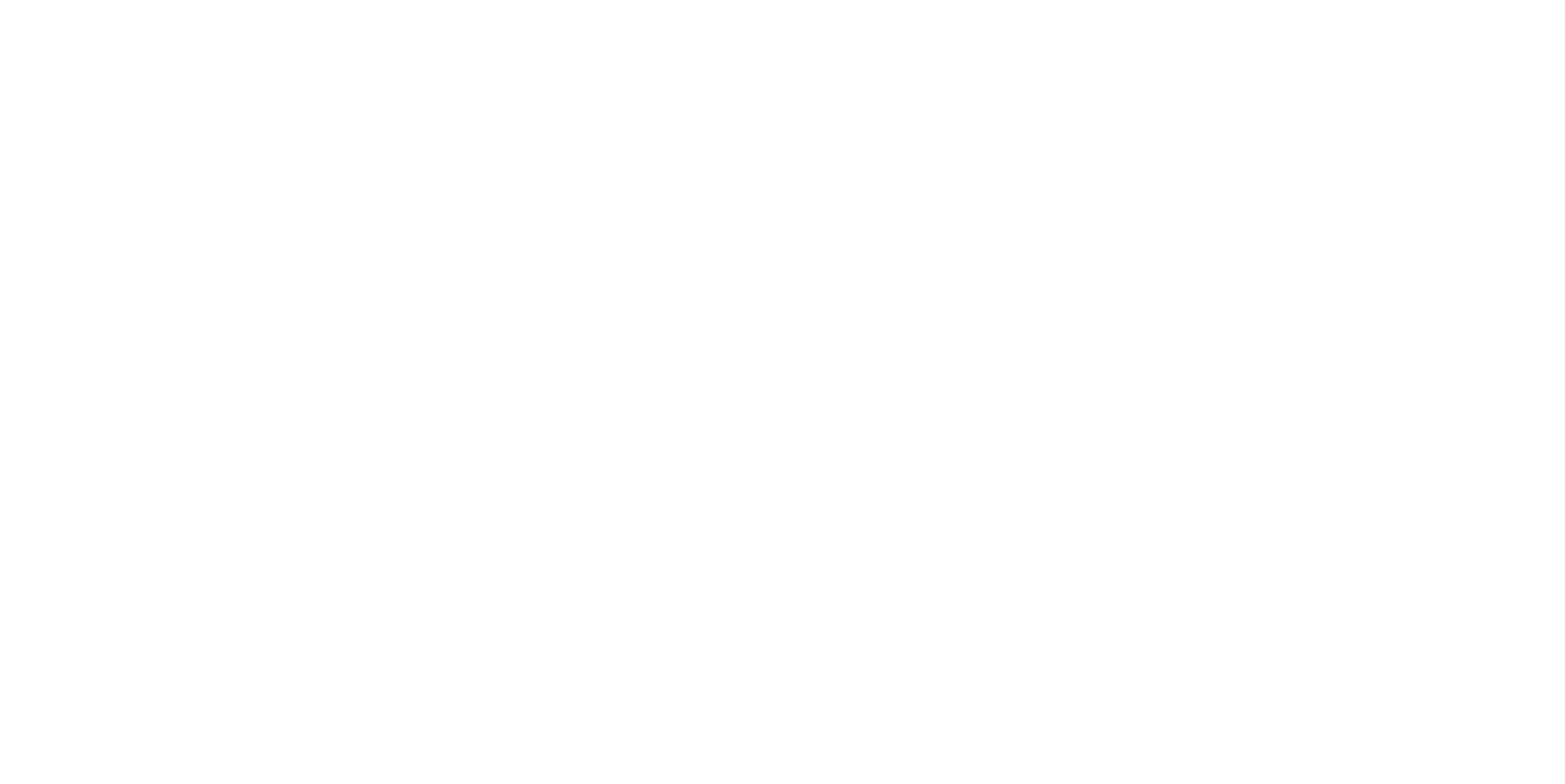}
			\subcaption{Step 0}
		\end{minipage}
		\begin{minipage}[b]{0.3\linewidth}
			\centering
			\includegraphics[width=\linewidth]{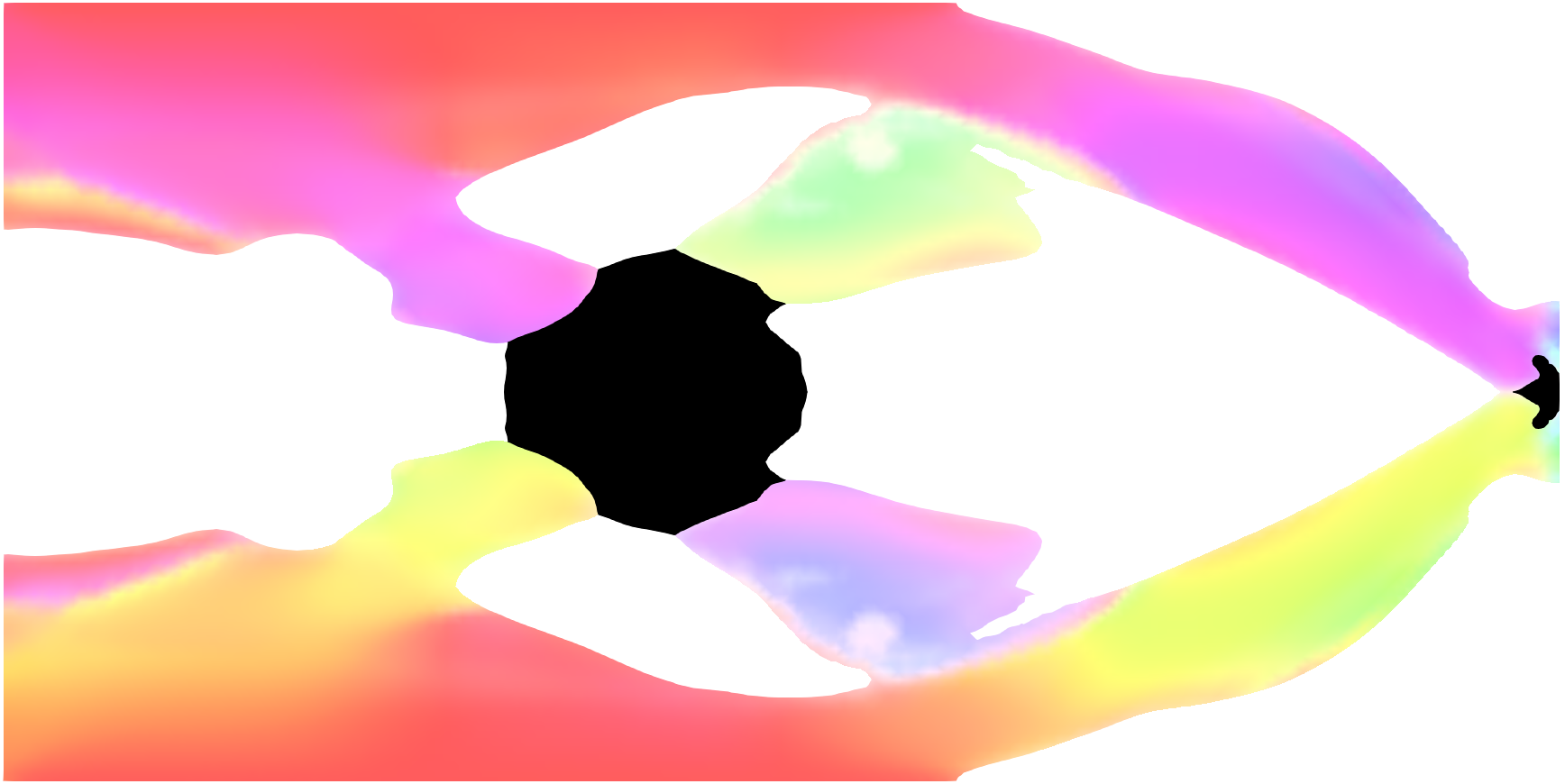}
			\subcaption{Step 10}
		\end{minipage}
		\begin{minipage}[b]{0.3\linewidth}
			\centering
			\includegraphics[width=\linewidth]{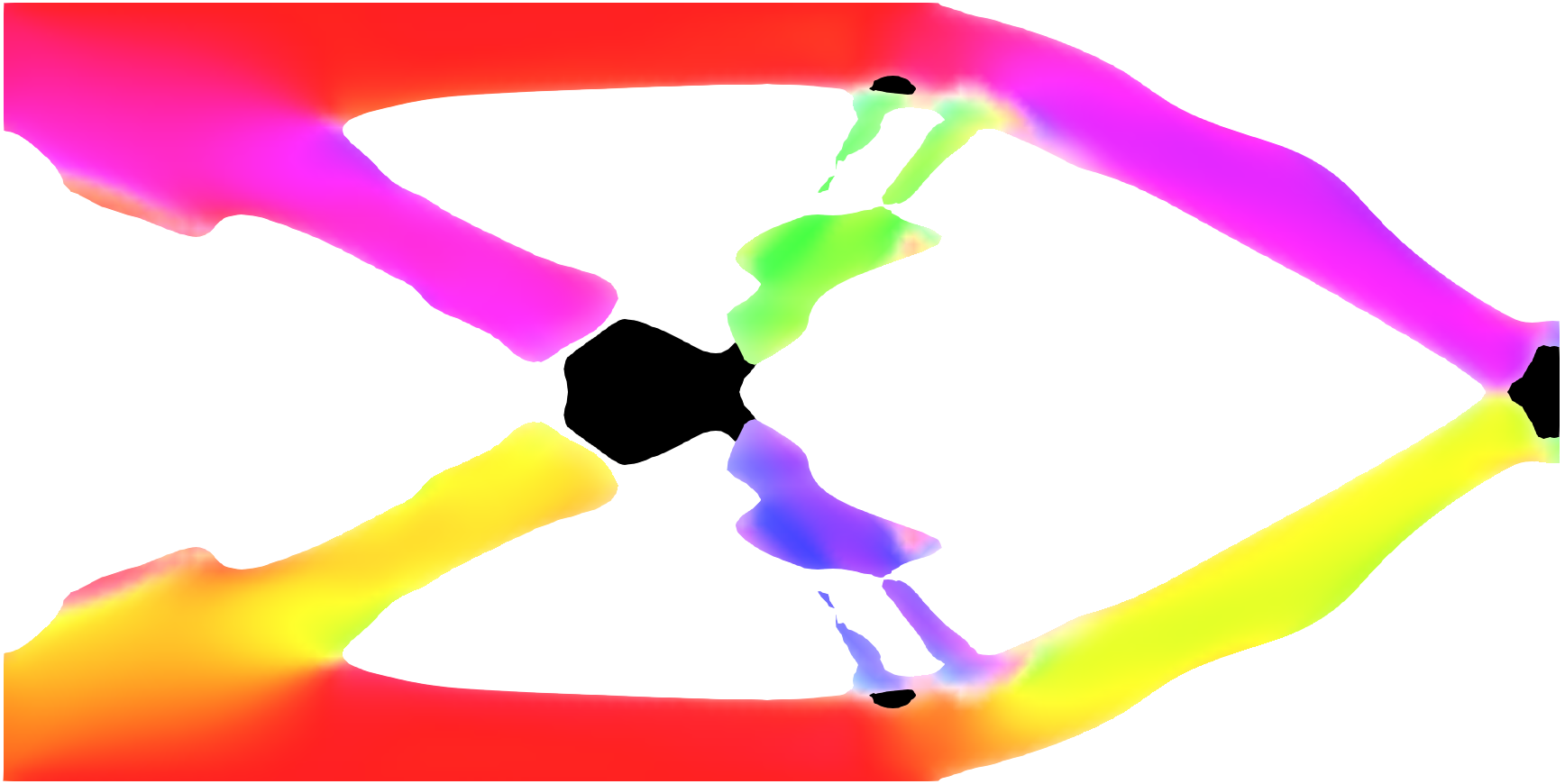}
			\subcaption{Step 20}
		\end{minipage}
		\begin{minipage}[b]{0.3\linewidth}
			\centering
			\includegraphics[width=\linewidth]{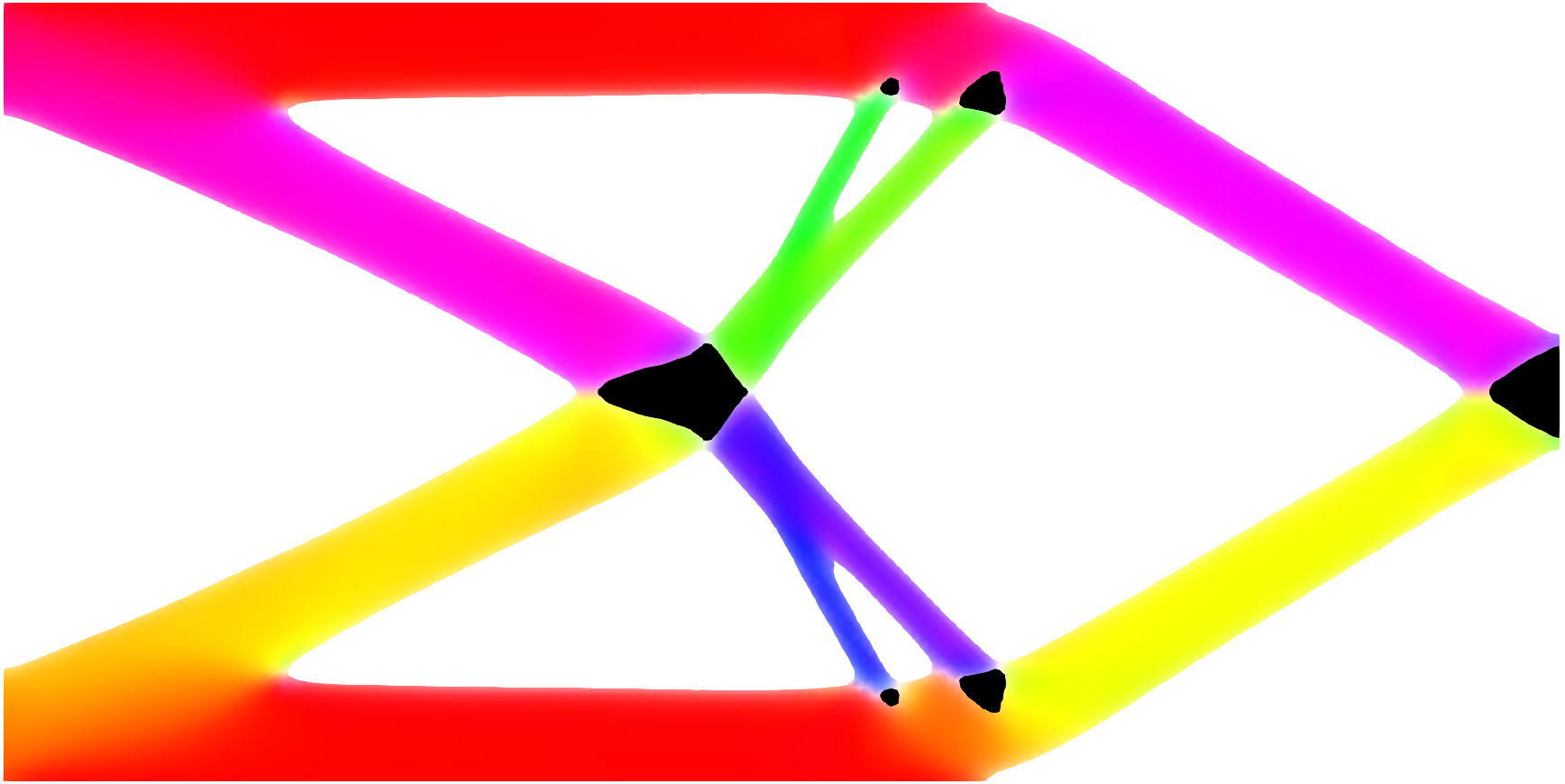}
			\subcaption{Step 50}
		\end{minipage}
		\begin{minipage}[b]{0.3\linewidth}
			\centering
			\includegraphics[width=\linewidth]{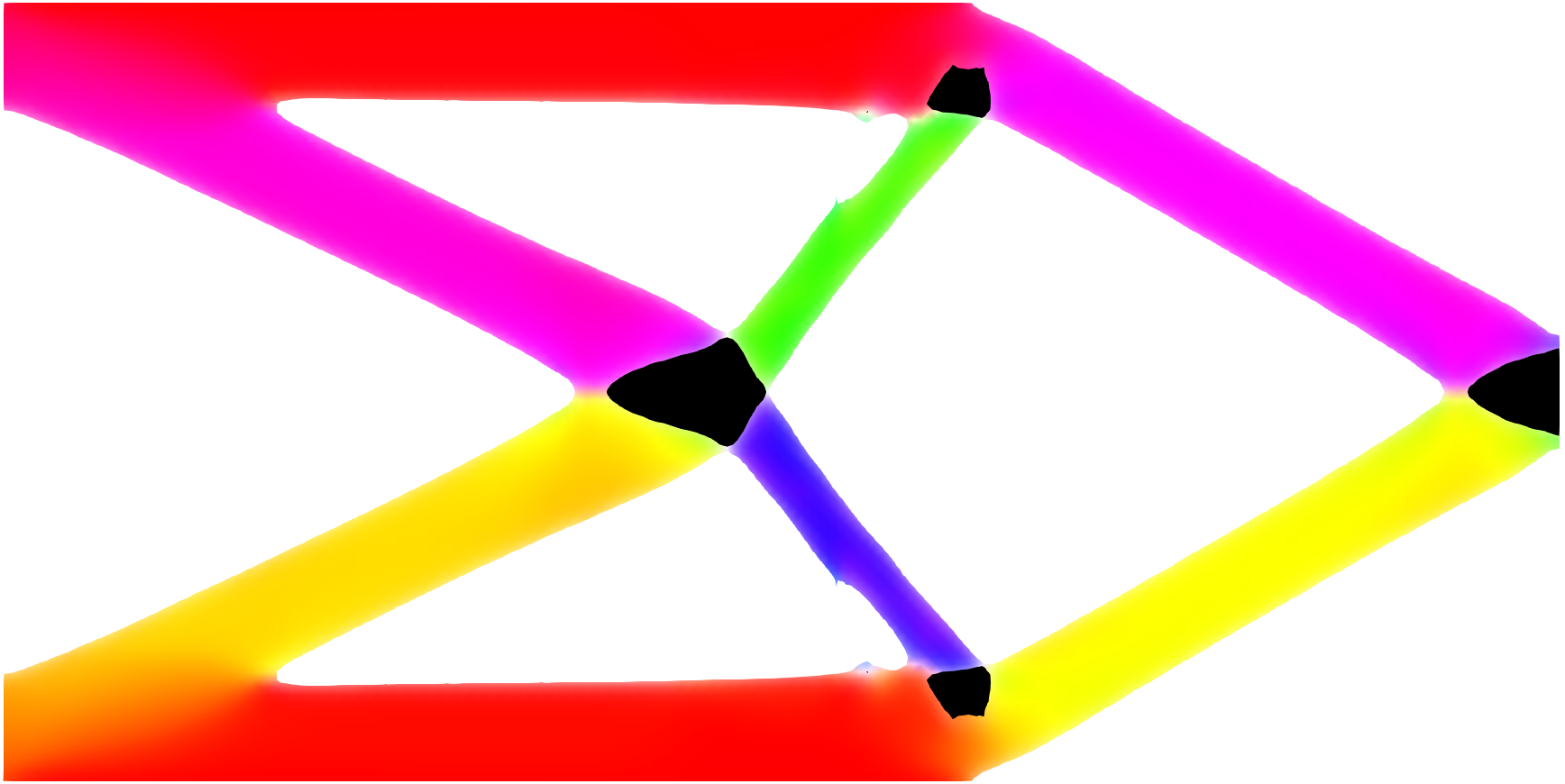}
			\subcaption{Step 200}
		\end{minipage}
		\begin{minipage}[b]{0.3\linewidth}
			\centering
			\includegraphics[width=\linewidth]{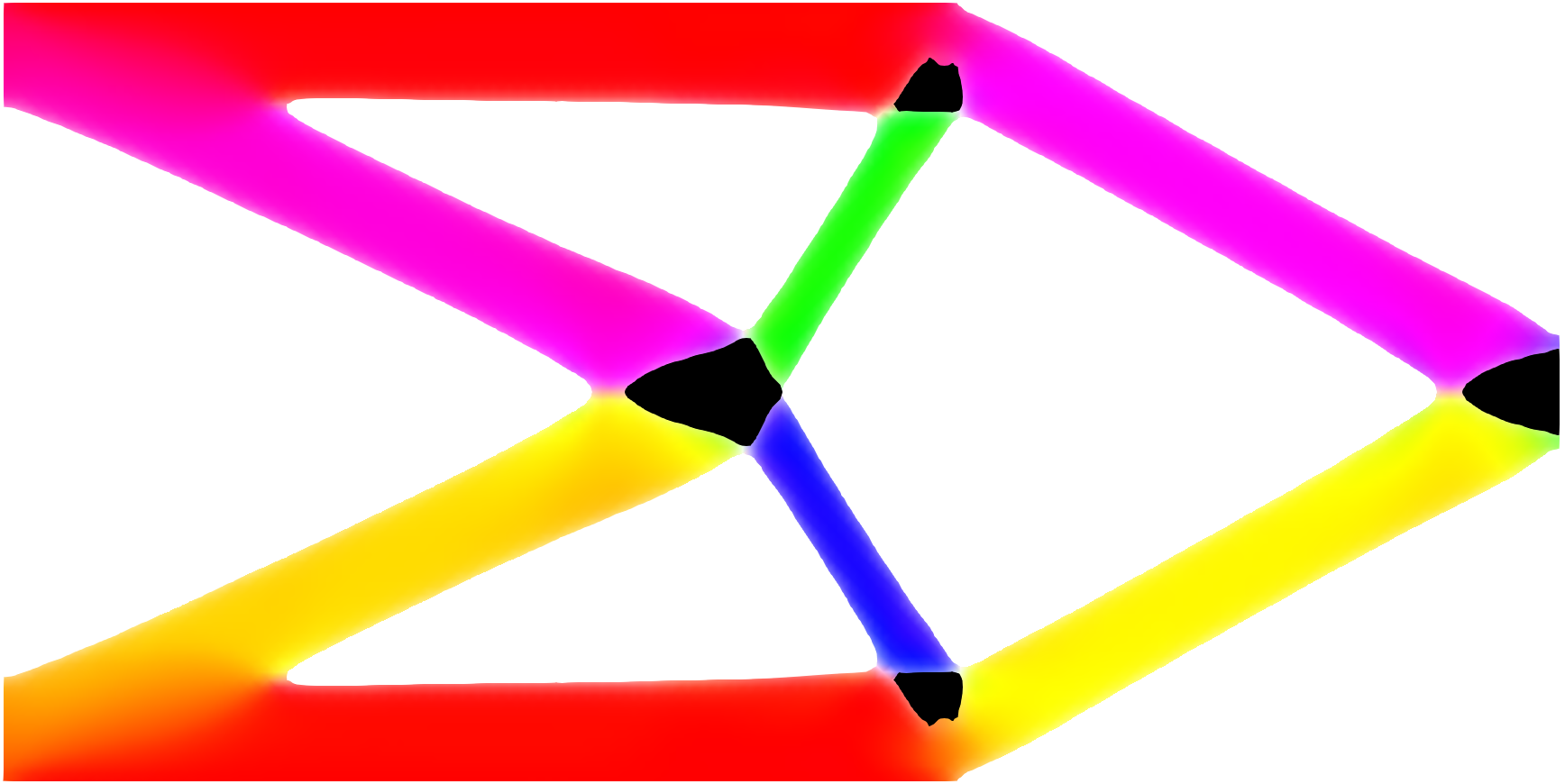}
			\subcaption{Optimal}
		\end{minipage}
		\caption{Optimization result for initial design A (defined in Eq.~\eqref{eq:4init0}).}\label{fig:4init0}
	\end{figure}

\begin{figure}
		\centering
		\begin{minipage}[b]{0.3\linewidth}
			\centering
			\includegraphics[width=\linewidth]{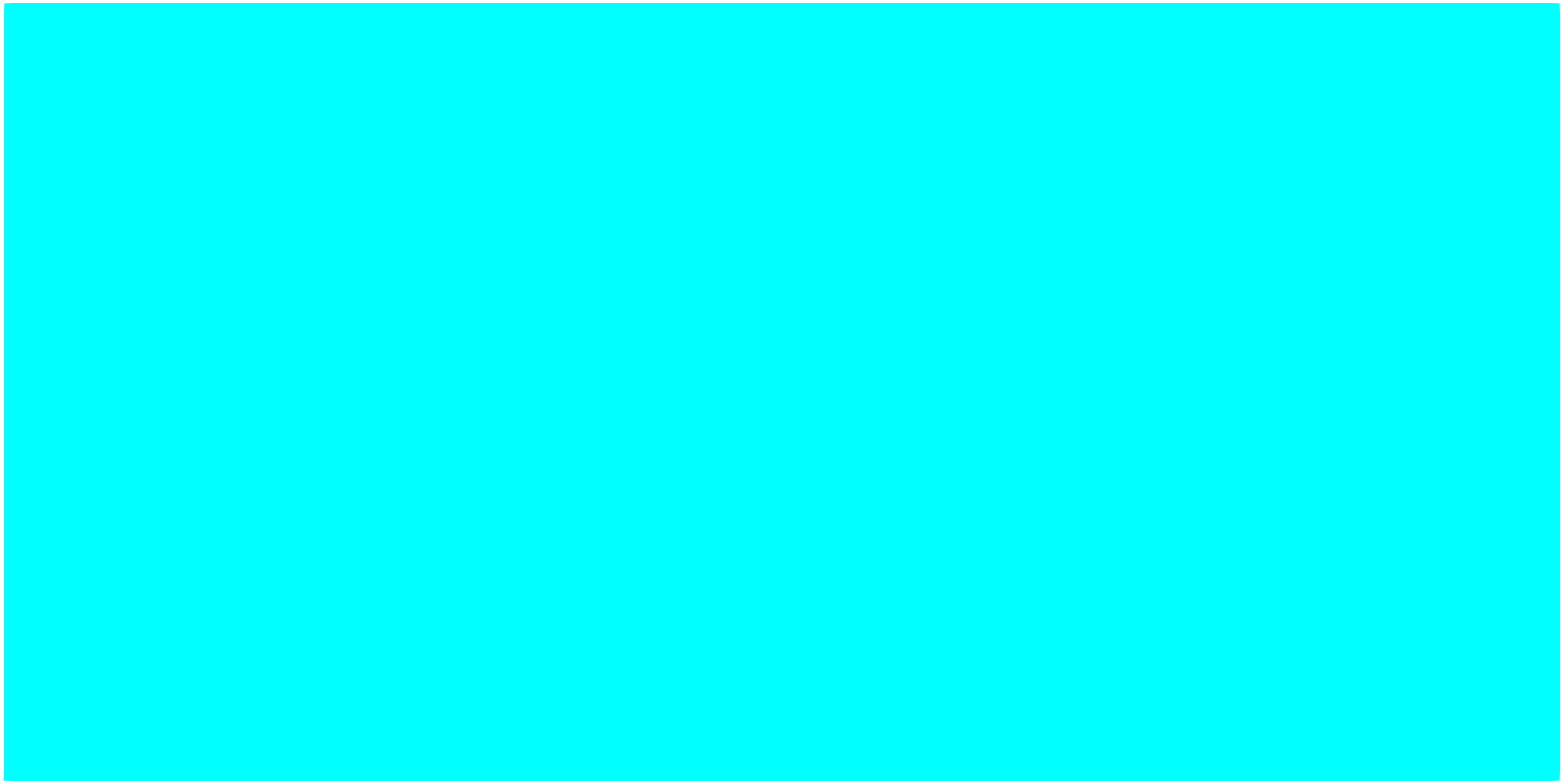}
			\subcaption{Step 0}
		\end{minipage}
		\begin{minipage}[b]{0.3\linewidth}
			\centering
			\includegraphics[width=\linewidth]{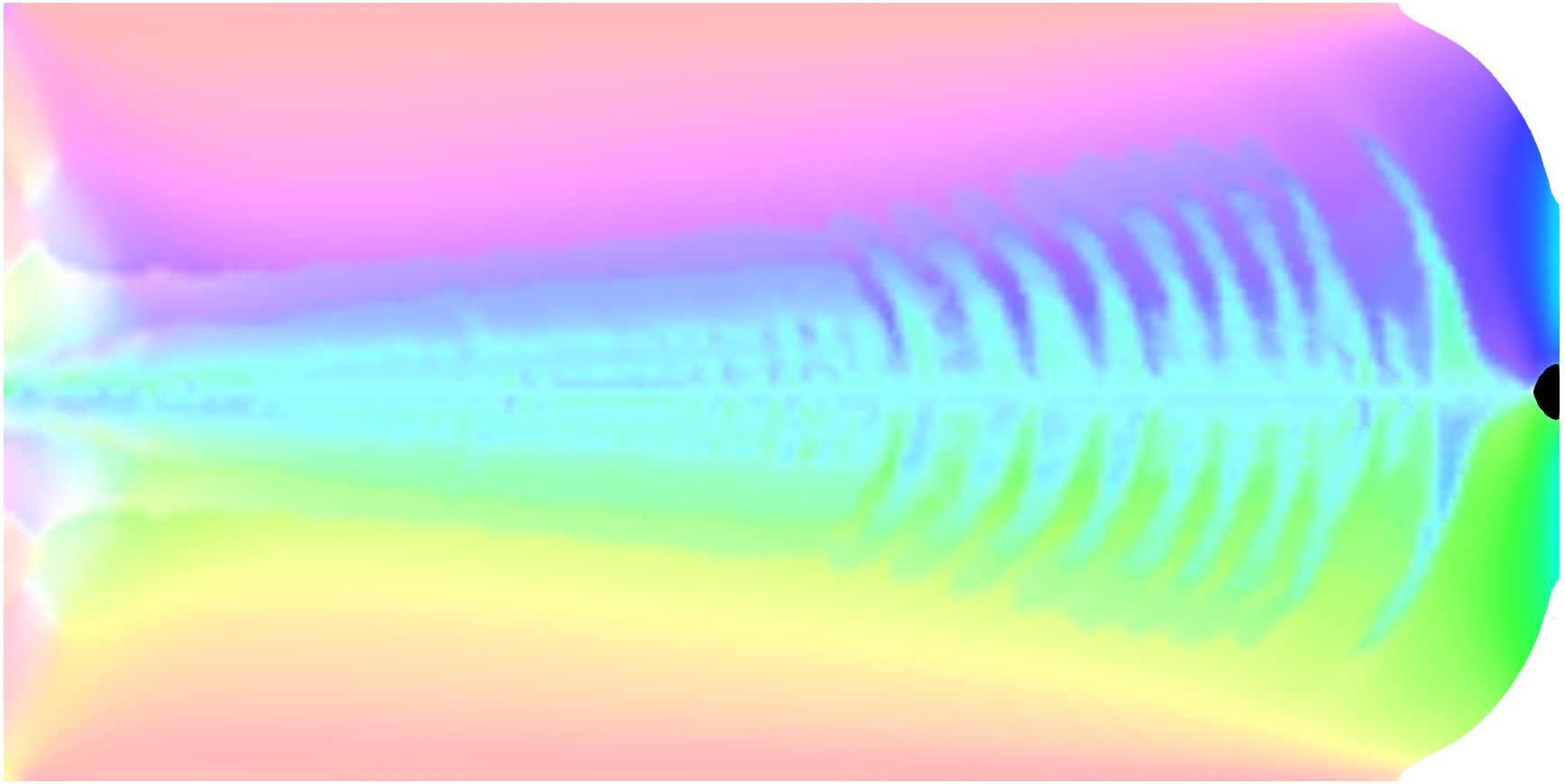}
			\subcaption{Step 10}
		\end{minipage}
		\begin{minipage}[b]{0.3\linewidth}
			\centering
			\includegraphics[width=\linewidth]{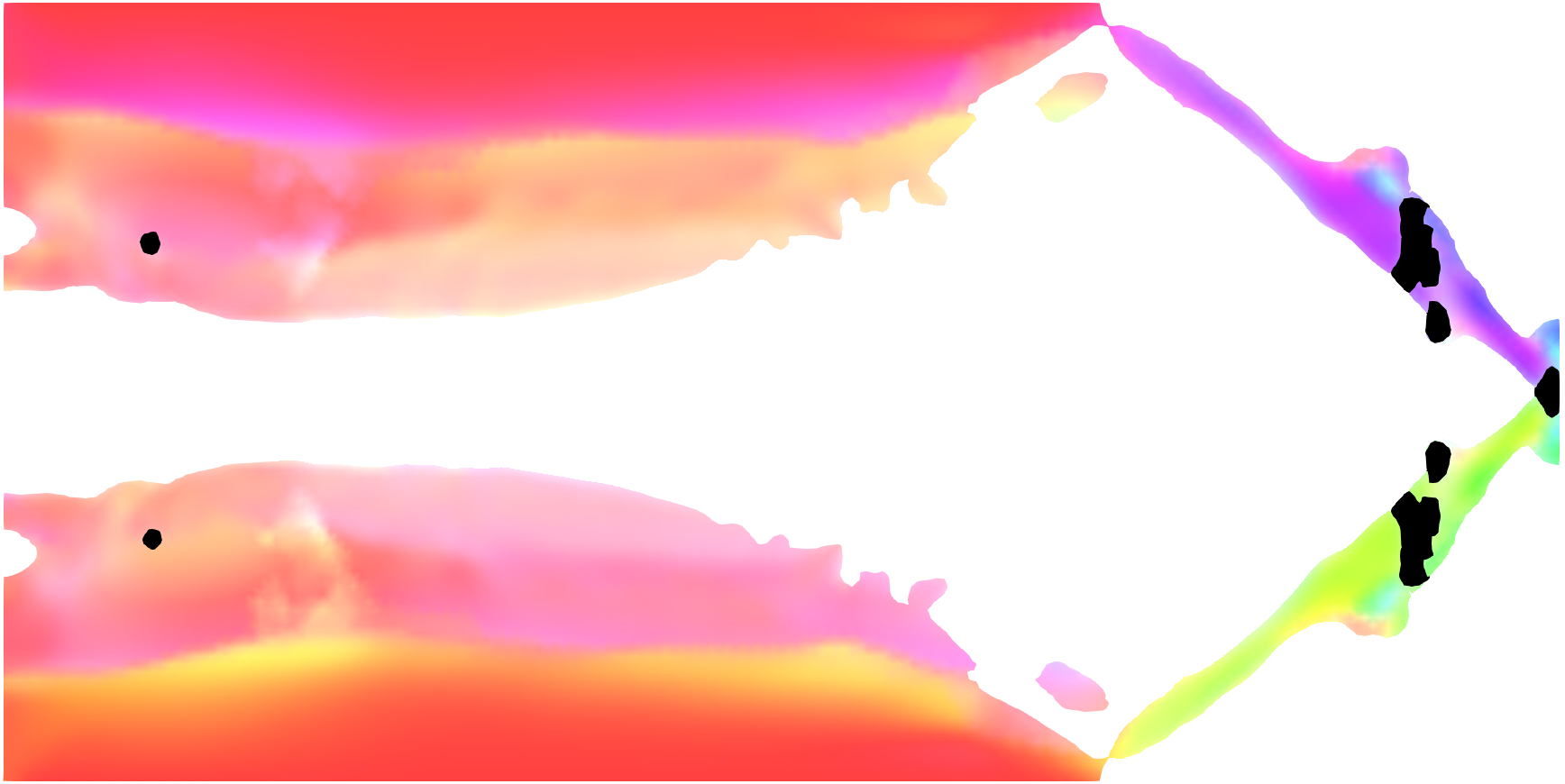}
			\subcaption{Step 20}
		\end{minipage}
		\begin{minipage}[b]{0.3\linewidth}
			\centering
			\includegraphics[width=\linewidth]{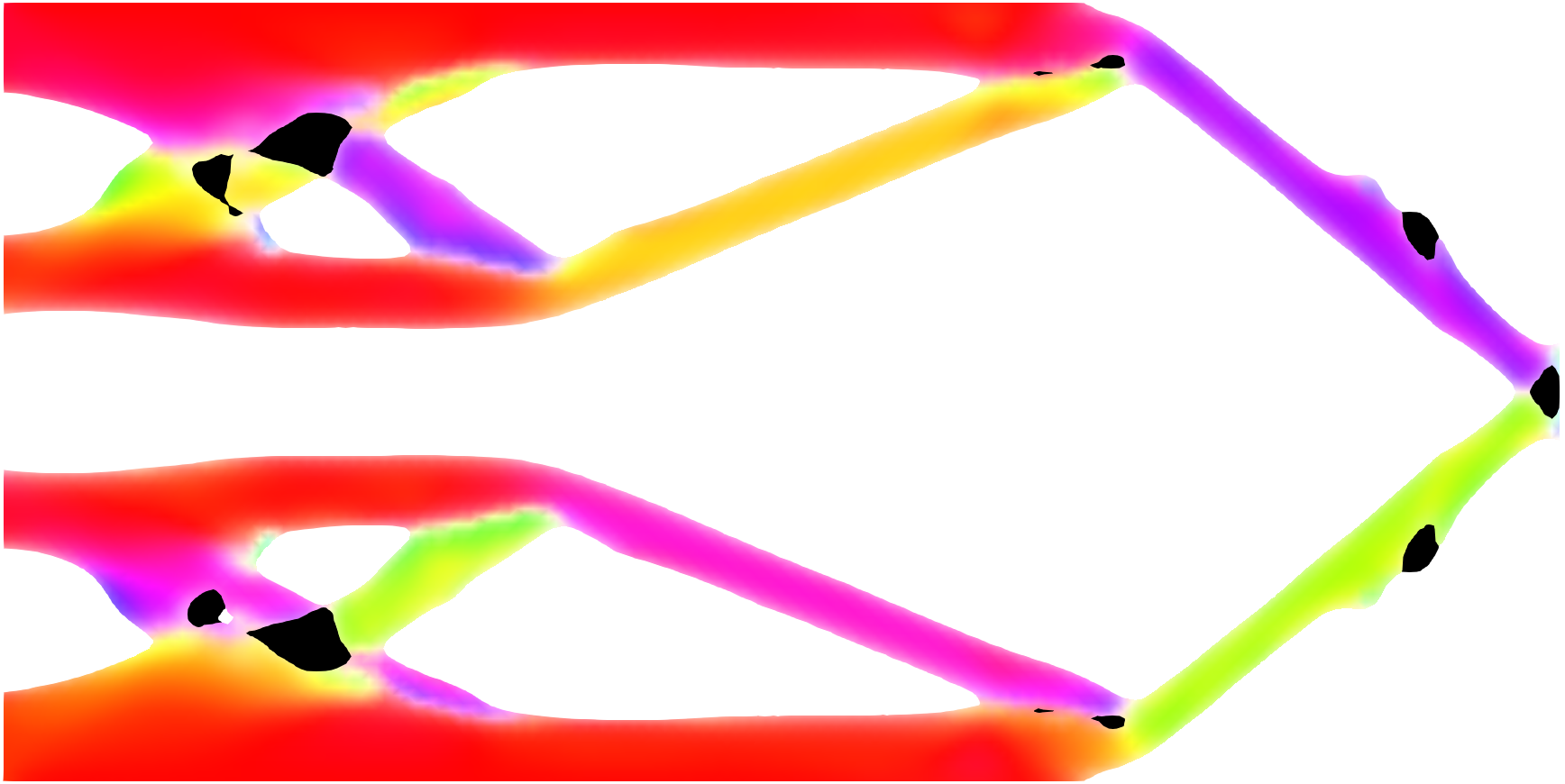}
			\subcaption{Step 50}
		\end{minipage}
		\begin{minipage}[b]{0.3\linewidth}
			\centering
			\includegraphics[width=\linewidth]{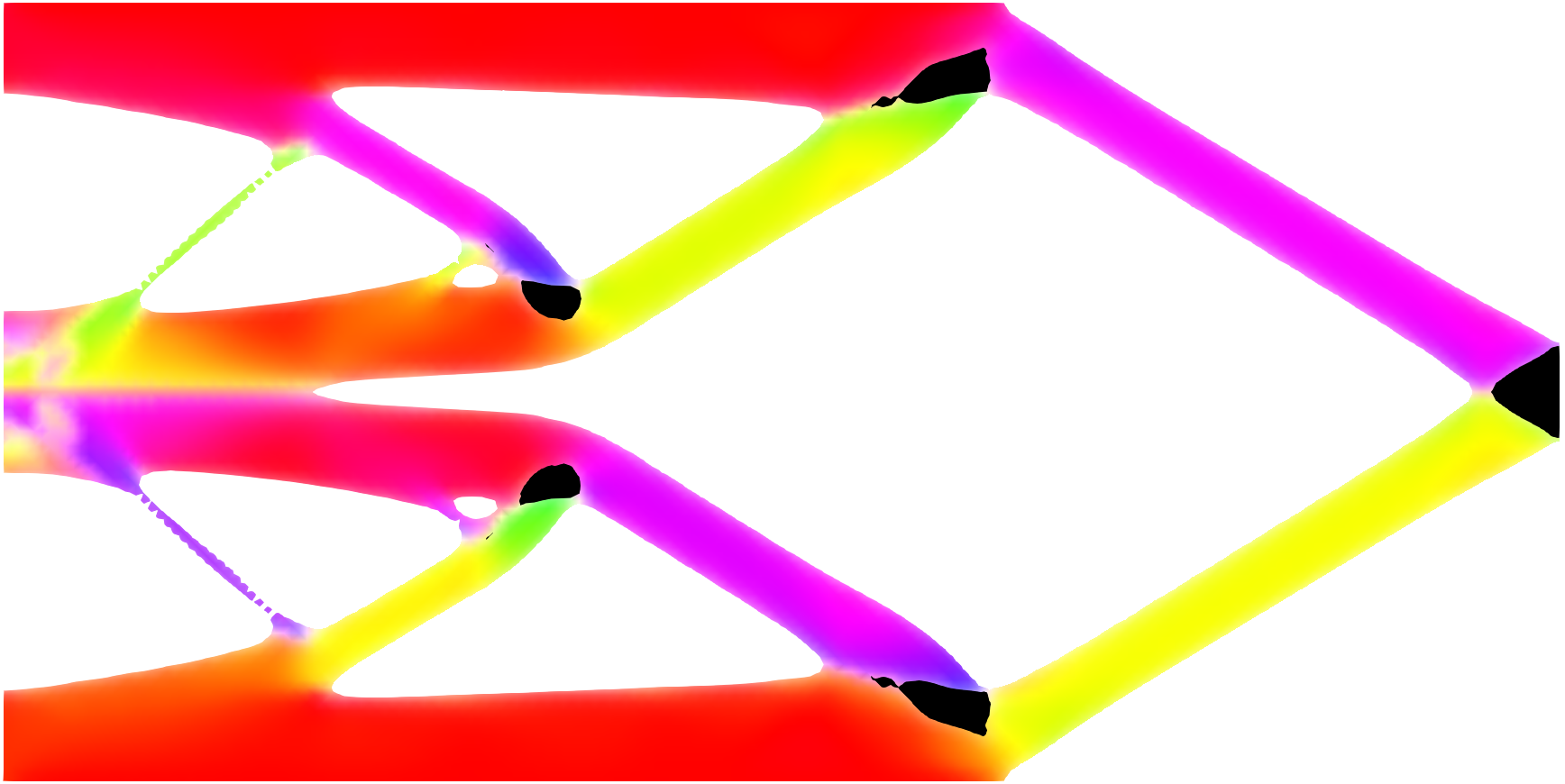}
			\subcaption{Step 200}
		\end{minipage}
		\begin{minipage}[b]{0.3\linewidth}
			\centering
			\includegraphics[width=\linewidth]{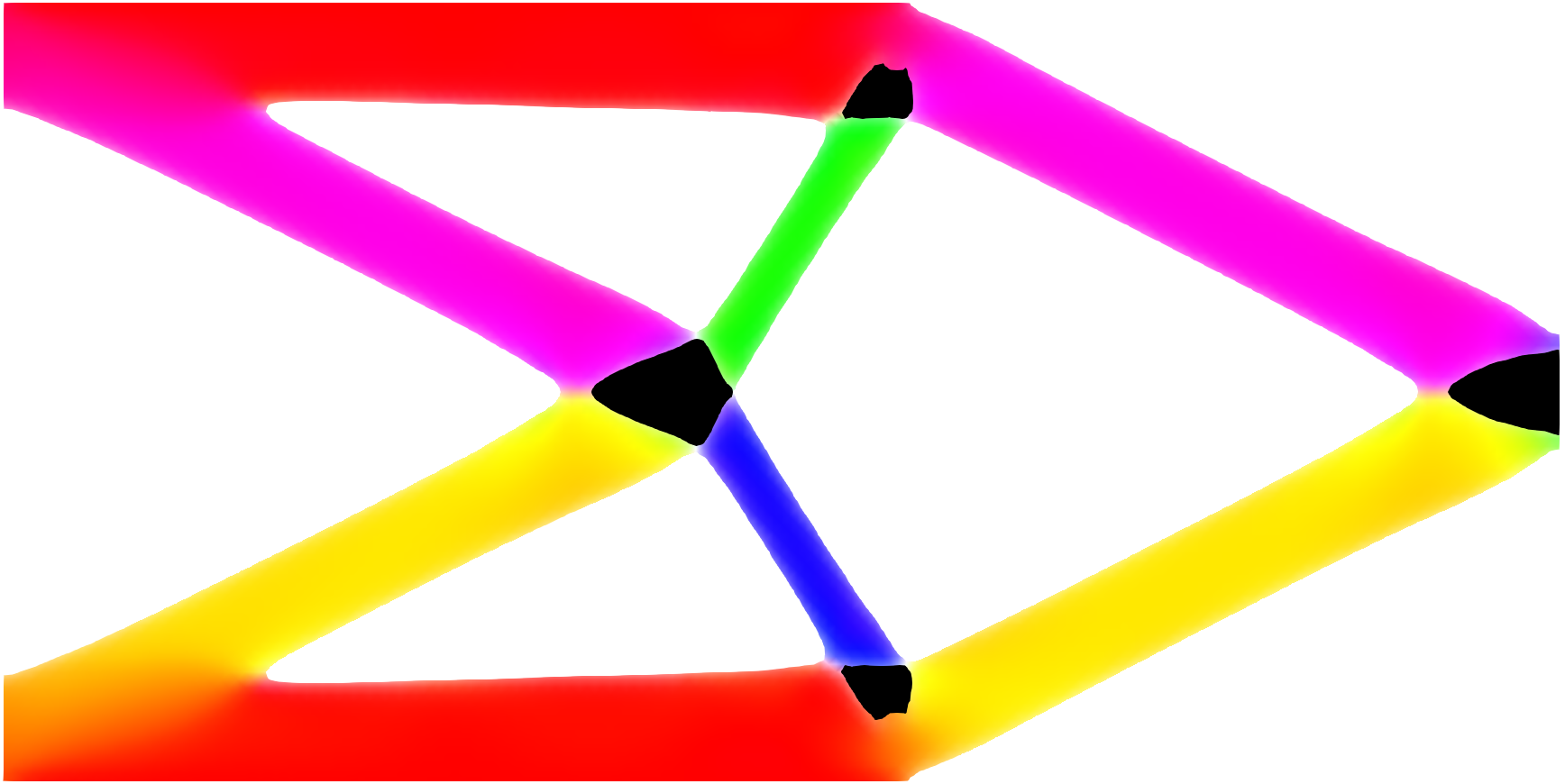}
			\subcaption{Optimal}
		\end{minipage}
		\caption{Optimization result for initial design B.}\label{fig:4initAniso0p5}
	\end{figure}

\begin{figure}
		\centering
		\begin{minipage}[b]{0.3\linewidth}
			\centering
			\includegraphics[width=\linewidth]{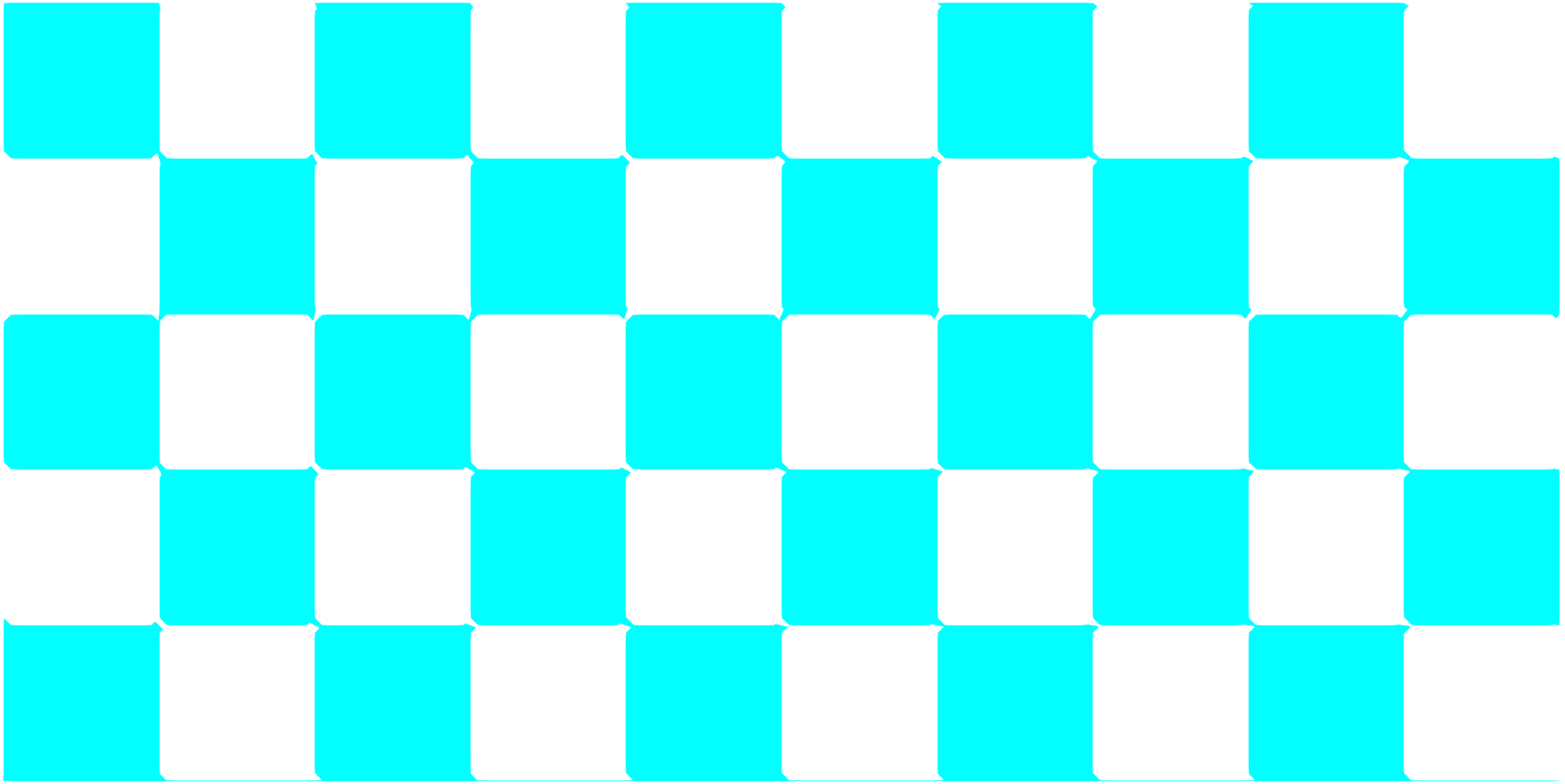}
			\subcaption{Step 0}
		\end{minipage}
		\begin{minipage}[b]{0.3\linewidth}
			\centering
			\includegraphics[width=\linewidth]{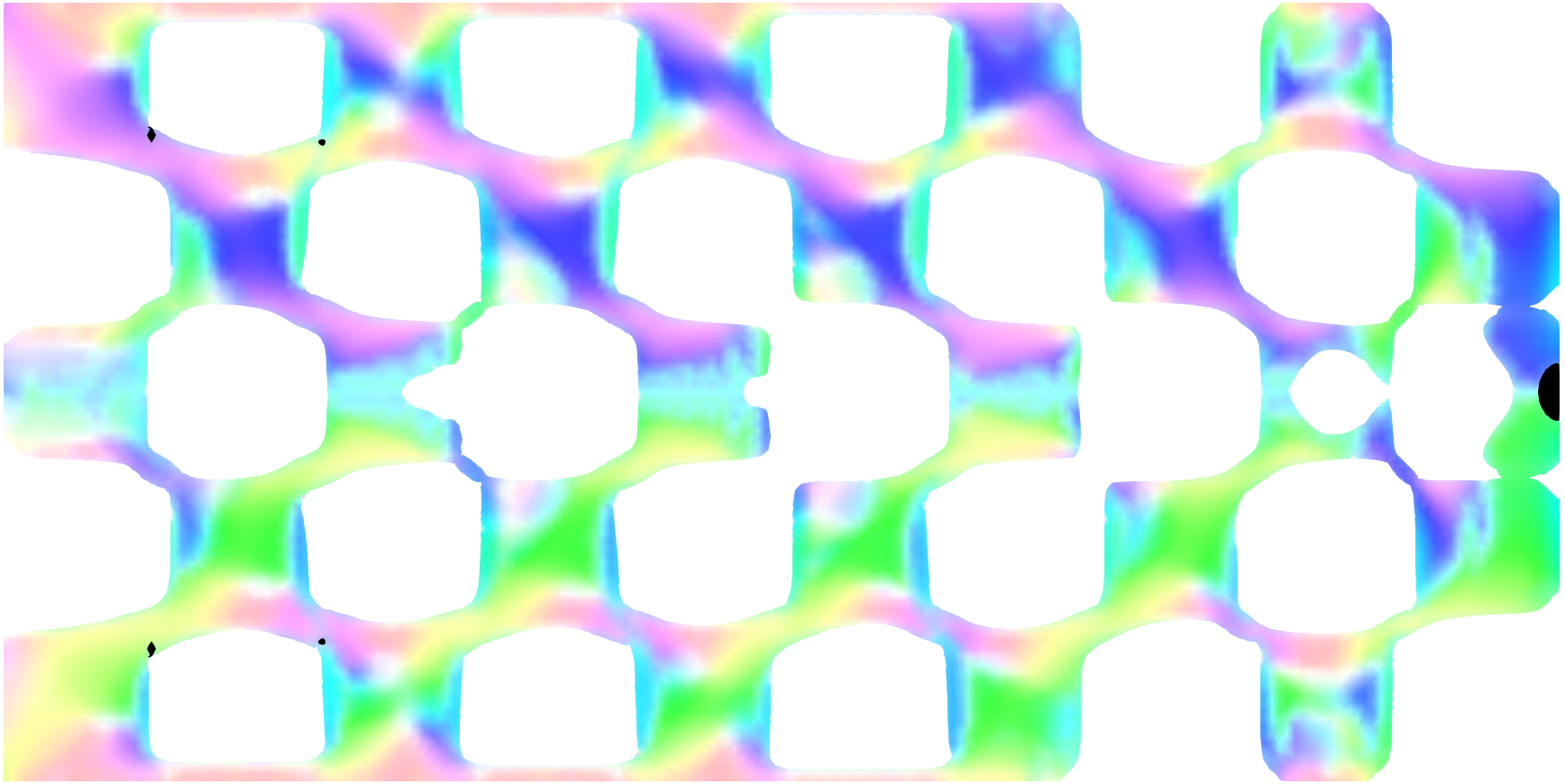}
			\subcaption{Step 10}
		\end{minipage}
		\begin{minipage}[b]{0.3\linewidth}
			\centering
			\includegraphics[width=\linewidth]{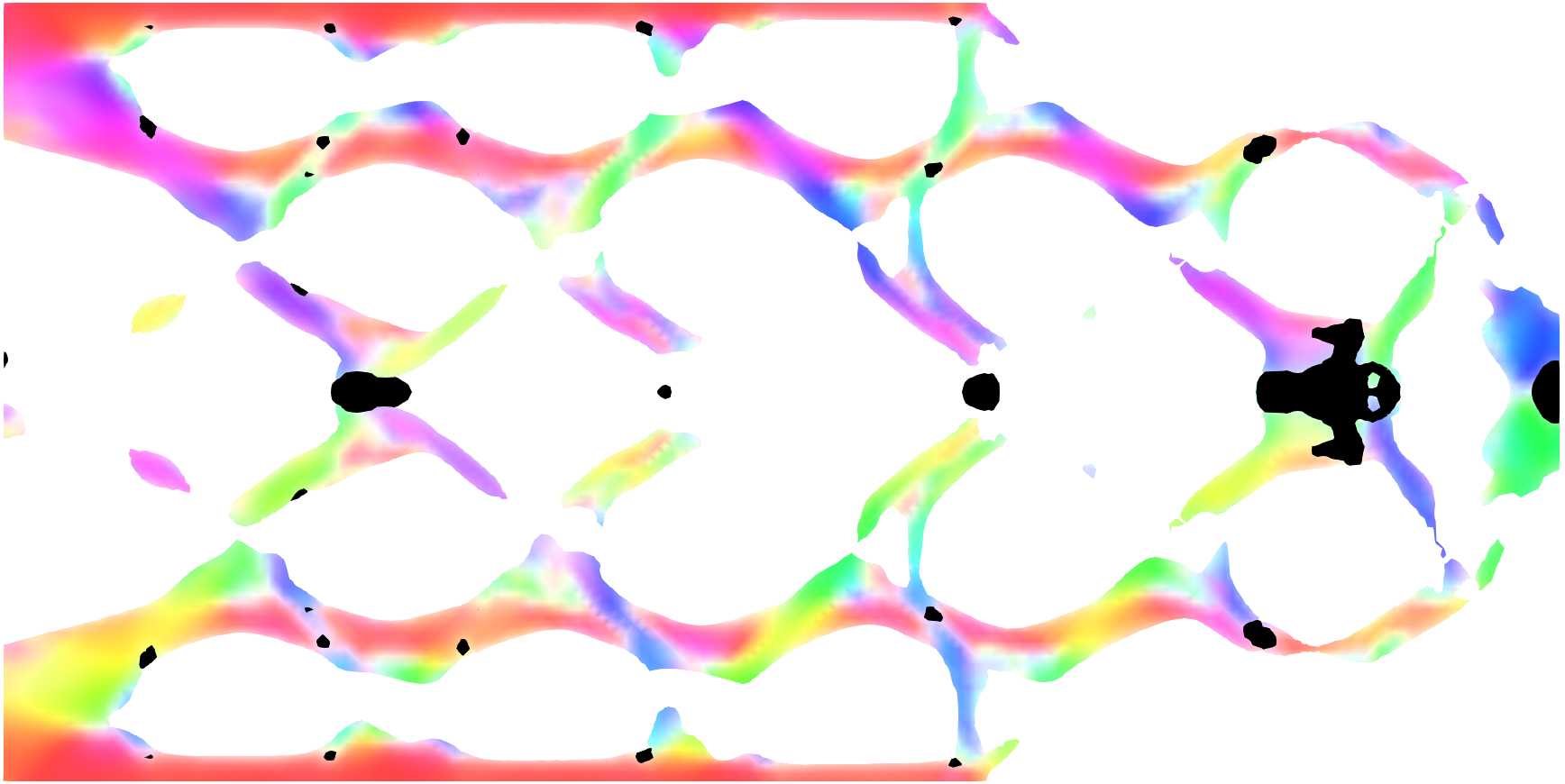}
			\subcaption{Step 20}
		\end{minipage}
		\begin{minipage}[b]{0.3\linewidth}
			\centering
			\includegraphics[width=\linewidth]{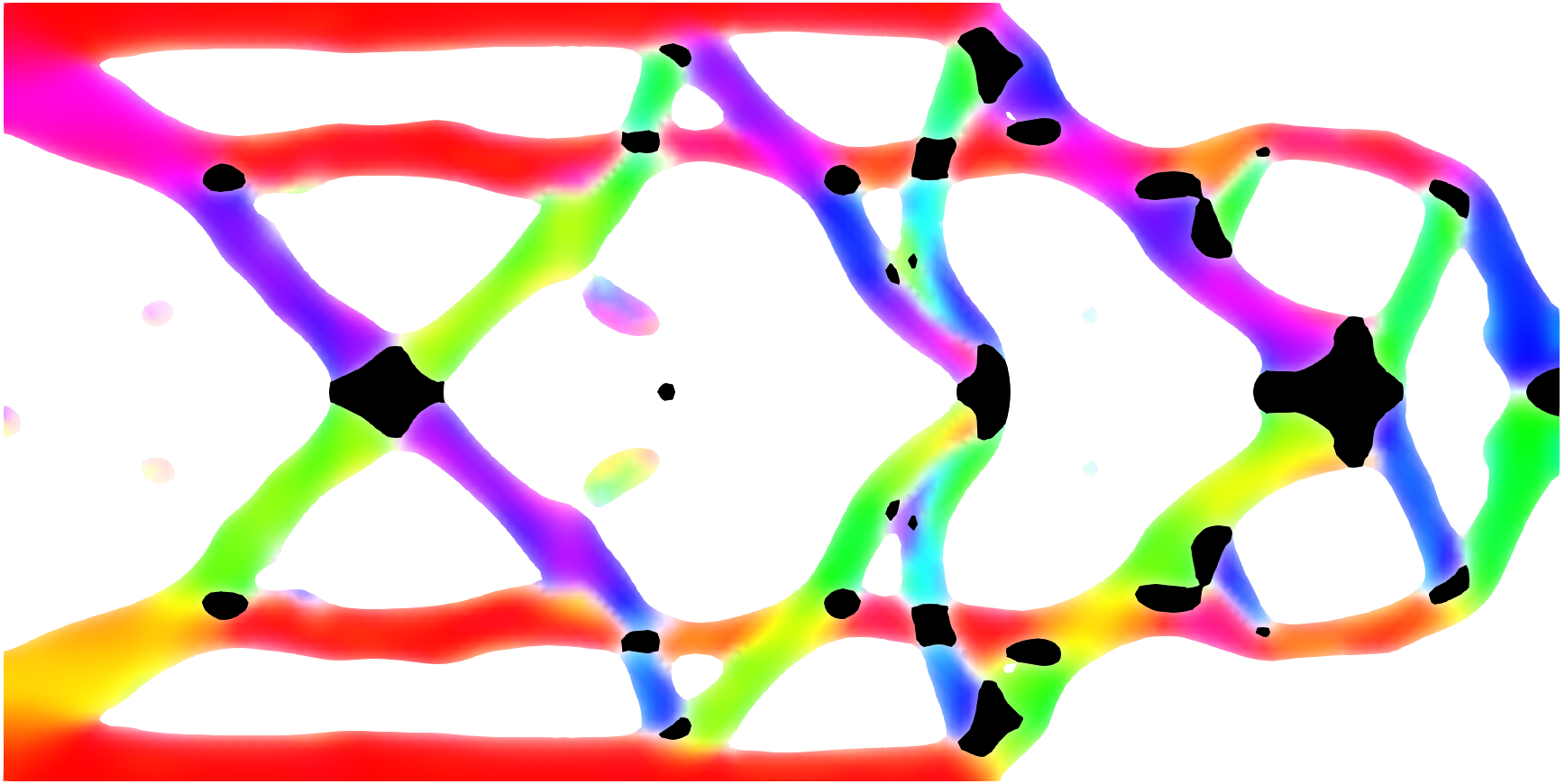}
			\subcaption{Step 50}
		\end{minipage}
		\begin{minipage}[b]{0.3\linewidth}
			\centering
			\includegraphics[width=\linewidth]{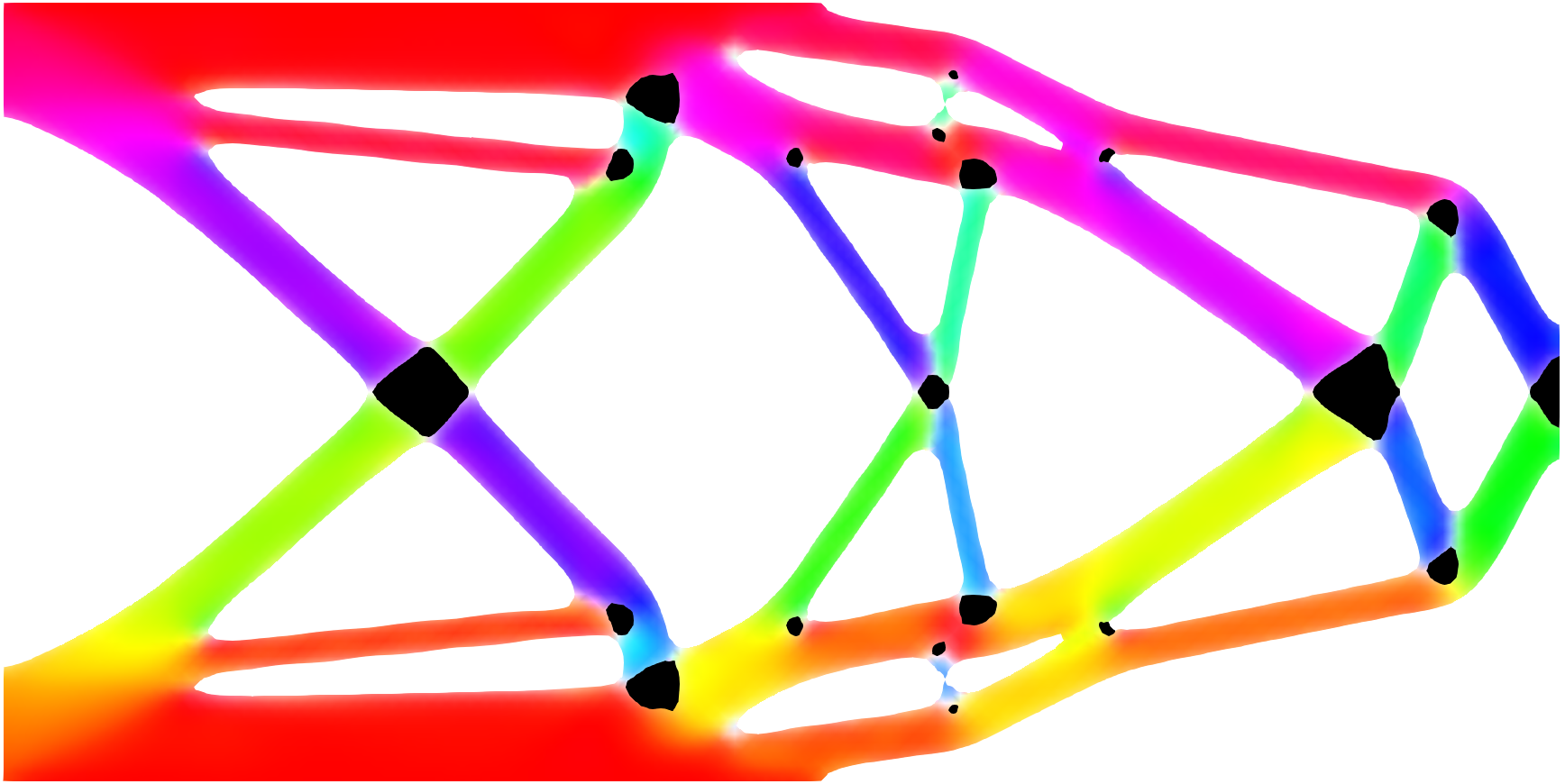}
			\subcaption{Step 200}
		\end{minipage}
		\begin{minipage}[b]{0.3\linewidth}
			\centering
			\includegraphics[width=\linewidth]{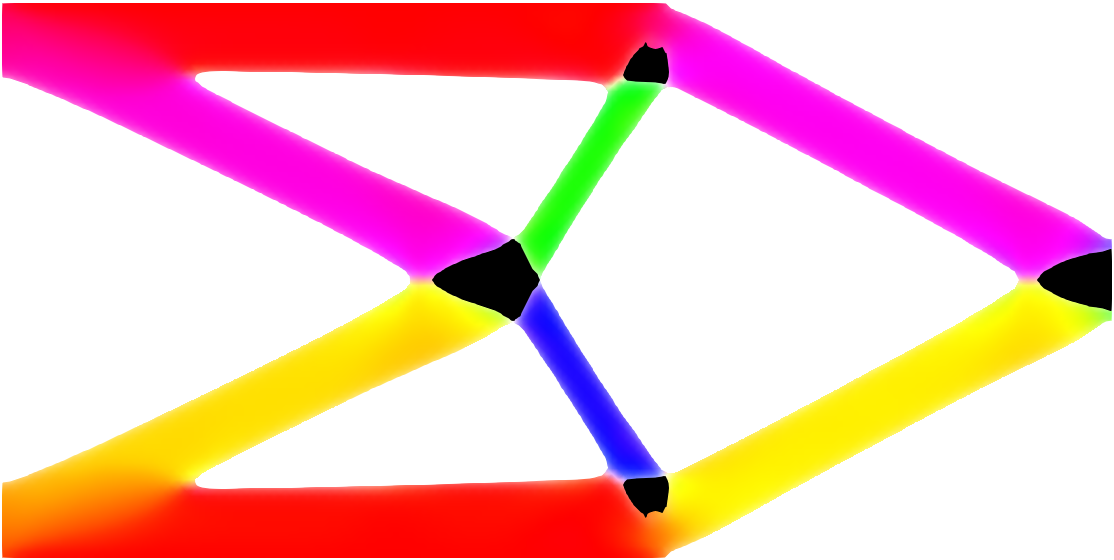}
			\subcaption{Optimal}
		\end{minipage}
		\caption{Optimization result for initial design C.}\label{fig:4initVoidAniso0p5}
	\end{figure}

\begin{figure}
		\centering
		\begin{minipage}[b]{0.3\linewidth}
			\centering
			\includegraphics[width=\linewidth]{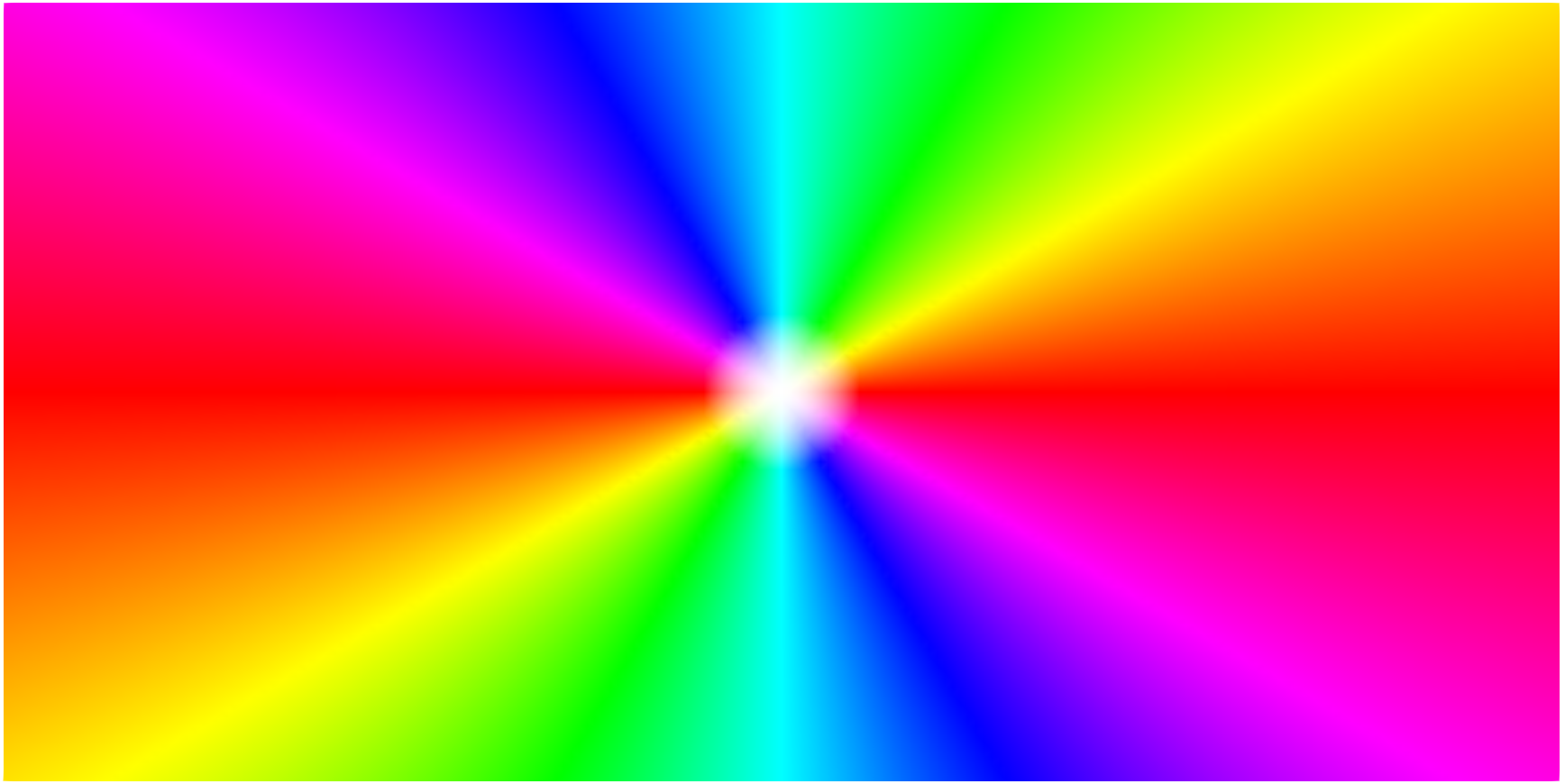}
			\subcaption{Step 0}
		\end{minipage}
		\begin{minipage}[b]{0.3\linewidth}
			\centering
			\includegraphics[width=\linewidth]{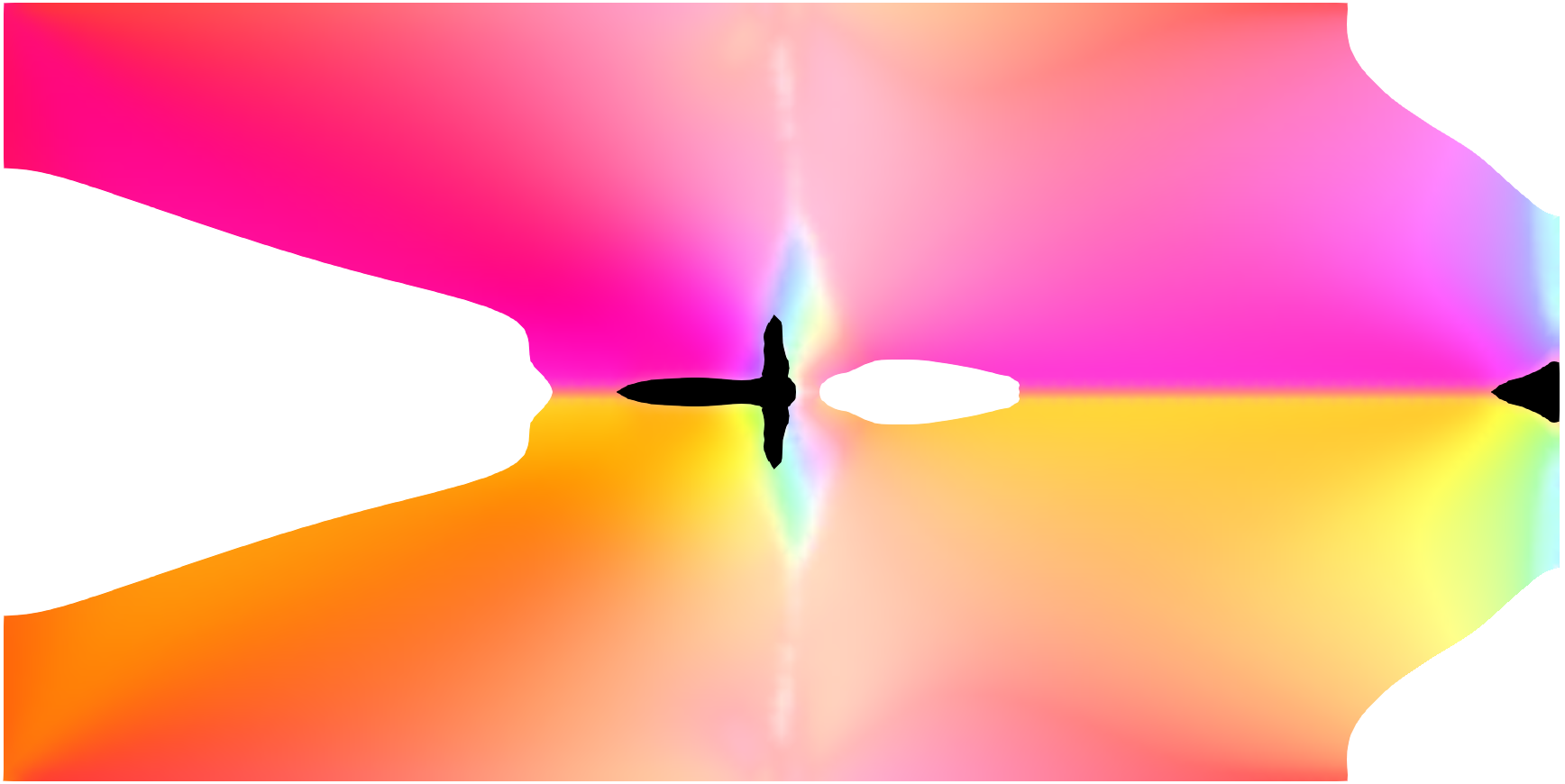}
			\subcaption{Step 10}
		\end{minipage}
		\begin{minipage}[b]{0.3\linewidth}
			\centering
			\includegraphics[width=\linewidth]{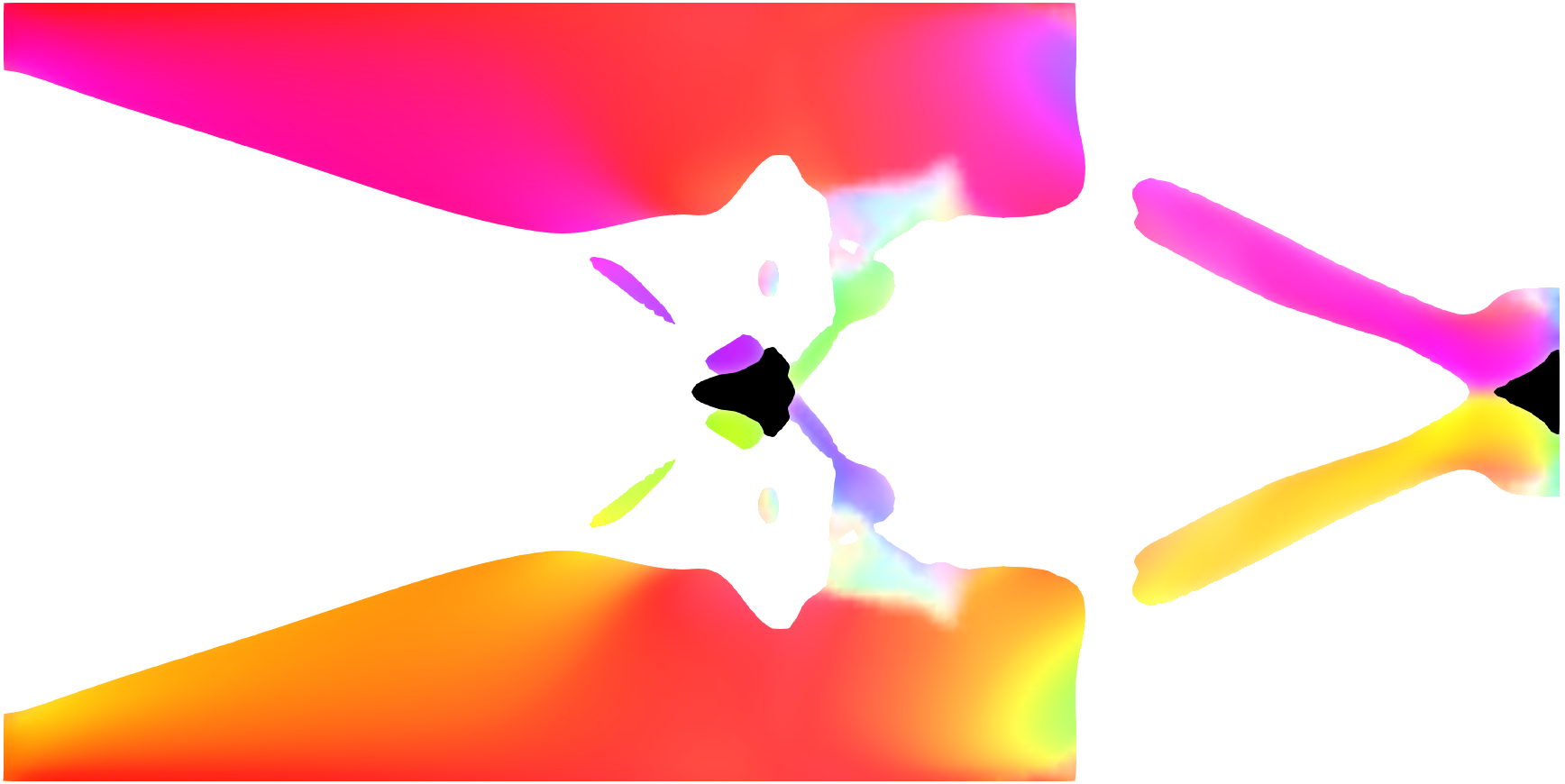}
			\subcaption{Step 20}
		\end{minipage}
		\begin{minipage}[b]{0.3\linewidth}
			\centering
			\includegraphics[width=\linewidth]{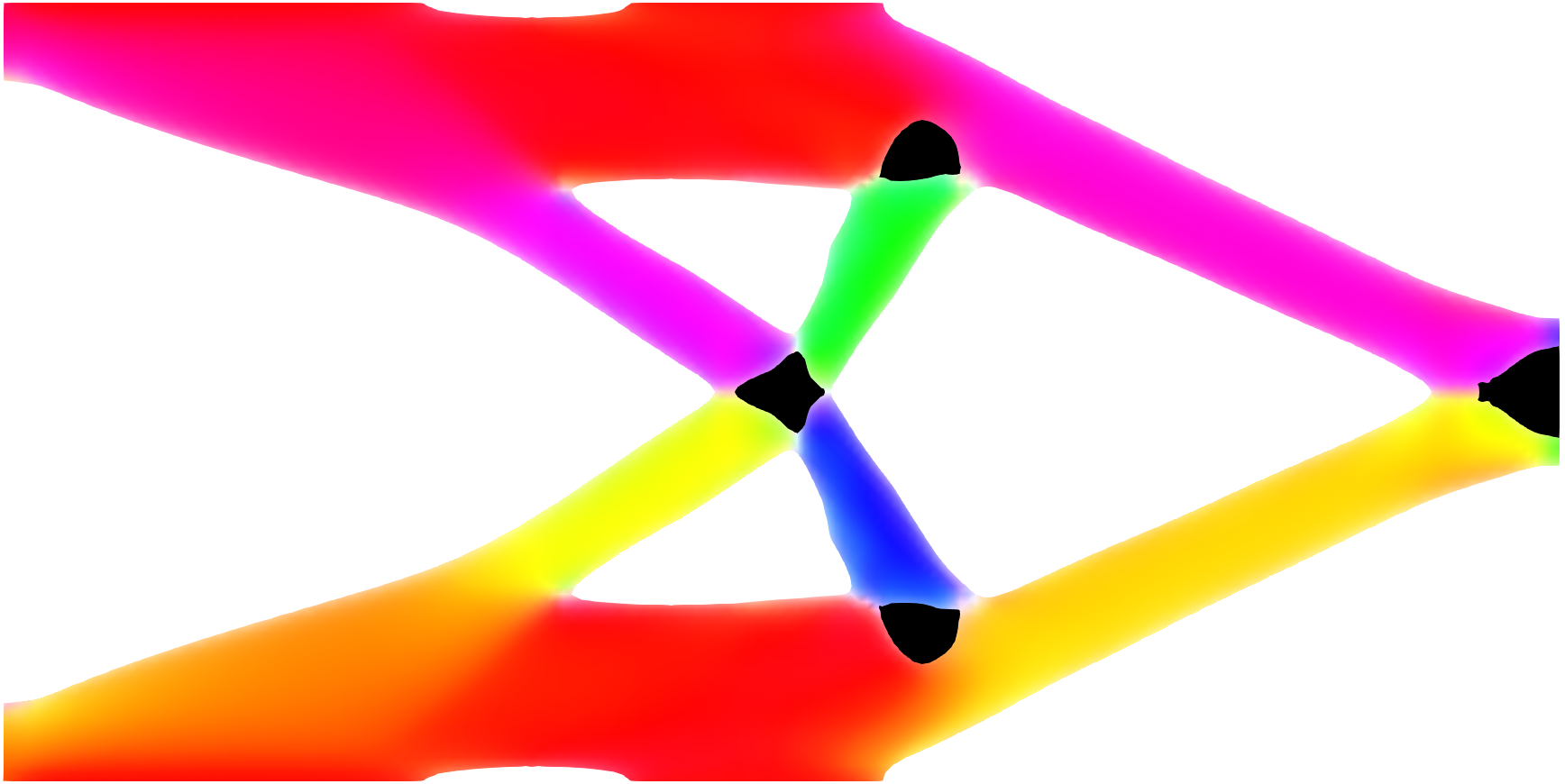}
			\subcaption{Step 50}
		\end{minipage}
		\begin{minipage}[b]{0.3\linewidth}
			\centering
			\includegraphics[width=\linewidth]{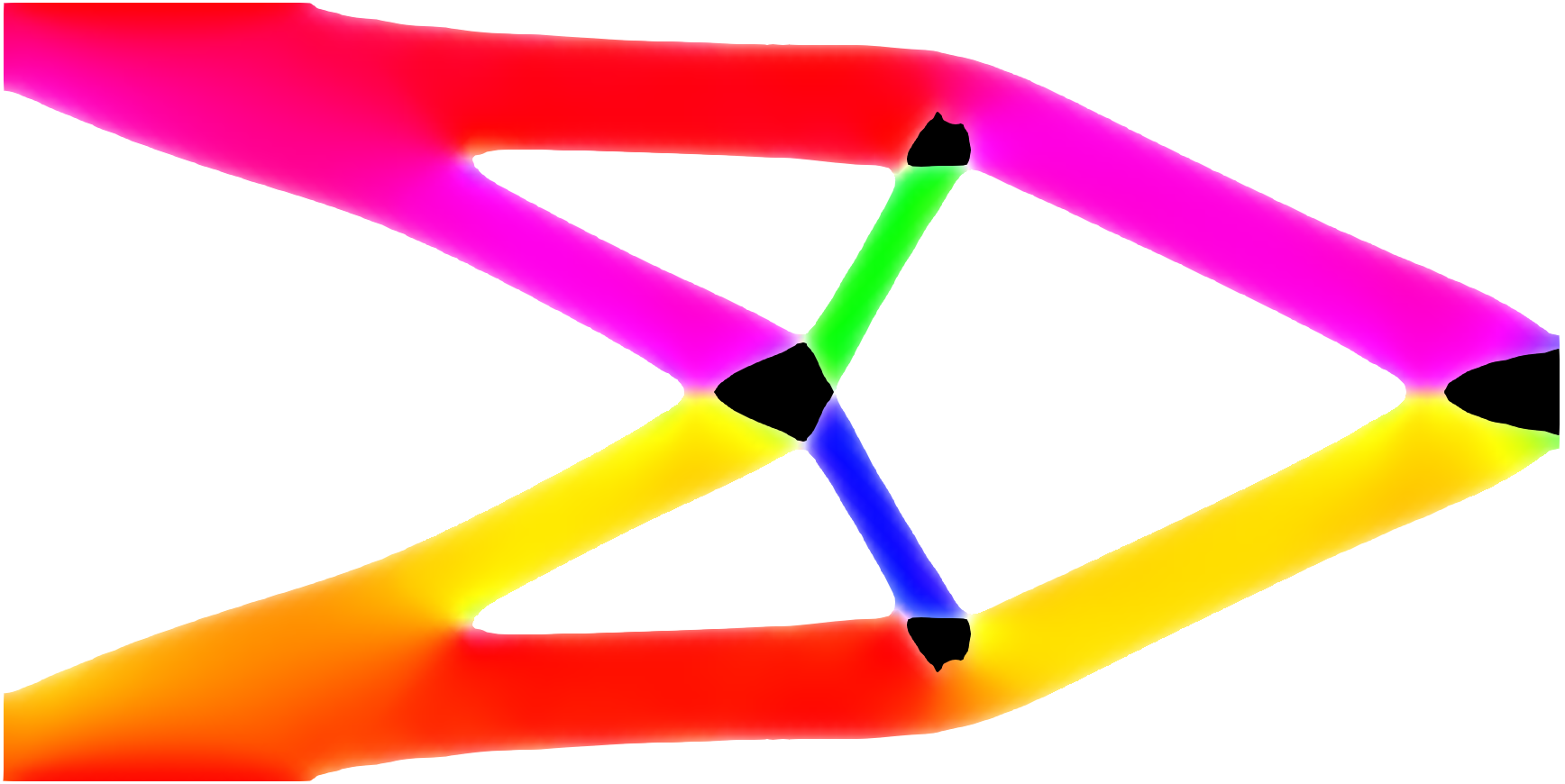}
			\subcaption{Step 200}
		\end{minipage}
		\begin{minipage}[b]{0.3\linewidth}
			\centering
			\includegraphics[width=\linewidth]{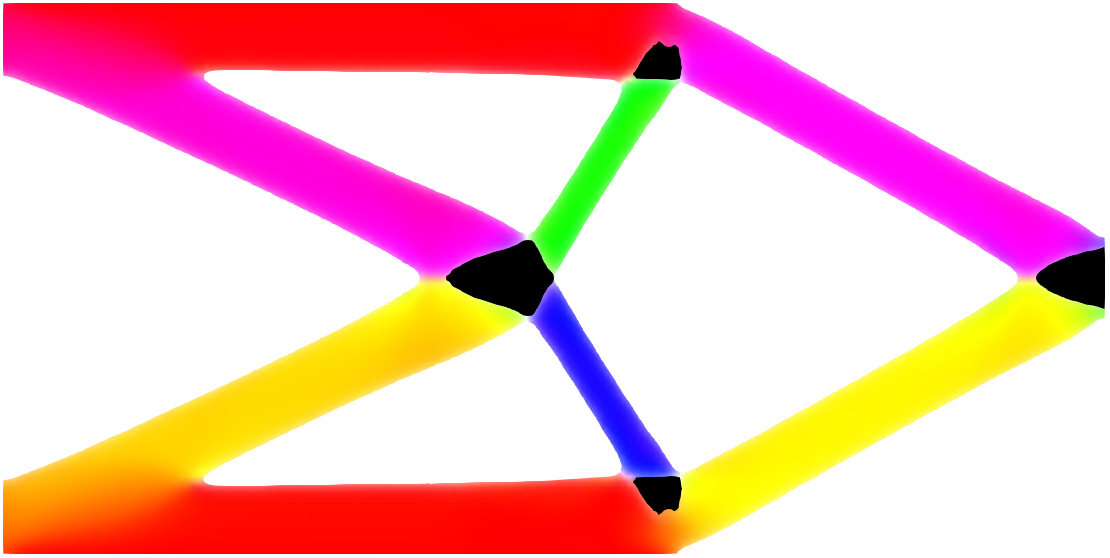}
			\subcaption{Optimal}
		\end{minipage}
		\caption{Optimization result for initial design D.}\label{fig:4initAnisoRadial}
	\end{figure}

\begin{figure}
		\centering
		\begin{minipage}[b]{0.3\linewidth}
			\centering
			\includegraphics[width=\linewidth]{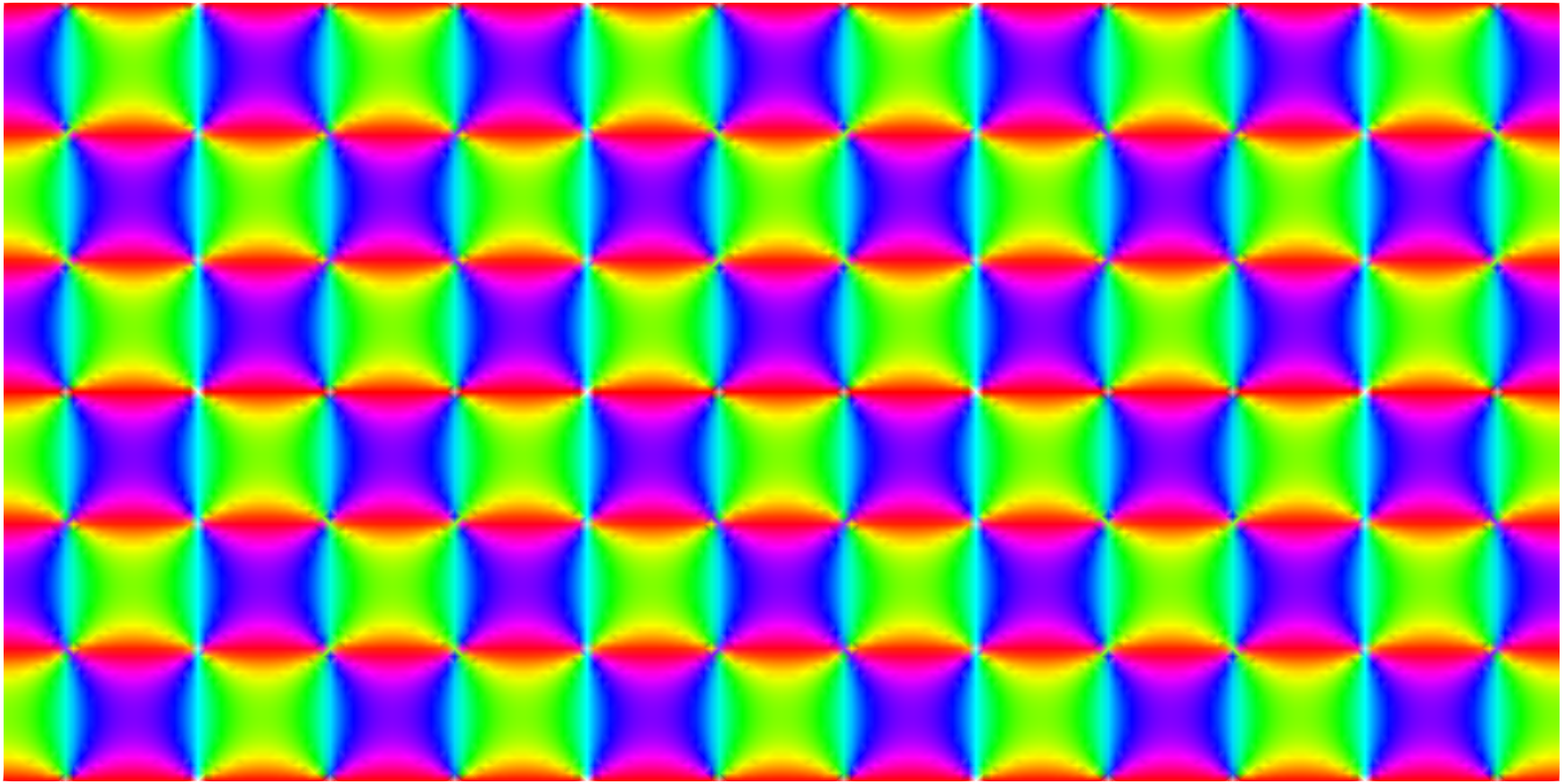}
			\subcaption{Step 0}
		\end{minipage}
		\begin{minipage}[b]{0.3\linewidth}
			\centering
			\includegraphics[width=\linewidth]{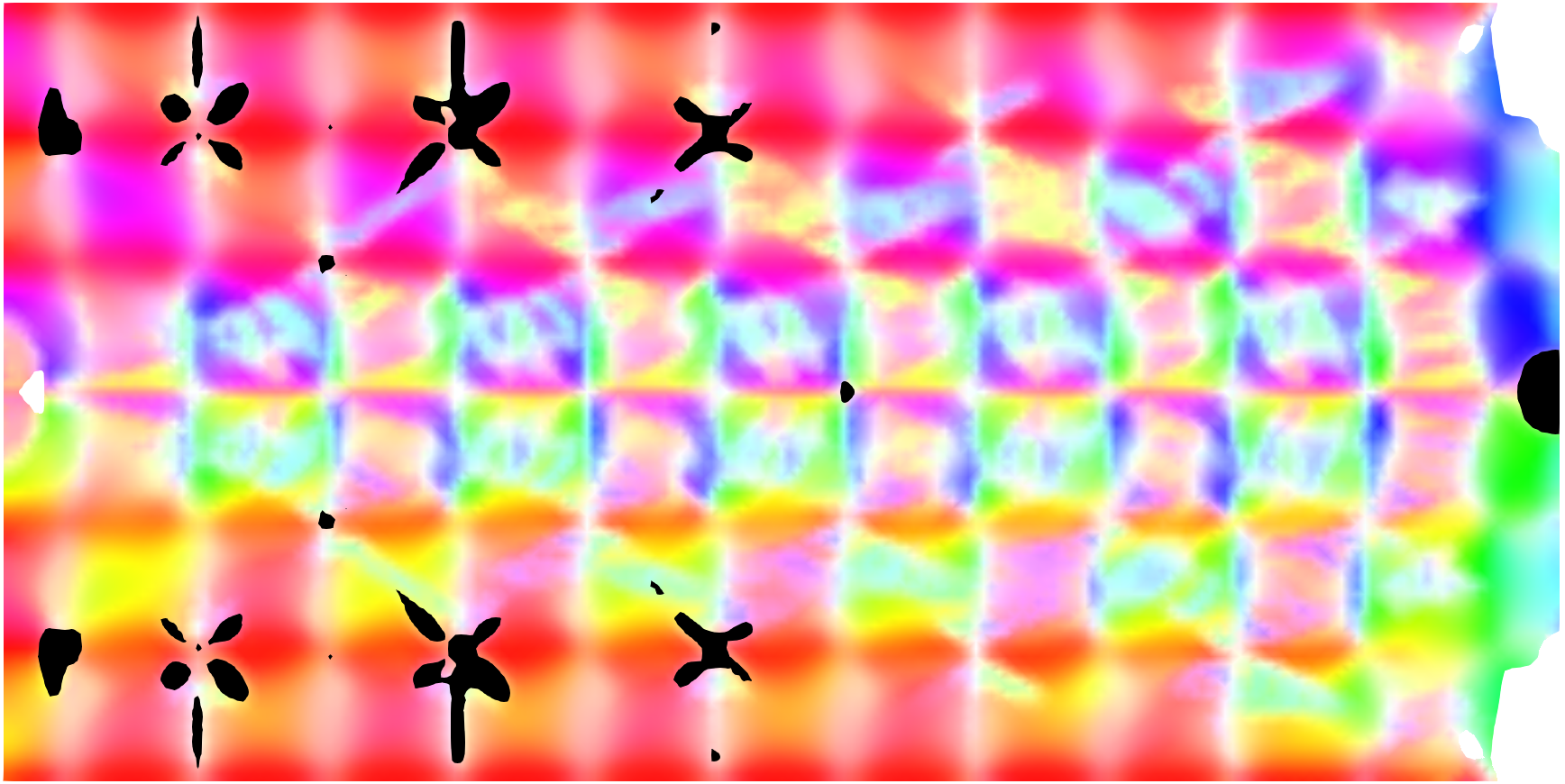}
			\subcaption{Step 10}
		\end{minipage}
		\begin{minipage}[b]{0.3\linewidth}
			\centering
			\includegraphics[width=\linewidth]{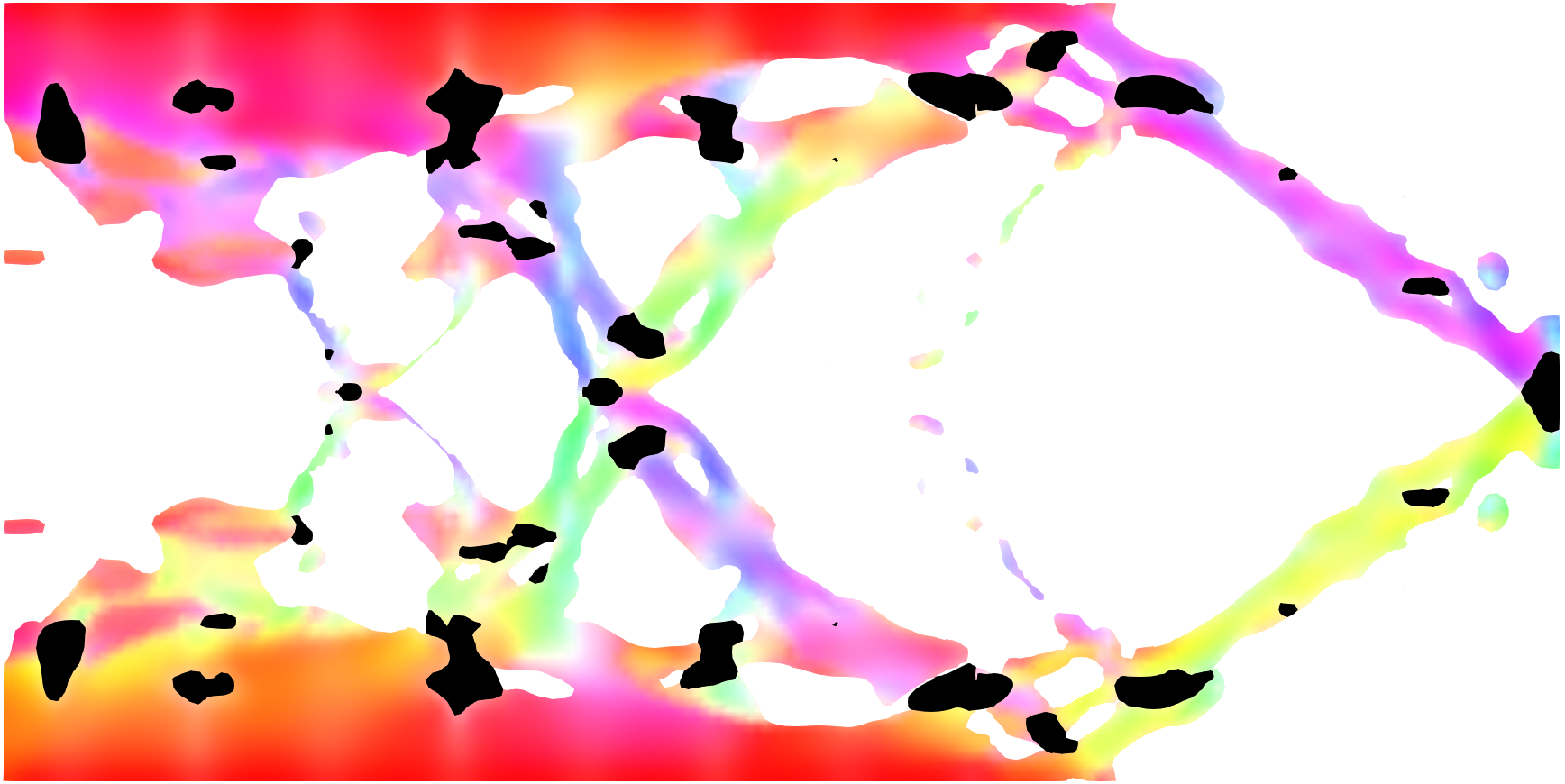}
			\subcaption{Step 20}
		\end{minipage}
		\begin{minipage}[b]{0.3\linewidth}
			\centering
			\includegraphics[width=\linewidth]{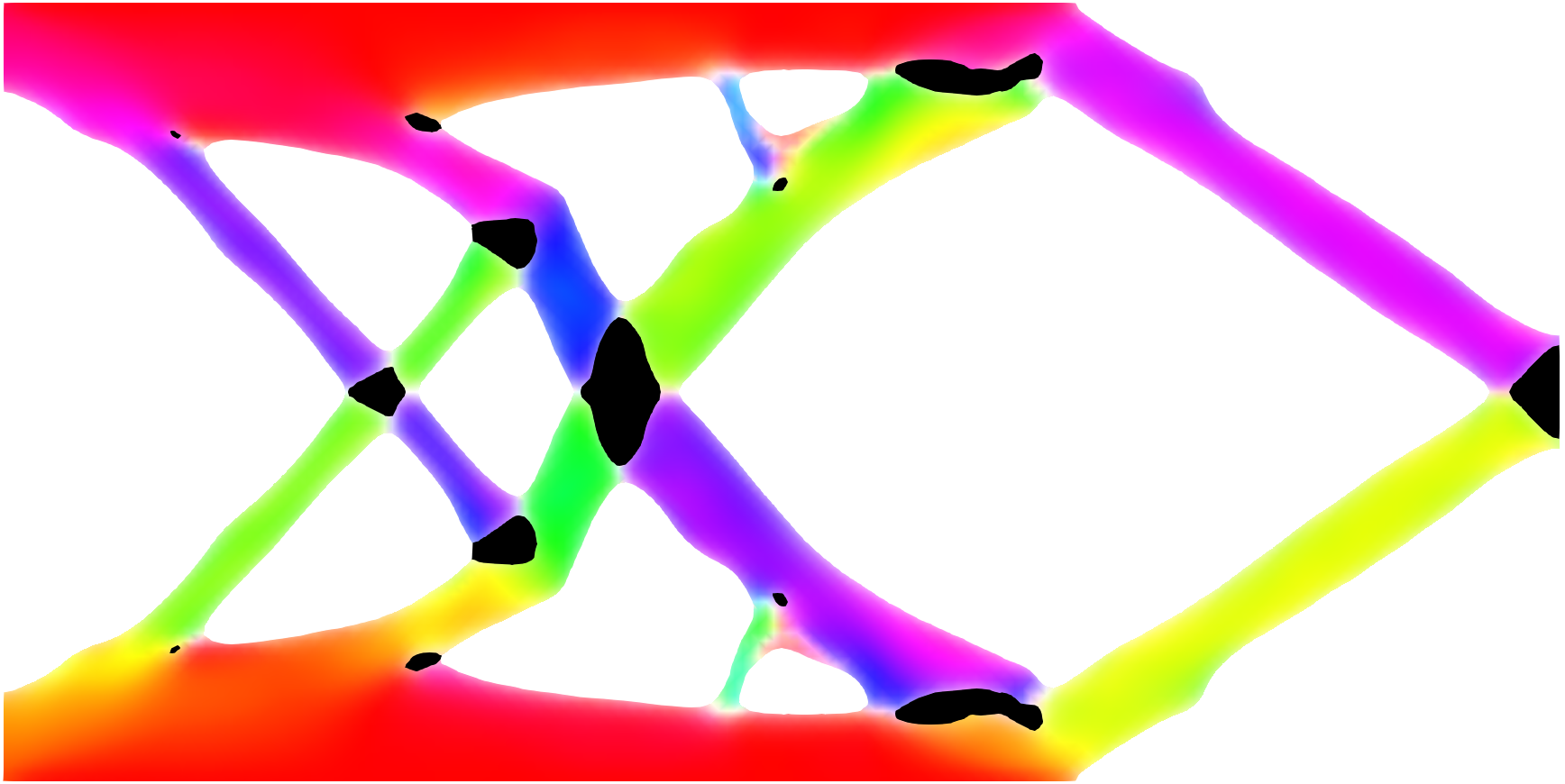}
			\subcaption{Step 50}
		\end{minipage}
		\begin{minipage}[b]{0.3\linewidth}
			\centering
			\includegraphics[width=\linewidth]{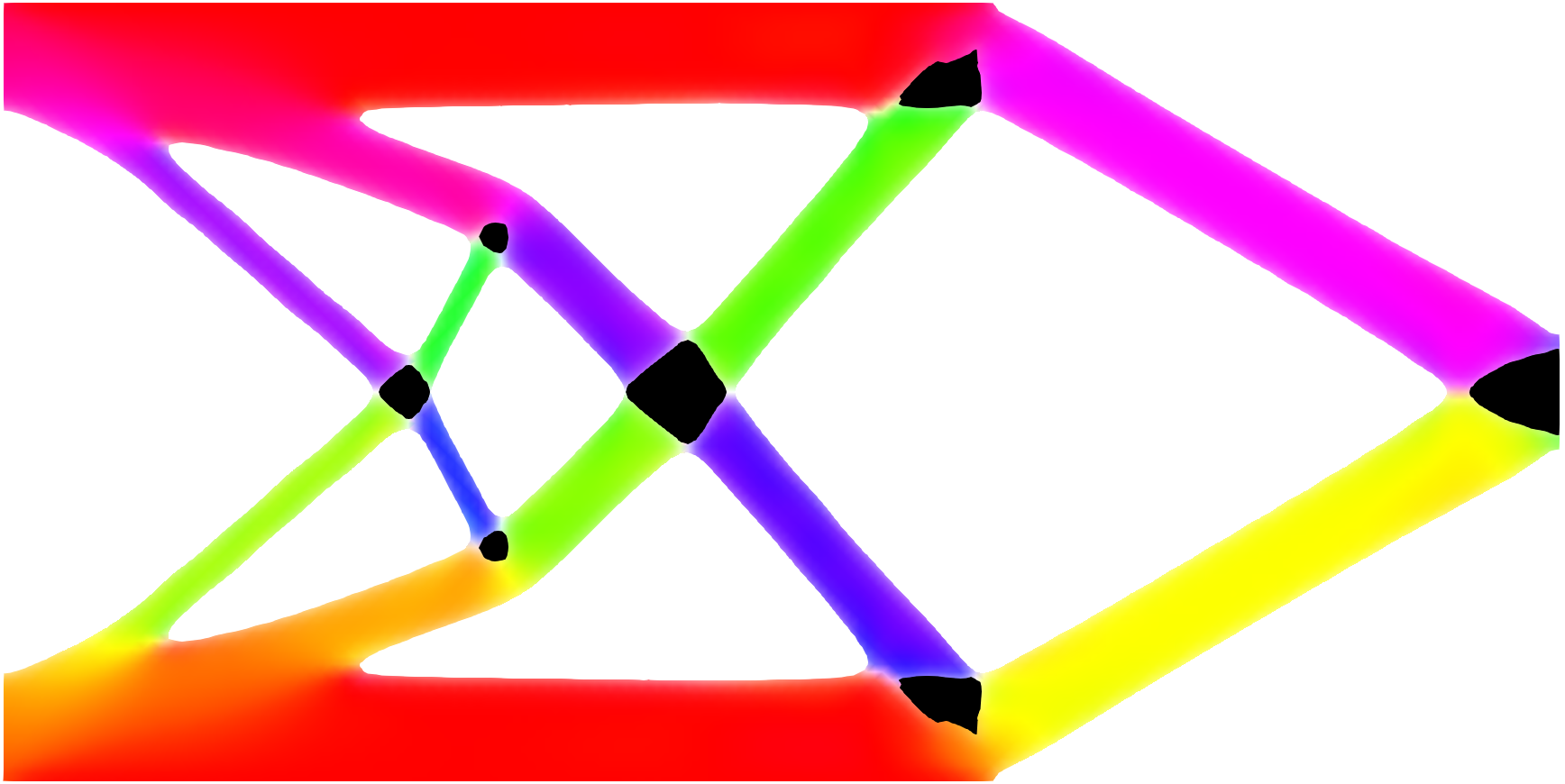}
			\subcaption{Step 200}
		\end{minipage}
		\begin{minipage}[b]{0.3\linewidth}
			\centering
			\includegraphics[width=\linewidth]{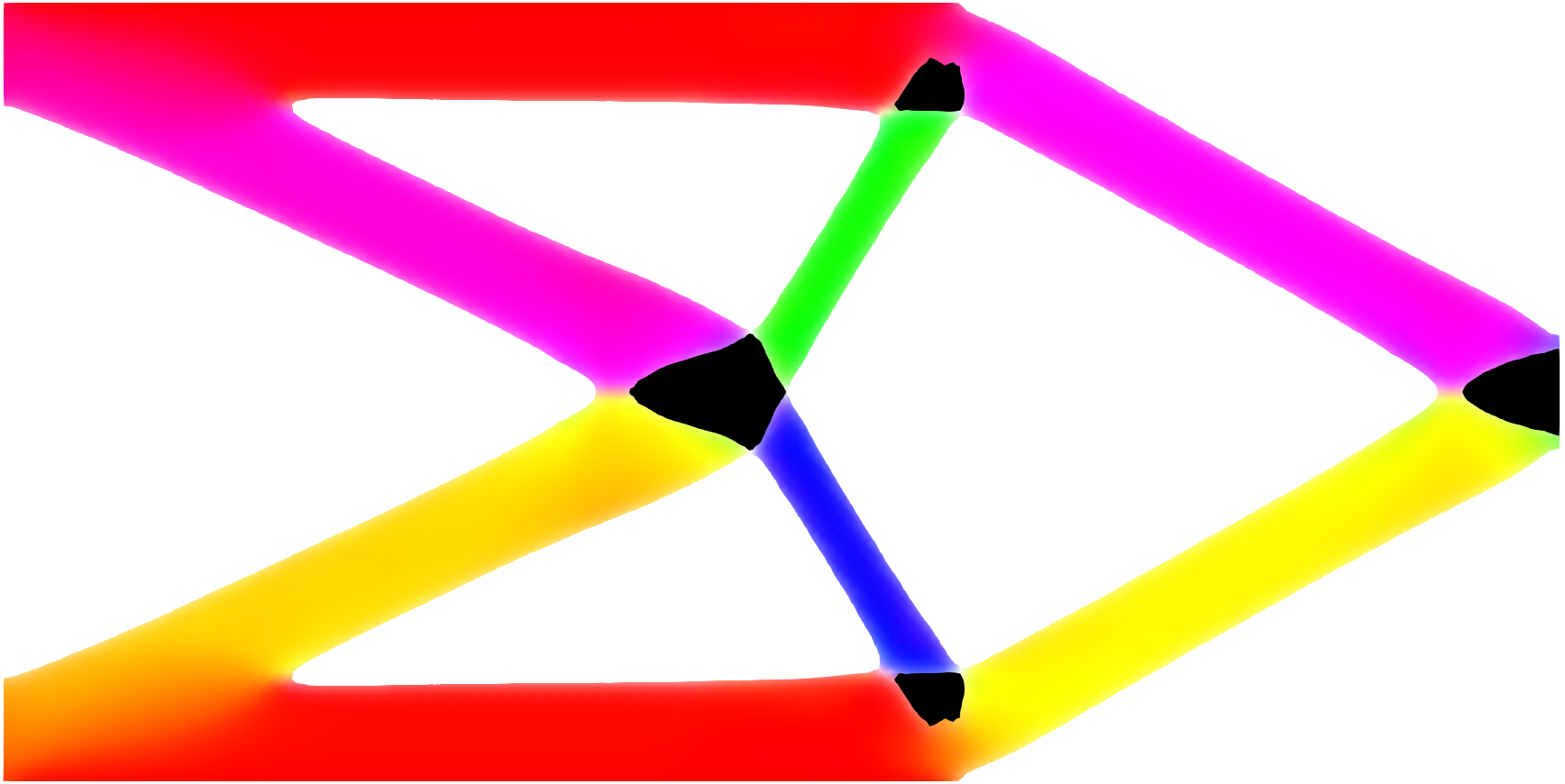}
			\subcaption{Optimal}
		\end{minipage}
		\caption{Optimization result for initial design E.}\label{fig:4initAnisoCase9}
	\end{figure}

	Initial designs B--E and the results are shown in Figs.~\ref{fig:4init0}--\ref{fig:4initAnisoCase9}.
	
	Figs.~\ref{fig:4init0}--\ref{fig:4initAnisoCase9} show that the material arrangement and orientation angle change drastically from the initial structure, converging to a similar optimal solution in all cases.
	
\begin{figure}
		\centering
		\begin{minipage}[b]{0.3\linewidth}
			\centering
			\includegraphics[width=\linewidth]{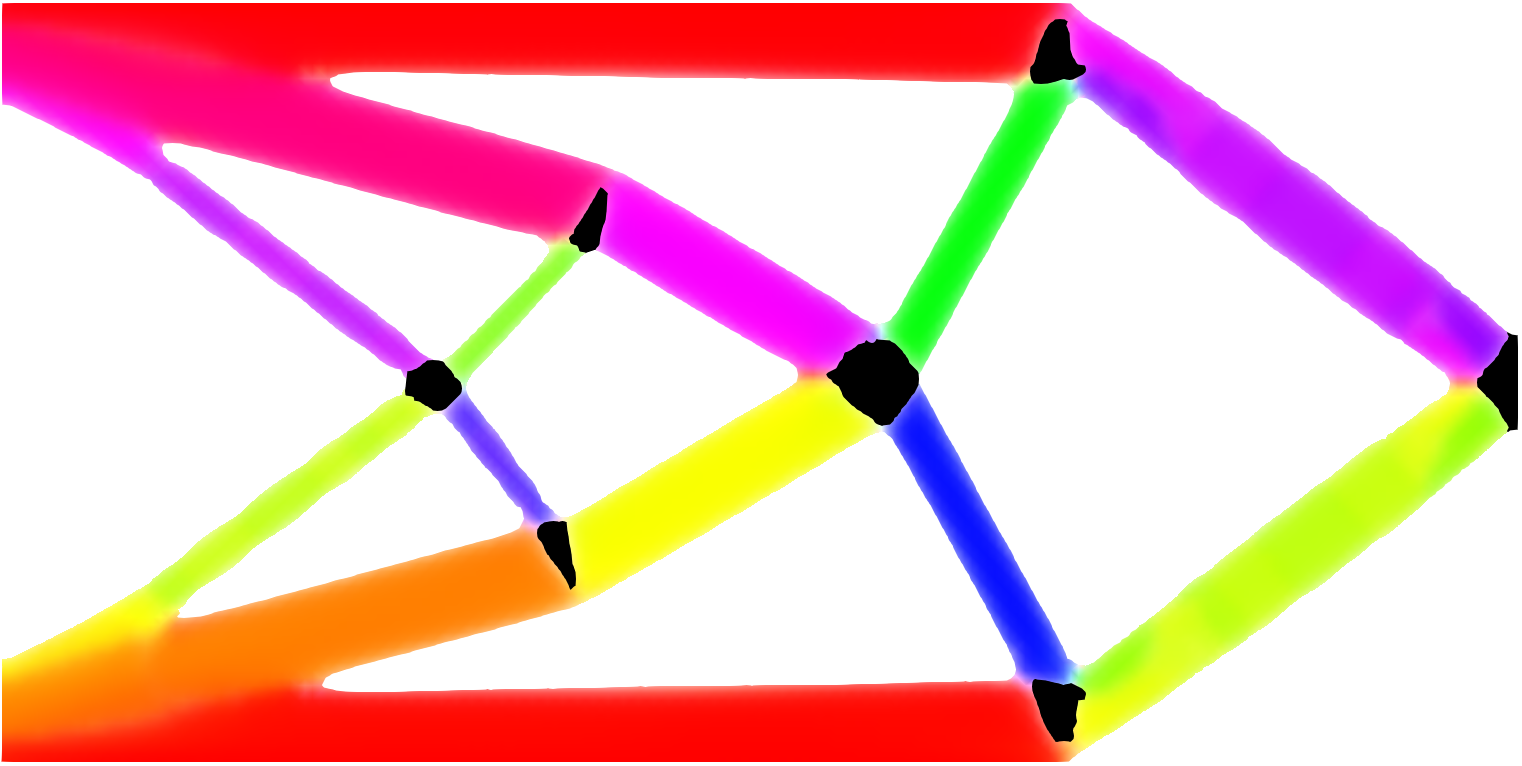}
			\subcaption{$E^\text{back}/E^\text{fib}=0.01$}
		\end{minipage}
		\begin{minipage}[b]{0.3\linewidth}
			\centering
			\includegraphics[width=\linewidth]{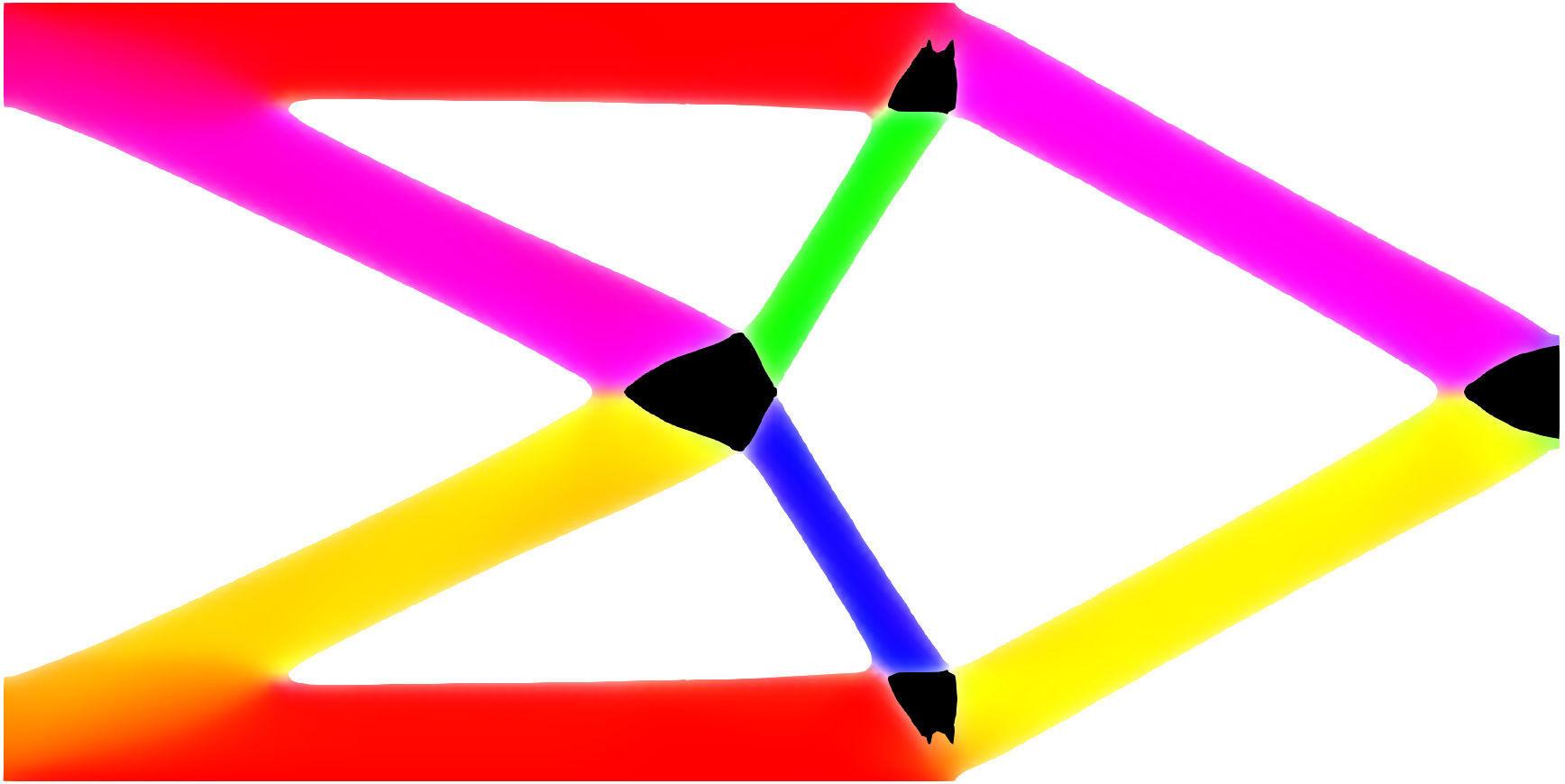}
			\subcaption{$E^\text{back}/E^\text{fib}=0.1$}
		\end{minipage}
		\begin{minipage}[b]{0.3\linewidth}
			\centering
			\includegraphics[width=\linewidth]{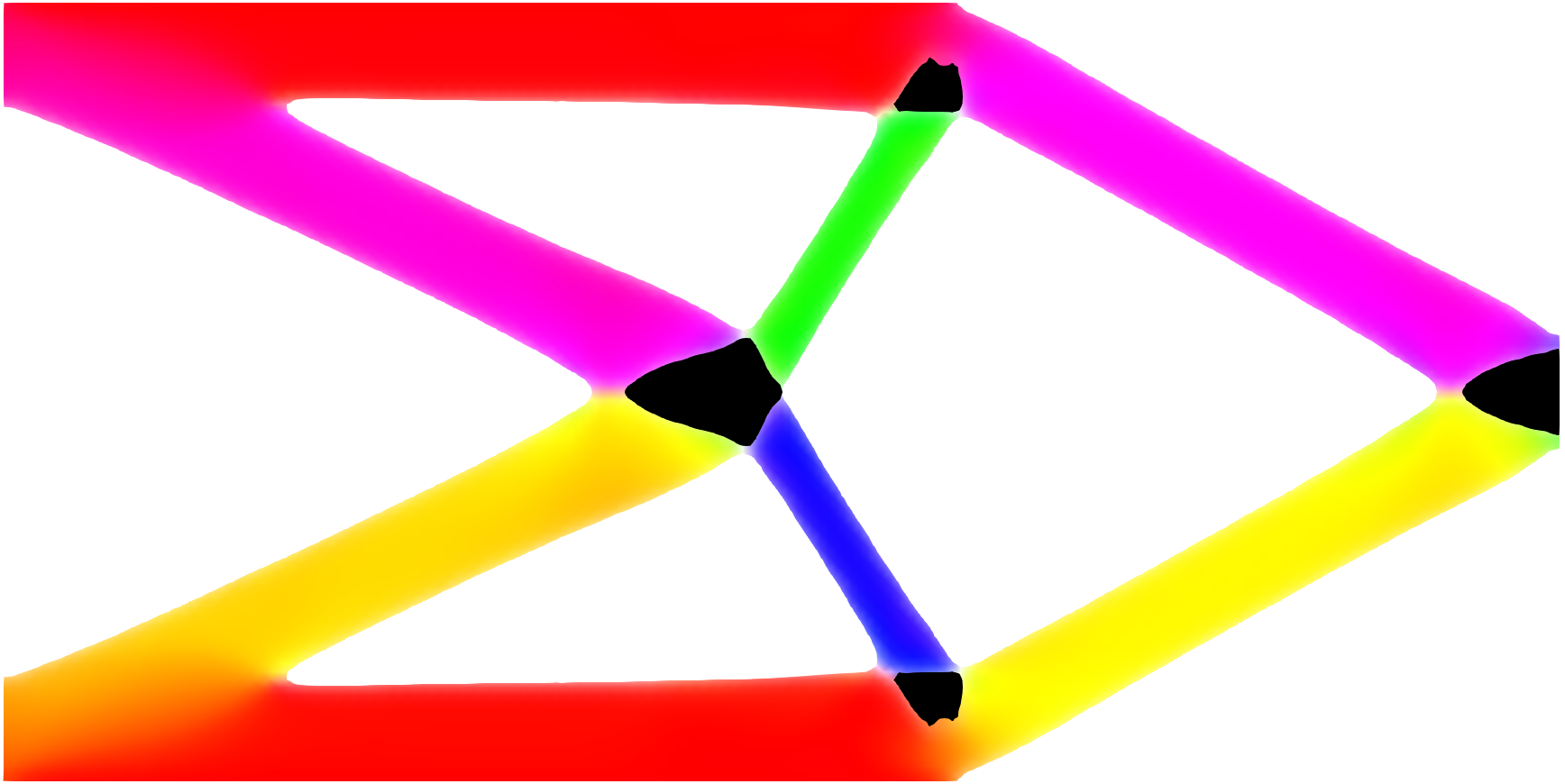}
			\subcaption{$E^\text{back}/E^\text{fib}=0.2$}
		\end{minipage}
		\begin{minipage}[b]{0.3\linewidth}
			\centering
			\includegraphics[width=\linewidth]{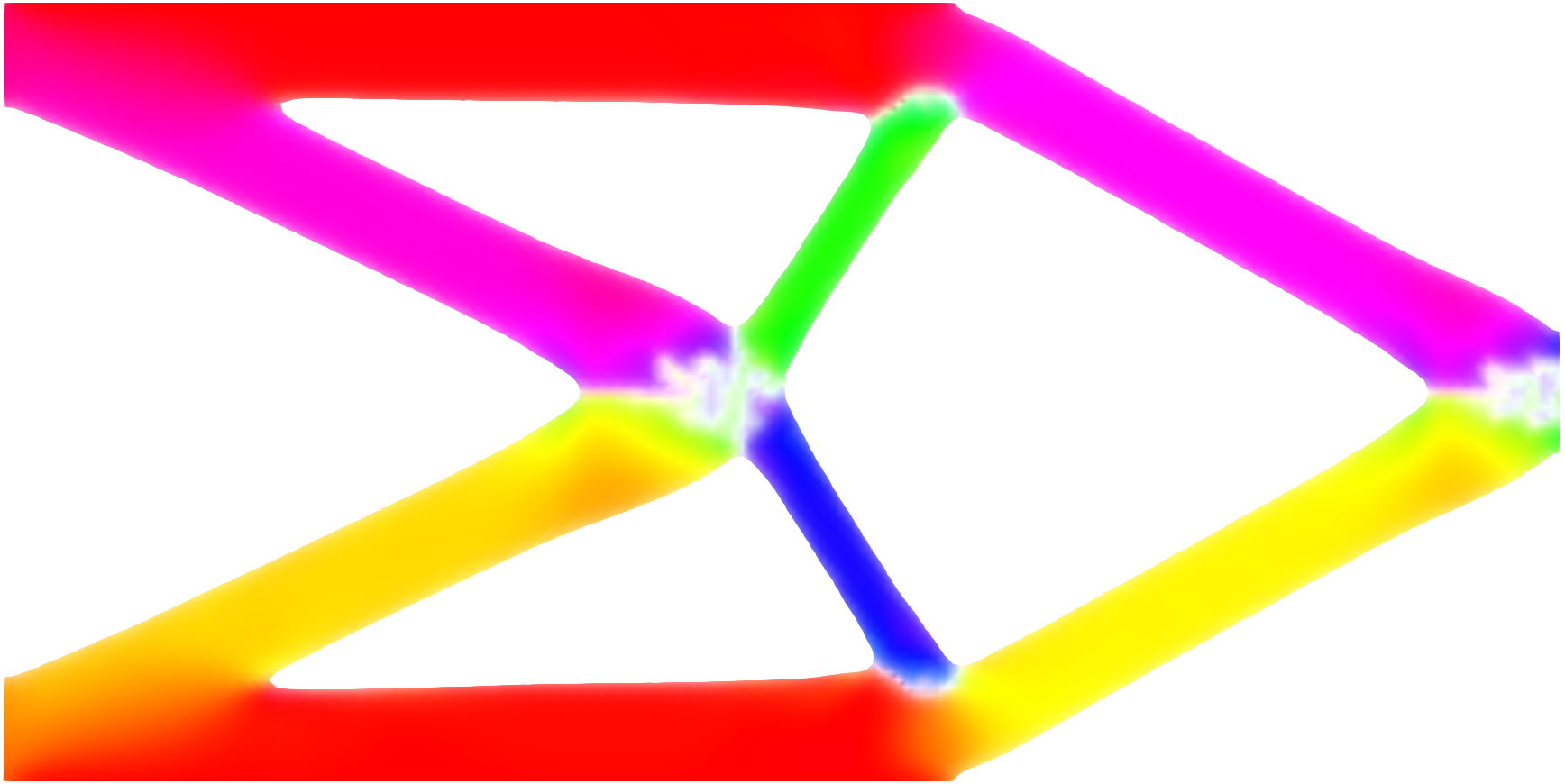}
			\subcaption{$E^\text{back}/E^\text{fib}=0.5$}
		\end{minipage}
		\begin{minipage}[b]{0.3\linewidth}
			\centering
			\includegraphics[width=\linewidth]{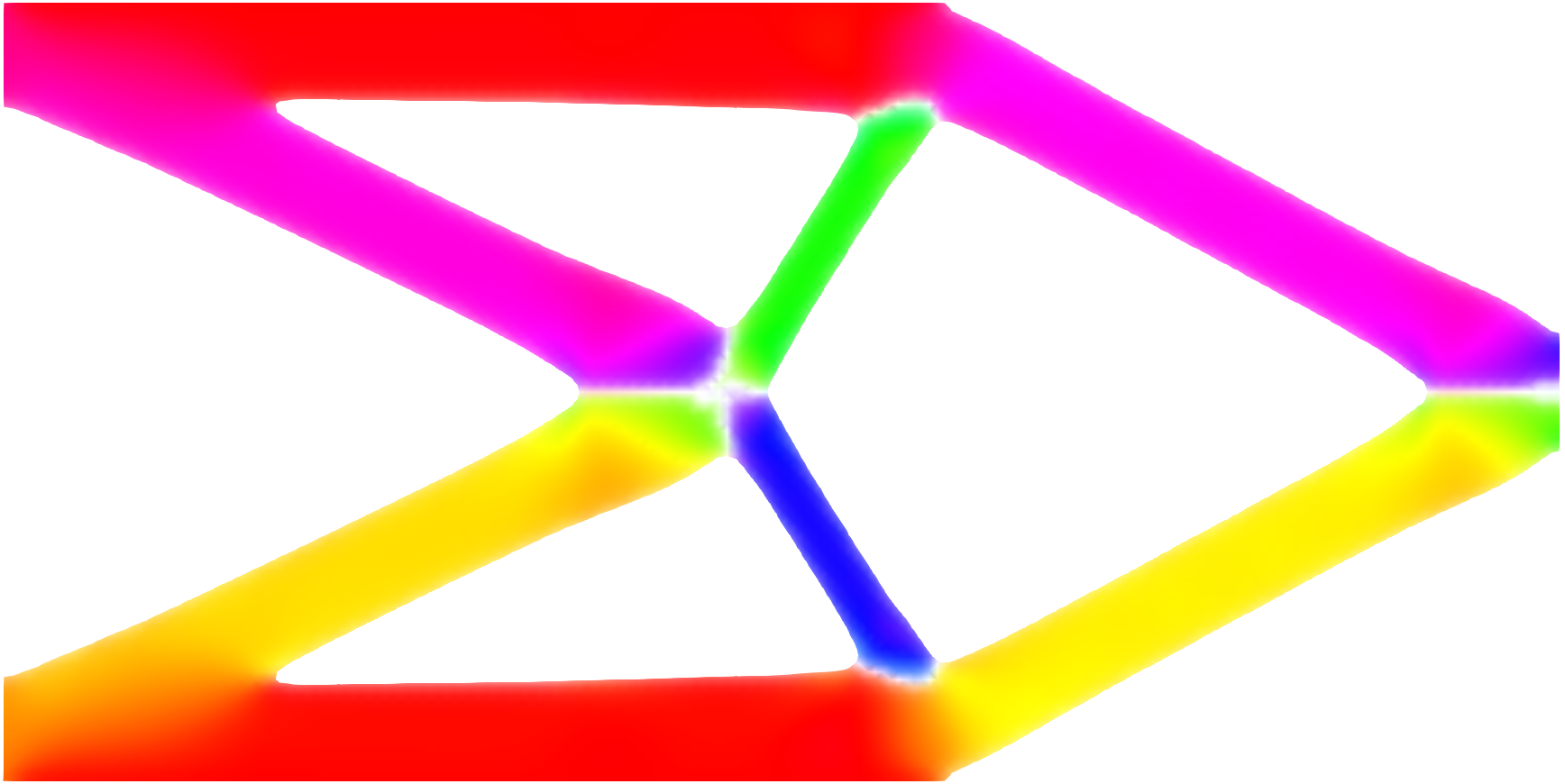}
			\subcaption{$E^\text{back}/E^\text{fib}=0.9$}
		\end{minipage}
		\caption{Optimization result for different value of contrast $E^\text{back}/E^\text{fib}$.}\label{fig:4varanisoContrast}
	\end{figure}

	Next, the initial design is set to initial design A defined in Eq.~\eqref{eq:4init0}, and for various anisotropic contrast $E^\text{back}/E^\text{fib}$, optimization is conducted. The optimization results are shown in Fig.~\ref{fig:4varanisoContrast}.
	In all cases shown in Fig.~\ref{fig:4varanisoContrast}, external shapes are almost the same. In Figs.~\ref{fig:4varanisoContrast}(a,b,c), in which the background material is weak, isotropic material is inserted at the connection point of rods. In Fig.  \ref{fig:4varanisoContrast}(d), the orientation is not obviously determined in the center point. This indicates that any orientation does not significantly affect the stiffness, and in such a region, the orientation angle can be determined appropriately from the viewpoint of manufacturability and other factors.
	
	Finally, the initial design is set to initial design A defined in Eq.~\eqref{eq:4init0}, and for various values of Young's modulus for isotropic material $E^\text{I}$, the material configurations are optimized. The optimization results are shown in Fig.~\ref{fig:4varanisoContrast}.
	\begin{figure}
		\centering
		\begin{minipage}[b]{0.3\linewidth}
			\centering
			\includegraphics[width=\linewidth]{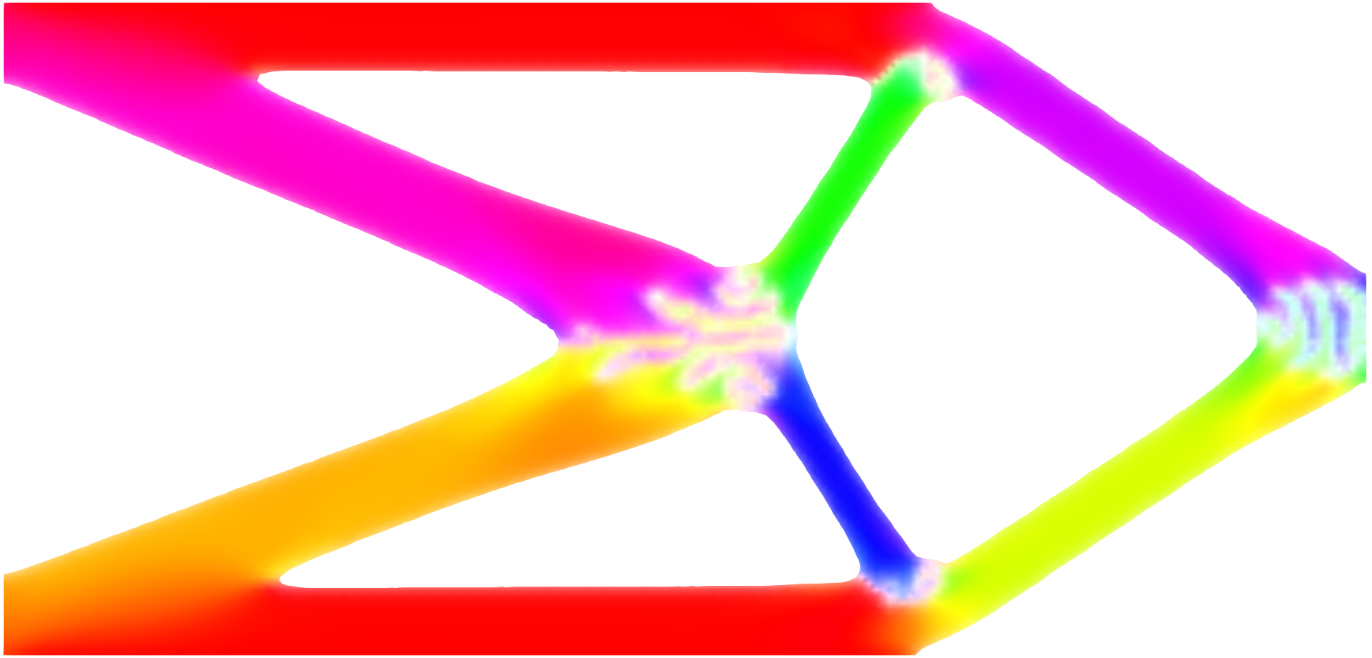}
			\subcaption{$E^\text{I}=50$[GPa]}
		\end{minipage}
		\begin{minipage}[b]{0.3\linewidth}
			\centering
			\includegraphics[width=\linewidth]{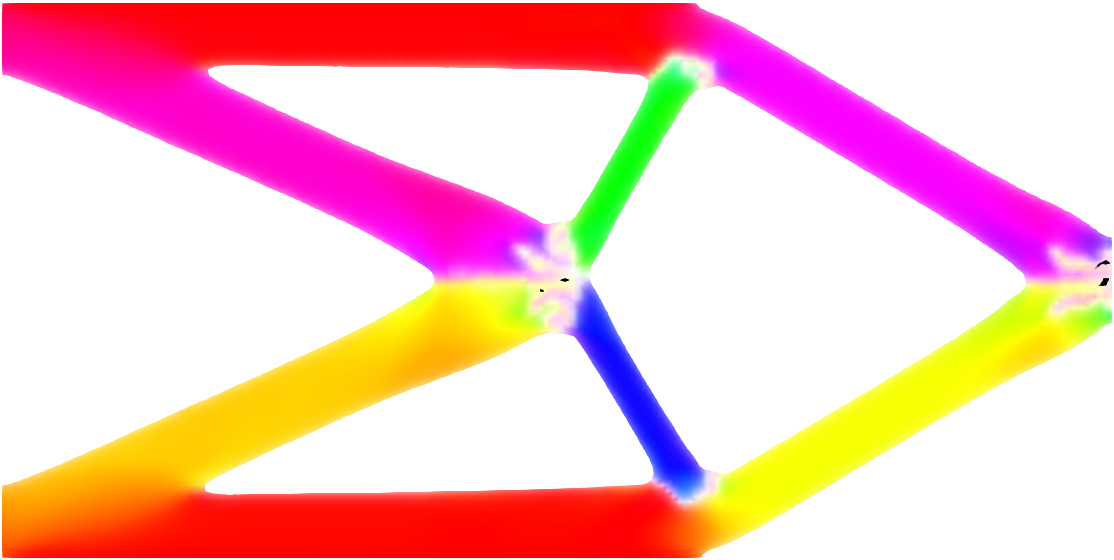}
			\subcaption{$E^\text{I}=70$[GPa]}
		\end{minipage}
		\begin{minipage}[b]{0.3\linewidth}
			\centering
			\includegraphics[width=\linewidth]{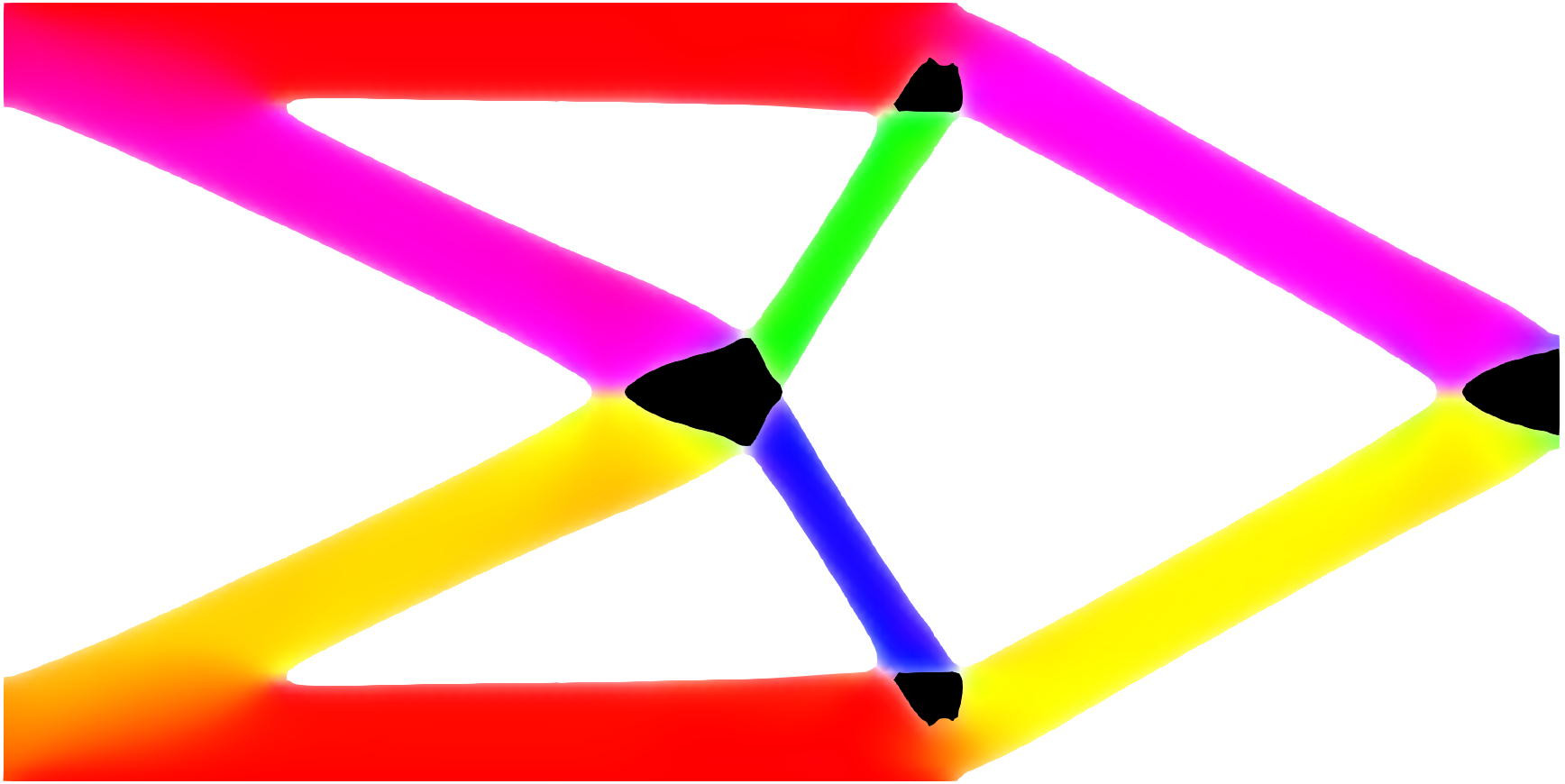}
			\subcaption{$E^\text{I}=80$[GPa]}
		\end{minipage}
		\begin{minipage}[b]{0.3\linewidth}
			\centering
			\includegraphics[width=\linewidth]{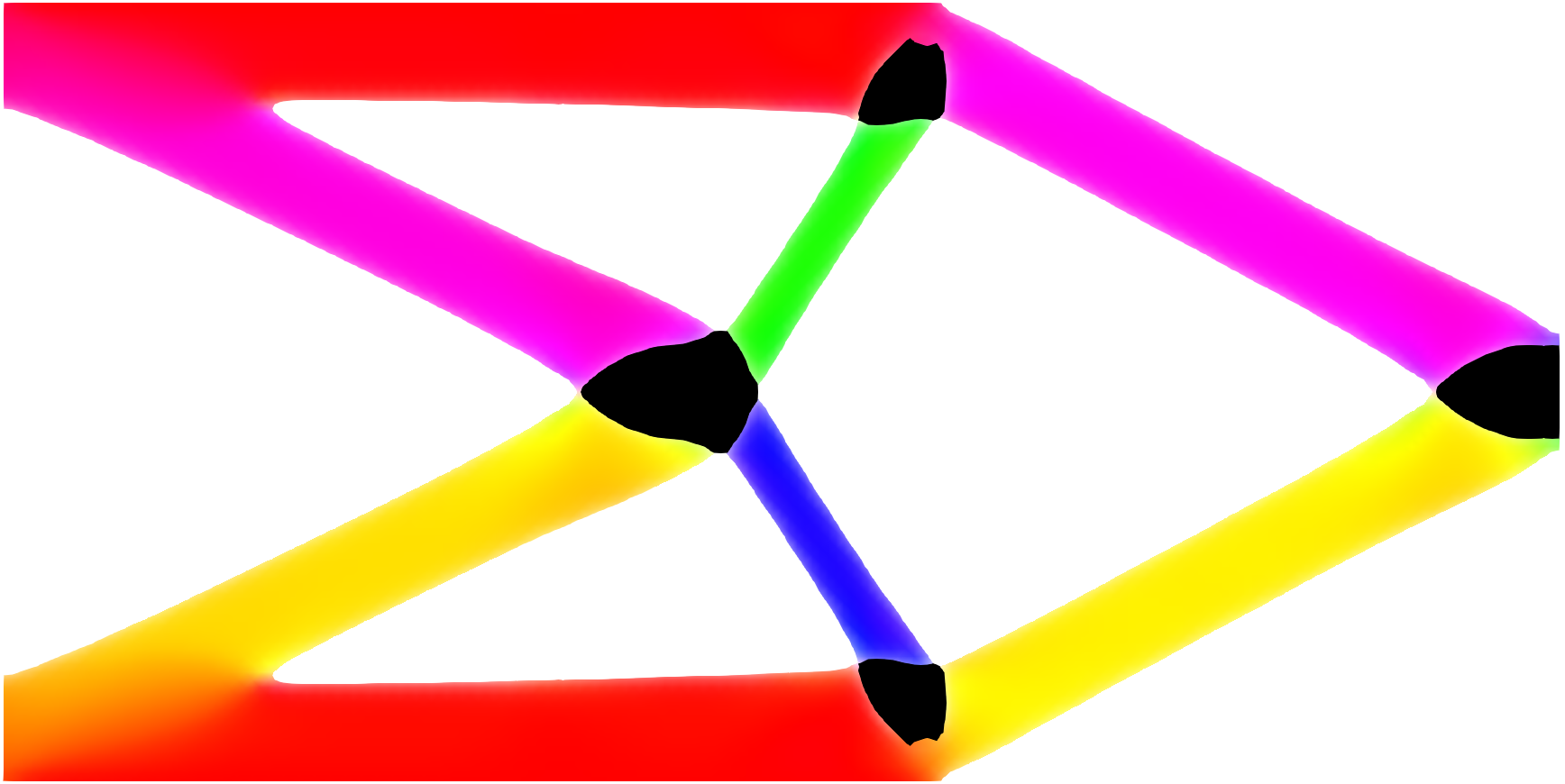}
			\subcaption{$E^\text{I}=90$[GPa]}
		\end{minipage}
		\begin{minipage}[b]{0.3\linewidth}
			\centering
			\includegraphics[width=\linewidth]{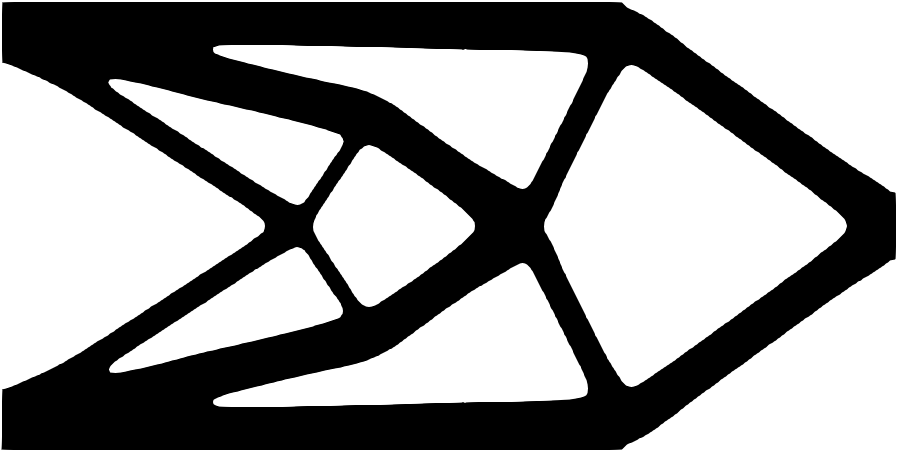}
			\subcaption{$E^\text{I}=100$[GPa]}
		\end{minipage}
		\caption{Optimization result for different values of Young's modulus for isotropic material $E^\text{I}$.}\label{fig:4varIso}
	\end{figure}
	As Young's modulus of the isotropic material increases, its area increases, which is reasonable in view of the objective of increasing stiffness.


	\section{Conclusion}\label{sec:4 Conclusion}
	In this paper, we proposed a new optimization method of multi-material topology and fiber orientation which avoids local optima.
	The study results are summarized below.
	\begin{enumerate}
		\item The topology of multiple materials is optimized based on the X-LS method.
		\item The orientation is expressed by relaxed Cartesian representation using auxiliary variables.
		\item Using the topological derivative for anisotropic materials, optimal orientation is estimated. 
		\item The X-LS functions and the auxiliary variables are optimized by optimally oriented topological derivatives and estimated optimal orientation.
		\item The proposed method is applied to the elasticity problem and it is shown that the proposed method can be used to obtain improved solutions independent of the initial design.
	\end{enumerate}

\section*{Acknowledgment}
Funding: 
This work was supported by JST FOREST Program (Grant Number JPMJFR202J, Japan).



\clearpage
\bibliographystyle{elsarticle-num} 
\bibliography{tex/reference.bib}

%
%
%
\end{document}